\documentclass[11pt,letterpaper]{article}
\pdfoutput=1 
\usepackage[T1]{fontenc}
\usepackage[utf8]{inputenc}
\usepackage{lmodern}
\usepackage[scaled=0.95]{helvet} 
\usepackage{textcomp}
\usepackage[margin=1in]{geometry}
\usepackage{amsmath,amssymb}
\usepackage[final]{microtype}
\usepackage{array}
\usepackage{calc}
\usepackage{longtable}
\usepackage{booktabs}
\usepackage{pdflscape}
\usepackage{needspace}
\usepackage{caption}
\usepackage{xcolor}
\usepackage{tikz}
\usetikzlibrary{arrows.meta}
\usepackage{xurl} 
\usepackage[pdfusetitle,colorlinks=true,linkcolor={blue!50!black},citecolor={blue!50!black},urlcolor={blue!60!black},bookmarksnumbered=false,breaklinks=true]{hyperref}
\usepackage{bookmark}
\providecommand{\tightlist}{\setlength{\itemsep}{1pt}\setlength{\parskip}{0pt}}
\newcommand{\cbr}{\penalty0\relax}
\newcommand{\cbrsoft}{\penalty300\relax}
\newlength{\cslhangindent}
\newlength{\csllabelwidth}
\newlength{\cslentryspacingunit}
\newenvironment{CSLReferences}[2]
 {\setlength{\parindent}{0pt}\raggedright
  \ifodd #1
  \let\oldpar\par
  \def\par{\hangindent=\cslhangindent\oldpar}
  \fi
  \setlength{\parskip}{#2\cslentryspacingunit}}
 {}

\hypersetup{
  pdftitle={Do Open-Source Video Ad Clients Follow VAST 4.3? An Empirical Study of Player Conformance},
  pdfauthor={Prashant Chaudhary},
  pdfsubject={cs.SE},
  pdfkeywords={}
}

\title{Do Open-Source Video Ad Clients Follow VAST 4.3? An Empirical Study of Player Conformance}
\author{Prashant Chaudhary\\[2pt]
  \normalsize Independent Researcher\\[1pt]
  \normalsize \href{mailto:prashant@prashantchaudhary.com}{prashant@prashantchaudhary.com}\\[1pt]
  \normalsize ORCID: \href{https://orcid.org/0009-0001-1980-109X}{0009-0001-1980-109X}}
\date{29 October 2026}

\begin{document}
\maketitle

\begin{abstract}

\textbf{Context.} Video ads reach streaming and connected-TV apps through VAST documents, and companies count ads from the HTTP requests the player's client library sends. A tracking error can leave playback intact and surface later as disagreeing counts.

\textbf{Objective.} To measure how far open-source VAST clients meet the player duties of VAST 4.3, whether verdicts depend on the runtime, what generated inputs add, and when deviations were introduced.

\textbf{Method.} A catalog of 108 requirements was extracted from the specification. Six open-source clients in eight configurations ran against 69 crafted test vectors and the IAB's sample tags, under Node.js and in five browsers on three engines, with every request recorded. Grids, fuzzing and a sweep over 36 releases extend the catalog; Google's closed IMA SDK is a comparison.

\textbf{Results.} The open-source clients have 78 findings and 3 to 21 must-level deviations per configuration. None of the tested open-source clients that follow redirect (Wrapper) chains fills the required macros in a Wrapper's request; the archived IMA 3.794.0 configuration tested does. One library's timeout throws an exception no caller can catch, ending a Node.js process. In a pod (ads played back to back), one player credits a finishing ad's completion to the next ad, fixed in release 5.3.0. A library can conform under jsdom and deviate in every tested Chrome- and Safari-family browser.

\textbf{Conclusions.} The tested clients diverge on the duties other companies use to count and diagnose ads. A shared conformance suite run in real browsers could expose most graded deviations; the harness and vectors are released.

\textbf{Keywords:} VAST; conformance testing; specification-based testing; test oracles; ad SDKs; empirical study

\end{abstract}

\hypertarget{introduction}{%
\section{1. Introduction}\label{introduction}}

Many standards are implemented many times over, often by small libraries with no published check against the text. When such a library is wrong, the consequence often falls on a party that never runs it. Video advertising is an extreme case. An ad in a streaming service or a connected-TV app reaches the screen through a VAST document (Video Ad Serving Template) (\protect\hyperlink{ref-iab2022vast43}{IAB Technology Laboratory 2022}). The ad servers in the chain are run by different companies. The first often does not have the ad and returns a Wrapper, a document that points to another company's server; the player follows the chain to an InLine document that holds the ad. As the ad plays, the player's client library sends HTTP requests back to every server in the chain. It sends an impression request when the ad starts, quartile and completion requests as it plays, and an error request when something fails (Figure 1). Advertisers pay, and ad servers diagnose their supply, from those requests. A library that sends them wrongly breaks nothing on screen; the ad plays and the content resumes.

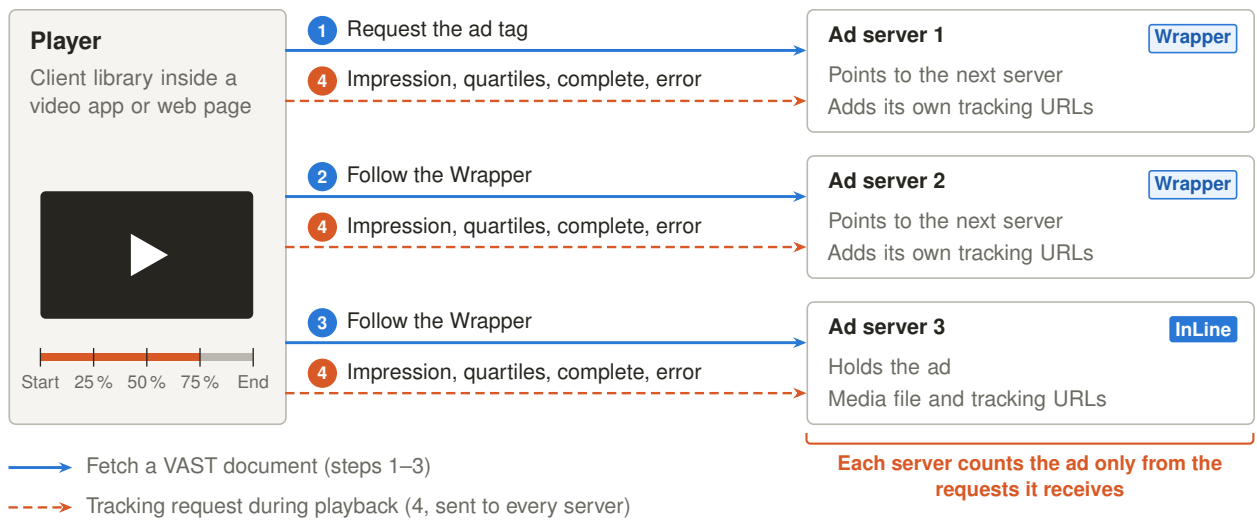
\begin{figure}[t]
\centering
\definecolor{mmBlue}{HTML}{2A78D6}\definecolor{mmBlueDk}{HTML}{1C5CAB}\definecolor{mmBlueLt}{HTML}{EAF2FC}
\definecolor{mmOrange}{HTML}{D95926}\definecolor{mmOrangeLt}{HTML}{FDEEE6}
\definecolor{mmInk}{HTML}{26251F}\definecolor{mmMuted}{HTML}{6B6A62}\definecolor{mmRule}{HTML}{B9B7AE}\definecolor{mmPanel}{HTML}{F5F4F0}
\begin{tikzpicture}[
  font=\sffamily\fontsize{8.5}{10.5}\selectfont, text=mmInk,
  x=1pt, y=1pt,
  card/.style={draw=mmRule, fill=white, line width=0.6pt, rounded corners=2.5pt, anchor=south west, inner sep=0pt},
  fetch/.style={-{Stealth[length=5pt,width=4pt]}, line width=1.1pt, draw=mmBlue},
  track/.style={-{Stealth[length=5pt,width=4pt]}, line width=0.9pt, draw=mmOrange, dash pattern=on 3.5pt off 2pt},
  num/.style={circle, inner sep=0pt, minimum size=11pt, text=white, font=\sffamily\bfseries\fontsize{7}{7}\selectfont},
  doc/.style={draw, line width=0.6pt, rounded corners=1pt, inner sep=2pt, font=\sffamily\bfseries\fontsize{7.5}{8}\selectfont}
]
\node[card, fill=mmPanel, minimum width=104pt, minimum height=156pt] (pl) at (0,0) {};
\node[anchor=north west, inner sep=0pt, font=\sffamily\bfseries\fontsize{9.5}{11}\selectfont] at (8,148) {Player};
\node[anchor=north west, inner sep=0pt, text width=90pt, text=mmMuted] at (8,134)
  {Client library inside a video app or web page};
\fill[mmInk, rounded corners=2pt] (12,40) rectangle (92,88);
\fill[white] (46,56) -- (46,72) -- (60,64) -- cycle;
\fill[mmRule] (12,24) rectangle (92,27);
\fill[mmOrange] (12,24) rectangle (72,27);
\foreach \q/\l in {0/0,20/25,40/50,60/75,80/100}
  \draw[mmInk, line width=0.5pt] (12+\q,22) -- (12+\q,29);
\foreach \q/\l in {0/Start,20/25\,\%,40/50\,\%,60/75\,\%,80/End}
  \node[anchor=north, inner sep=0pt, text=mmMuted, font=\sffamily\fontsize{7}{8}\selectfont] at (12+\q,19) {\l};
\def\S#1#2#3#4#5#6{%
  \node[card, minimum width=168pt, minimum height=46pt] (#1) at (300,#2) {};
  \node[anchor=north west, inner sep=0pt, font=\sffamily\bfseries\fontsize{8.5}{10}\selectfont] at (308,#2+40) {#3};
  \node[doc, #5, anchor=north east] at (462,#2+41) {#4};
  \node[anchor=north west, inner sep=0pt, text width=154pt, text=mmMuted] at (308,#2+25) {#6};}
\S{s1}{110}{Ad server 1}{Wrapper}{draw=mmBlue, text=mmBlueDk, fill=mmBlueLt}{Points to the next server\\[1.5pt] Adds its own tracking URLs}
\S{s2}{55}{Ad server 2}{Wrapper}{draw=mmBlue, text=mmBlueDk, fill=mmBlueLt}{Points to the next server\\[1.5pt] Adds its own tracking URLs}
\S{s3}{0}{Ad server 3}{InLine}{draw=mmBlue, fill=mmBlue, text=white}{Holds the ad\\[1.5pt] Media file and tracking URLs}
\def\R#1#2#3#4{%
  \draw[fetch] (104,#1+31) -- (300,#1+31);
  \node[num, fill=mmBlue] at (118,#1+38.5) {#2};
  \node[anchor=west, inner sep=0pt] at (127,#1+38.5) {#3};
  \draw[track] (104,#1+12) -- (300,#1+12);
  \node[num, fill=mmOrange] at (118,#1+19.5) {4};
  \node[anchor=west, inner sep=0pt] at (127,#1+19.5) {#4};}
\R{110}{1}{Request the ad tag}{Impression, quartiles, complete, error}
\R{55}{2}{Follow the Wrapper}{Impression, quartiles, complete, error}
\R{0}{3}{Follow the Wrapper}{Impression, quartiles, complete, error}
\draw[mmOrange, line width=0.8pt, rounded corners=1.5pt] (300,-3) -- (300,-7) -- (468,-7) -- (468,-3);
\node[anchor=north, inner sep=0pt, text=mmOrange, font=\sffamily\bfseries\fontsize{8}{9.5}\selectfont] at (384,-11)
  {Each server counts the ad only from the};
\node[anchor=north, inner sep=0pt, text=mmOrange, font=\sffamily\bfseries\fontsize{8}{9.5}\selectfont] at (384,-21)
  {requests it receives};
\draw[fetch] (0,-16) -- (24,-16);
\node[anchor=west, inner sep=0pt, text=mmMuted] at (29,-16) {Fetch a VAST document (steps 1--3)};
\draw[track] (0,-31) -- (24,-31);
\node[anchor=west, inner sep=0pt, text=mmMuted] at (29,-31) {Tracking request during playback (4, sent to every server)};
\end{tikzpicture}
\vspace{6pt}
\caption{\textbf{Figure 1.} How a VAST ad reaches the screen and is counted. The player follows a chain of Wrappers to the InLine ad, then sends impression, progress and error requests to every server in the chain; each server counts the ad only from the requests it receives.}
\end{figure}

A pod is a group of ads played back to back, each with its own completion URL. A player receives a pod of three ads and plays all three without a visible problem. On the wire, the first ad's completion request is never sent, the second ad's is sent when the second ad starts, and the third ad's is sent twice. The total number of completions is right, but the first ad is never credited, the second is credited early and the third twice, and a check of the pod's total alone shows nothing wrong. The player is Shaka Player 5.2.12; its 5.3.0 release fixes this case and adds one regression (Section 5.1.3). A second case is a crash. Some companies resolve VAST chains on a server, for example for server-side ad insertion, which stitches ads into the video stream itself. There, a slow server anywhere in a chain makes one widely used library throw an exception that no caller can catch, and the whole Node.js process exits.

The libraries are widely used: in September 2026 the most downloaded of the dedicated VAST libraries, vast-client-js, was downloaded about 237,000 times from npm, and Shaka Player, whose ad module is one of the clients tested, about 1.6 million times. Downloads count package fetches by developers and build systems, not players in the field, so they indicate use, not deployment. VAST is an open specification, but its conformance is self-declared. IAB Technology Laboratory (IAB Tech Lab), the industry body that maintains VAST, publishes the specification, sample tags and a tag validator; a validator checks the document an ad server sends. This study found no published suite that checks what a player does with that document, and no academic study known to the author has tested VAST clients against the specification (Section 3). This paper provides that study for six open-source VAST client libraries used in web and Node.js players. It asks four research questions:

\begin{itemize}
\tightlist
\item
  \textbf{RQ1 (Conformance).} To what extent do open-source VAST clients meet the VAST 4.3 player duties they take on (some libraries leave duties such as rendering to the integrating player), where do their deviations cluster, and how does a closed production SDK compare?
\item
  \textbf{RQ2 (Runtime).} Which verdicts depend on the runtime and platform the client runs in, and why?
\item
  \textbf{RQ3 (Beyond the catalog).} Does one test vector per requirement predict a client's behavior on generated variants of the same failure, and what does mutation fuzzing find that the catalog does not?
\item
  \textbf{RQ4 (History).} When were the deviations introduced, and how long have they persisted?
\end{itemize}

\textbf{Study design.} The oracle is the specification text. The player obligations of VAST 4.3 and of the IAB macro list (the list of URL placeholders a player fills; Section 2) were extracted into a catalog of 108 requirements, each with the section it rests on and its strength as written (must, should or may; Appendix A). A recording server serves crafted test documents (the test vectors) and logs every request a client makes, and each requirement is graded mostly from what reaches the network (Figure 2). Each client runs through its documented integration in two layers: under Node.js (the Node.js layer) and in real browsers (the browser layer). The open-source clients are analyzed white-box: every deviation is traced to the source line that produces it, or to the absence of any code for the duty. Google's closed IMA SDK is a black-box reference, graded only from its network requests and the events of its public API (Section 5.1.5). Generated error and macro grids, a mutation fuzzer and a sweep over earlier releases test the catalog's coverage and date its findings (Section 4).

\textbf{Contributions.}

\begin{enumerate}
\def\labelenumi{\arabic{enumi}.}
\tightlist
\item
  The first empirical conformance study of VAST clients, to the author's knowledge: six open-source clients in eight configurations and one closed SDK, graded on 108 requirements in each layer the client runs in (Node.js, and five browsers on three engines), with 86 findings (78 in the open-source clients, 8 in IMA). Each finding is tied to its requirements where the catalog has one, to a source location in the open-source clients or to the observable evidence for IMA, and, where the sweep covers it, to a release range (Appendix B).
\item
  A cross-client analysis showing that deviations concentrate on the requests other companies rely on (macros in redirect requests, error codes, pods, progress events and verification), plus robustness defects outside the specification (Sections 5.1 and 5.3).
\item
  Evidence that a library's verdict can depend on the XML parser of the runtime it runs in. A test suite that runs only under jsdom, as those of vast-client-js and video-ad-sdk do, therefore cannot see a class of defects (Section 5.2).
\item
  A released harness, vector set, requirement catalog and result data, packaged to rerun every layer, the grids and the access check on another machine (Statements and Declarations).
\end{enumerate}

\textbf{Scope.} The study covers client-side insertion of linear video ads by JavaScript libraries, which is how web players and many connected-TV web apps play VAST. Server-side ad insertion, interactive formats, companion and overlay ads (shown beside or over the content), and native mobile and TV SDKs are outside it; the coverage map in the replication package records each duty left out and why.

\hypertarget{what-vast-asks-of-a-player}{%
\section{2. What VAST asks of a player}\label{what-vast-asks-of-a-player}}

VAST 4.3 states that compliance is a two-party effort: ad servers send valid responses, and media players play the ad as instructed, honor the instructions of every server in a redirect chain, and send tracking information back (§2.1, §2.2). The duties this study tests fall into eight groups; Section 4.2 regroups them into nine for reporting, adding robustness.

\textbf{Wrappers (§2.3.5, §3.19).} The player follows each Wrapper's VASTAdTagURI until it reaches an InLine document. It is required to accept five Wrappers (§3.19.1) and may reject more, reporting error 302 to every Wrapper's Error URI (§2.3.5.1). Wrapper attributes constrain the chain. With \texttt{follow\cbrsoft{}Additional\cbrsoft{}Wrappers=\cbr{}\textquotedbl{}false\textquotedbl{}}, a further Wrapper is to be ignored. allowMultipleAds defaults to false, which means that only the first stand-alone ad (one that is not part of an ad pod, described below) of the next response may be used. fallbackOnNoAd asks the player to try another ad when the chain yields none. When the InLine has no ClickThrough URL, the one in the Wrapper closest to the InLine is used (§2.3.5.2).

\textbf{Errors (Error Reporting section, §3.2.1, §3.4.11).} When an ad cannot be served, the player requests the Error URIs and replaces the {[}ERRORCODE{]} macro with a code from the specification's table. The codes include 100 for an XML parsing error, 301 for an unavailable or timed-out Wrapper URI, 302 for the Wrapper limit, 303 for no ad after Wrappers and 403 for no supported media file. An error in an InLine ad reached through Wrappers is reported to every Wrapper as well. A response with no ad may carry document-level Error URIs, which the player must then request.

\textbf{Tracking (§2.3.4, §3.4.4, §3.14).} All Impression URIs of the InLine and of every Wrapper before it are requested at the same time, or as close to it as possible. Linear tracking events (start, firstQuartile, midpoint, thirdQuartile, complete) fire when playback reaches 0, 25, 50, 75 and 100 percent; progress events fire at an offset given as a time or a percentage. Pause, resume, mute, unmute, rewind, skip and loaded each have an event. Wrappers can carry their own tracking elements, which fire alongside the InLine ones.

\textbf{Macros (§6 and the IAB macro list (\protect\hyperlink{ref-iab2026vastmacros}{IAB Technology Laboratory 2026})).} URLs in a VAST document carry macros, bracketed names such as {[}TIMESTAMP{]} that the requesting party replaces with a value. The list marks five as required: {[}TIMESTAMP{]} (ISO 8601), {[}CACHEBUSTING{]} (a random 8-digit integer), {[}ERRORCODE{]}, {[}LIMITADTRACKING{]} (whether the user has limited ad tracking) and {[}REASON{]}. CACHEBUSTING and TIMESTAMP apply to VAST request URIs as well as to tracking URLs, and the party that makes the request, usually the player, replaces them (§6.1). Values are percent-encoded one by one. Listed optional macros that the player does not fill become \(-1\) (value unknown) or \(-2\) (value known but withheld by policy), but macros the player does not know are not to be replaced with \(-1\) (§6.1).

\textbf{XML and parsing (§2.2, §2.3.2).} The response is well-formed XML 1.0, and comments and character entities are honored. From VAST 4.2 a creative can carry several UniversalAdId elements, and from 4.1 a MediaFiles element several ClosedCaptionFile elements.

\textbf{Ad pods.} Ads with a sequence attribute form a pod. A player that plays a pod ``must play the ads in the Pod in the prescribed sequence and should play as many of the ads as possible.'' It moves on when one fails and should substitute an unplayed stand-alone ad from the same response. Pod support is optional, as is the skippable linear ad type (§2.3.1).

\textbf{Verification and icons (§3.11, §3.16, §3.17).} Verification vendors are third parties that measure an ad with their own code run in the player. A player that does not run that code must request that vendor's verificationNotExecuted URI with a {[}REASON{]} code. The code is 1 when the player declines the resource, 2 when it does not support the API framework (the interface that code expects, such as the Open Measurement Interface Definition, OMID) and 3 when the resource fails to load. When icons (small images shown over the ad) are present, the player displays them after the ad starts and requests their view and click URLs.

\textbf{Media and versions (MediaFiles, §4.3).} The player chooses a media file by its attributes, not its position, and VAST 4 players should keep accepting older documents.

\hypertarget{related-work}{%
\section{3. Related work}\label{related-work}}

\textbf{Specification-derived oracles.} In the taxonomy of Barr et al. (\protect\hyperlink{ref-barr2015oracle}{2015}), this study's catalog is a specified oracle, its crash detection is an implicit oracle, and its differential pass is a derived one. The closest methodological precedent is RFCcert (\protect\hyperlink{ref-chen2018rfc}{Chen et al. 2018}; \protect\hyperlink{ref-tian2019rfc}{Tian et al. 2019}), which extracts certificate rules from the RFC text and runs targeted certificates through SSL/TLS libraries. Like RFCcert, this study takes the expected behavior from the standard instead of from agreement between implementations. It differs in three ways. Requirements are extracted by hand and kept as a citable catalog with their obligation strengths. The observed behavior is network traffic during playback, where RFCcert observes an accept or reject verdict. No prior test corpus exists. Conformance testing of many independent implementations of one protocol is established practice in security research (\protect\hyperlink{ref-deruiter2015protocol}{de Ruiter and Poll 2015}; \protect\hyperlink{ref-somorovsky2016tls}{Somorovsky 2016}; \protect\hyperlink{ref-sivakorn2017hvlearn}{Sivakorn et al. 2017}). This study transfers it to a domain whose implementations are small, whose specification is prose with a partial schema, and whose harm is economic.

\textbf{Placing the method.} Grading a client from a record of the requests it sends is log-file analysis used as a test oracle (\protect\hyperlink{ref-andrews2003logfile}{Andrews and Zhang 2003}): each oracle reads the request log of a run, not the client's internal state. The strength labels use the requirement levels must, should and may as RFC 2119 defines them (\protect\hyperlink{ref-bradner1997rfc2119}{Bradner 1997}); VAST 4.3 does not invoke RFC 2119, so the rule of Section 4.2 decides where its text has no such word. RQ3's question, whether a client gives the same code for the same failure at every depth and level, is an invariance relation of the kind metamorphic testing uses (\protect\hyperlink{ref-segura2016metamorphic}{Segura et al. 2016}), with the expected value taken from the specification. The robustness oracle follows Ballista (\protect\hyperlink{ref-kropp1998ballista}{Kropp et al. 1998}) in classifying how off-the-shelf components fail under exceptional inputs, from crashes and hangs to wrong error reports. A shared suite of player vectors, as proposed in Section 6.2, has a precedent in web-platform-tests, a test suite written to run in every browser that browser projects use to check their implementations of the Web platform specifications (\protect\hyperlink{ref-wpt2026}{web-platform-tests contributors 2026}).

\textbf{Differential testing.} Differential testing (\protect\hyperlink{ref-mckeeman1998differential}{McKeeman 1998}) takes disagreement between implementations as the bug signal, from compilers (\protect\hyperlink{ref-yang2011csmith}{Yang et al. 2011}) and virtual machines (\protect\hyperlink{ref-chen2016jvm}{Chen et al. 2016}) to certificate validation (\protect\hyperlink{ref-brubaker2014frankencerts}{Brubaker et al. 2014}) and guided input generation (\protect\hyperlink{ref-petsios2017nezha}{Petsios et al. 2017}). Herbold and Tunkel (\protect\hyperlink{ref-herbold2023differential}{2023}) found across machine-learning frameworks that equivalent configurations are hard to establish and that most raw differences are noise. VAST clients raise a sharper form of that problem: they take on different shares of a player's duties, and several share code (one player in this study bundles another library). A majority vote would hide a defect two clients share, so this study grades against the text and uses disagreement only to triage mutated documents.

\textbf{Conformance and consistency studies of library families.} Kuchta et al. (\protect\hyperlink{ref-kuchta2018pdf}{2018}) study inconsistencies among PDF readers, the closest precedent in form: many consumers of one document format, with divergence as the object of study. Harrand et al. (\protect\hyperlink{ref-harrand2021json}{2021}) measure how Java JSON libraries differ on well-formed and ill-formed input; Davis et al. (\protect\hyperlink{ref-davis2019regex}{2019}) show that regular expressions are not portable across engines; Lima et al. (\protect\hyperlink{ref-lima2021javascript}{2021}) transplant test suites between JavaScript engines and find bugs each engine's own suite missed. Wang et al. (\protect\hyperlink{ref-wang2026sbom}{2026}) study whether software bill of materials (SBOM) tools adhere to their standards, with a longitudinal stage that parallels this study's version sweep, and Prado et al. (\protect\hyperlink{ref-prado2026sbom}{2026}) attribute most divergence among SBOM generators to design choices and specification ambiguity rather than implementation errors. In HTTP, differential fuzzing of parsers has found discrepancies whose victim is a third party (\protect\hyperlink{ref-jabiyev2021treqs}{Jabiyev et al. 2021}; \protect\hyperlink{ref-kallus2024httpgarden}{Kallus et al. 2024}), the same structure as a VAST client whose errors land on an ad server's counts. This study adds a requirement-level catalog, observes side effects instead of outputs, and dates each finding by release.

\textbf{Ad SDKs and ad measurement.} Ahasanuzzaman et al.~study how mobile ad libraries are released and updated (\protect\hyperlink{ref-ahasanuzzaman2020longitudinal}{2020}) and how apps integrate them (\protect\hyperlink{ref-ahasanuzzaman2022integration}{2022}); Gui et al. (\protect\hyperlink{ref-gui2015truth}{2015}) measure the hidden costs ads impose on apps and users. FraudDroid (\protect\hyperlink{ref-dong2018frauddroid}{Dong et al. 2018}) and MAdFraud (\protect\hyperlink{ref-crussell2014madfraud}{Crussell et al. 2014}) detect deliberate ad fraud from network traffic, and Wu et al. (\protect\hyperlink{ref-wu2026splash}{2026}) show that mobile ad SDKs decide which interactions count as billable. Measurement studies of smart TVs and streaming devices map which tracking services apps contact (\protect\hyperlink{ref-mohajerimoghaddam2019watching}{Mohajeri Moghaddam et al. 2019}; \protect\hyperlink{ref-varmarken2020tv}{Varmarken et al. 2020}), and Marciel et al. (\protect\hyperlink{ref-marciel2016viewfraud}{2016}) show that video view counting can be inflated on the platform side. These studies share this paper's premise that counting is decided in client code; they look for fraud or for which hosts are contacted, while this study grades whether libraries that try to comply send the requests the specification requires.

\textbf{VAST.} Published work on VAST is descriptive or architectural (\protect\hyperlink{ref-seeliger2016hbbtv}{Seeliger et al. 2016}; \protect\hyperlink{ref-smith2016ott}{Smith 2016}). Chaudhary and Khandelwal (\protect\hyperlink{ref-chaudhary2026trigger}{2026}) separate trigger, readiness and playback events in accounting for dynamic ad insertion (ads inserted into a stream at playback time) on synthetic data; the present study checks whether real clients emit those events at all. The nearest work is Sekowski's VASTlint validator and its preprint (\protect\hyperlink{ref-sekowski2026vastxml}{Sekowski 2026a}), which measures whether structural validation of VAST tags fits inside a real-time bidding pipeline (the auction that selects an ad for each request). This study adds the client side: a validator grades the tag an ad server sends and cannot see a client that drops a correct request. The same project has since released vastlint-client (\protect\hyperlink{ref-sekowski2026vastlintclient}{Sekowski 2026b}), a session runtime that this study tests as a client. Some findings were known to maintainers in part: the vast-client-js timeout exception had been reported as a browser error (\protect\hyperlink{ref-8bitsten2026issue510}{8bitsten 2026}), and the missing macro replacement in Wrapper requests had been raised for vast-client-js and iab-vast-loader (\protect\hyperlink{ref-hwde2019issue70}{hwde 2019}; \protect\hyperlink{ref-doca2019issue331}{Doca 2019}). This study adds the wire evidence, the release range and the Node.js consequence.

\hypertarget{study-design}{%
\section{4. Study design}\label{study-design}}

The study follows the engineering-research and benchmarking standards of the ACM SIGSOFT Empirical Standards (\protect\hyperlink{ref-ralph2020standards}{Ralph et al. 2020}). Figure 2 shows the pipeline.

\begin{figure}[!htb]
\centering
\definecolor{mmBlue}{HTML}{2A78D6}\definecolor{mmBlueDk}{HTML}{1C5CAB}\definecolor{mmBlueLt}{HTML}{EAF2FC}
\definecolor{mmSlate}{HTML}{3F4A5A}\definecolor{mmSlateLt}{HTML}{F1F2F4}
\definecolor{mmInk}{HTML}{26251F}\definecolor{mmMuted}{HTML}{5E5D56}\definecolor{mmRule}{HTML}{B9B7AE}
\begin{tikzpicture}[
  font=\sffamily\fontsize{8.5}{10.5}\selectfont, text=mmInk,
  x=1pt, y=1pt,
  arr/.style={-{Stealth[length=5pt,width=4pt]}, line width=0.9pt, draw=mmInk},
  lab/.style={inner sep=1pt, text=mmMuted, font=\sffamily\fontsize{7.5}{9}\selectfont},
  chip/.style={draw=mmBlue, fill=mmBlueLt, text=mmBlueDk, line width=0.5pt, rounded corners=1.5pt,
               inner sep=0pt, minimum height=15pt, text height=6pt, text depth=1.5pt, font=\sffamily\fontsize{7.5}{8}\selectfont, anchor=south west}
]
\long\def\B#1#2#3#4#5#6#7#8{%
  \fill[white, rounded corners=2.5pt] (#2,#3) rectangle +(#4,#5);
  \fill[#6, rounded corners=2.5pt] (#2,#3+#5-15) rectangle +(#4,15);
  \fill[#6] (#2,#3+#5-15) rectangle +(#4,4);
  \draw[mmRule, line width=0.6pt, rounded corners=2.5pt] (#2,#3) rectangle +(#4,#5);
  \coordinate (#1-sw) at (#2,#3); \coordinate (#1-ne) at (#2+#4,#3+#5);
  \coordinate (#1) at (#2+#4/2,#3+#5/2);
  \node[anchor=base west, inner sep=0pt, text=white, font=\sffamily\bfseries\fontsize{8.5}{10}\selectfont] at (#2+6,#3+#5-10.5) {#7};
  \node[anchor=north west, inner sep=0pt, text width={#4pt-12pt}, align=left,
        execute at begin node={\hyphenpenalty=10000\exhyphenpenalty=10000\setlength{\parskip}{2.5pt}\everypar{\hangindent=6pt\hangafter=1\relax}}] at (#2+6,#3+#5-22) {#8\par};}
\fill[mmBlueLt, rounded corners=3pt] (0,96) rectangle (468,212);
\fill[mmSlateLt, rounded corners=3pt] (0,0) rectangle (468,84);
\node[rotate=90, text=mmBlueDk, font=\sffamily\bfseries\fontsize{8.5}{10}\selectfont] at (9,154) {Execution};
\node[rotate=90, text=mmSlate, font=\sffamily\bfseries\fontsize{8.5}{10}\selectfont] at (9,42) {Grading};
\B{in}{22}{104}{116}{100}{mmBlue}{Test inputs}{%
  69 crafted vectors\par 49 IAB sample tags\par Generated error and macro grids\par 1,442 fuzz cases}
\B{srv}{160}{104}{116}{100}{mmBlue}{Recording server}{%
  Serves the test documents\par Rewrites URLs to itself\par Logs every request\par
  Injects errors, delays, stalls and redirects}
\B{cli}{320}{104}{140}{100}{mmBlue}{Client adapters}{%
  6 open-source clients\par Google IMA (black box)\par
  {\color{mmMuted}Load script and 8 playback scripts (full, skip, click, pod, \dots)\par}}
\node[chip, minimum width=61pt] at (326,110) {Node.js + jsdom};
\node[chip, minimum width=61pt] at (393,110) {5 browsers};
\B{log}{22}{8}{74}{62}{mmSlate}{Request log}{Every request, timestamped}
\B{orc}{113}{8}{74}{62}{mmSlate}{Oracles}{One per catalog requirement (108)}
\B{ver}{204}{8}{74}{62}{mmSlate}{Verdicts}{Conforms, deviates, delegated, n/a}
\B{trc}{295}{8}{74}{62}{mmSlate}{Tracing}{Each open-source deviation traced to its source line}
\B{fnd}{386}{8}{74}{62}{mmSlate}{Findings}{Defects\par Release ranges\par Issue reports}
\draw[arr] (138,154) -- (160,154);
\draw[arr] (276,170) -- node[lab, above=1pt] {documents} (320,170);
\draw[arr] (320,138) -- node[lab, below=1pt] {requests} (276,138);
\foreach \a/\b in {96/113, 187/204, 278/295, 369/386}
  \draw[arr] (\a,39) -- (\b,39);
\draw[arr, rounded corners=3pt] (218,104) -- (218,90) -- (59,90) -- (59,70);
\end{tikzpicture}
\vspace{6pt}
\caption{\textbf{Figure 2.} Study pipeline. In the execution band, the recording server serves the test documents to each client, which runs under a load or playback script, and logs every request the client sends back. In the grading band, one oracle per catalog requirement reads the log and gives a verdict for each cell (one requirement on one client configuration in one runtime); every deviation of an open-source client is then traced to client source; IMA's are described from its requests and public API.}
\end{figure}
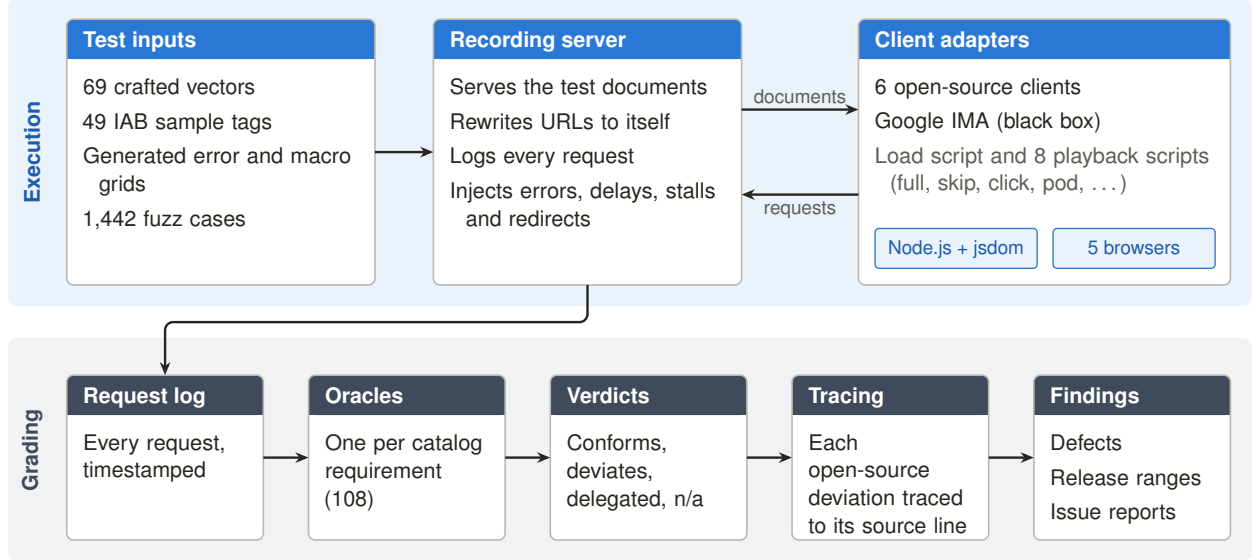

\hypertarget{subjects}{%
\subsection{4.1 Subjects}\label{subjects}}

\textbf{Survey.} Sixteen queries against the npm registry on October 16, 2026 returned 267 packages whose name, description or keywords mention VAST, the related IAB formats VMAP, VPAID and SIMID, or IMA; Online Resource 1 lists the queries, the packages and every decision. Each plausible client was classified from its registry metadata; the ten whose role the metadata left unclear were unpacked and searched for their own handling of VASTAdTagURI, TrackingEvents and {[}ERRORCODE{]}. A package counts as a VAST client when it (1) carries an OSI-approved license, (2) has its own VAST implementation rather than wrapping vast-client-js, video-ad-sdk or IMA, (3) resolves Wrappers or fires tracking, and (4) had a release in the 24 months before October 16, 2026.

\textbf{Sampling.} The sample is purposive. Three clients were chosen before the survey because they are the longest-standing independent libraries: vast-client-js, video-ad-sdk and iab-vast-loader, the last being the library behind IAB Tech Lab's original VAST Tester. video-ad-sdk and iab-vast-loader have since aged out of condition (4) and are kept as reference points for the players built on them; no current deployment is claimed. Shaka Player was added by hand, because it is a widely used media player with its own VAST parser that the npm queries did not return under these keywords. Of the candidates the survey found, vastlint-client (a session runtime from the VASTlint project) was tested, and rmp-vast, a player built on vast-client-js that therefore fails condition (2), was added as a real integration of vast-client-js; its release 17.2.0 is distributed through GitHub, and its npm package stops at 6.0.0. Four small player applications that pass all four conditions were not tested: itube-modern-player, streamit-player, ts-video-player and stormcloud-video-player. They embed VAST code in a larger player instead of offering it as a library; their adoption data are in Online Resource 1, and testing them is left to future work. IMA fails condition (1) and is kept as a closed comparison. Download counts were not used to select clients, because three small packages in the survey show nearly equal monthly counts (about 23,000 each) that look like a mirror or an automated client.

The clients are vast-client-js (\protect\hyperlink{ref-dailymotion2026vastclient}{Dailymotion 2026}), video-ad-sdk (\protect\hyperlink{ref-rambler2024videoadsdk}{Rambler Digital Solutions 2024}), iab-vast-loader (\protect\hyperlink{ref-zentrick2021iabvast}{Zentrick 2021}), vastlint-client (\protect\hyperlink{ref-sekowski2026vastlintclient}{Sekowski 2026b}), Shaka (\protect\hyperlink{ref-shaka2026player}{Shaka Project 2026}), rmp-vast (\protect\hyperlink{ref-radiant2026rmpvast}{Radiant Media Player 2026}) and IMA (\protect\hyperlink{ref-google2026ima}{Google 2026}). iab-vast-loader runs in two configurations (library defaults, and its \texttt{no\cbrsoft{}Single\cbrsoft{}Ad\cbrsoft{}Pods} option, which stops it from filing a lone ad as a one-ad pod) and rmp-vast in two (defaults, and \texttt{omid\cbrsoft{}Support}, the setting under which it handles verification at all), which makes eight open-source configurations: seven in the browsers and five under Node.js. All versions are pinned. video-ad-sdk ships directory imports meant for bundlers, and iab-vast-loader depends on a module loader that fails on Node.js 22; both are bundled unchanged with esbuild. rmp-vast runs from its vendored distribution file, whose hash is recorded.

\begingroup\small\setlength{\tabcolsep}{4pt}\Needspace{14\baselineskip}

\begin{longtable}[]{@{}
  >{\raggedright\arraybackslash}p{(\columnwidth - 10\tabcolsep) * \real{0.1786}}
  >{\raggedright\arraybackslash}p{(\columnwidth - 10\tabcolsep) * \real{0.1042}}
  >{\raggedright\arraybackslash}p{(\columnwidth - 10\tabcolsep) * \real{0.1726}}
  >{\raggedright\arraybackslash}p{(\columnwidth - 10\tabcolsep) * \real{0.1726}}
  >{\raggedright\arraybackslash}p{(\columnwidth - 10\tabcolsep) * \real{0.1935}}
  >{\raggedright\arraybackslash}p{(\columnwidth - 10\tabcolsep) * \real{0.1786}}@{}}
\caption{\textbf{Table 1.} Clients and configurations. ``Claimed VAST'' is the version the client's documentation states; ``none'' means no claim was found.}\tabularnewline
\toprule\noalign{}
\begin{minipage}[b]{\linewidth}\raggedright
Client
\end{minipage} & \begin{minipage}[b]{\linewidth}\raggedright
Version
\end{minipage} & \begin{minipage}[b]{\linewidth}\raggedright
Style
\end{minipage} & \begin{minipage}[b]{\linewidth}\raggedright
Claimed VAST
\end{minipage} & \begin{minipage}[b]{\linewidth}\raggedright
Duties it leaves to the integrator
\end{minipage} & \begin{minipage}[b]{\linewidth}\raggedright
Layers
\end{minipage} \\
\midrule\noalign{}
\endfirsthead
\multicolumn{6}{@{}l}{\textbf{Table 1} (continued)}\tabularnewline
\toprule\noalign{}
\begin{minipage}[b]{\linewidth}\raggedright
Client
\end{minipage} & \begin{minipage}[b]{\linewidth}\raggedright
Version
\end{minipage} & \begin{minipage}[b]{\linewidth}\raggedright
Style
\end{minipage} & \begin{minipage}[b]{\linewidth}\raggedright
Claimed VAST
\end{minipage} & \begin{minipage}[b]{\linewidth}\raggedright
Duties it leaves to the integrator
\end{minipage} & \begin{minipage}[b]{\linewidth}\raggedright
Layers
\end{minipage} \\
\midrule\noalign{}
\endhead
\bottomrule\noalign{}
\endlastfoot
vast-client-js & 6.5.0 & parser, resolver, tracker & 4.3 & media, rendering, ad fallback, verification decisions & Node.js, browsers \\
video-ad-sdk & 4.0.1 & full player & none & none & Node.js, browsers \\
iab-vast-loader (with iab-vast-parser 1.2.1) & 2.5.1 & loader & 2.0 to 4.0 (its model package) & everything after loading & Node.js, browsers \\
vastlint-client & 0.15.0 & session runtime & none (its validator: 2.0 to 4.4) & error codes, media and playback, skip-control availability, verification & Node.js \\
Shaka Player & 5.2.12 & media player & none (``basic support'') & none & browsers \\
rmp-vast (on vast-client-js 6.4.2) & 17.2.0 & full player & 3 and 4 up to 4.3 & none & browsers \\*
Google IMA HTML5 SDK & 3.794.0 & closed full player & 2 and 3; 4 in part & none & browsers (comparison) \\
\end{longtable}

\endgroup

\textbf{Version yardstick.} Every client is graded against VAST 4.3, the current published version, because a player in the field receives whatever version the ad servers send. Four requirements on which a client deviates are newer than its claim: iab-vast-loader's rejection of repeated UniversalAdId (X05, added in 4.2) and ClosedCaptionFile (X06, 4.1) and its unreplaced macros in a Wrapper's request (M12; macro-based ad requests were added in 4.1), against its 4.0 claim, and IMA's ignoring of the Wrapper ClickThrough (W17, 4.2). They are reported with that note. The clients without a claim are graded the same way.

\hypertarget{requirement-catalog}{%
\subsection{4.2 Requirement catalog}\label{requirement-catalog}}

\textbf{Extraction.} The catalog was built from an inventory of the 122 lines of the VAST 4.3 text, from §2.1 on, that contain must, should, required or may and mention the player or Player Support, plus the IAB macro list. Each duty was mapped to a requirement or to a recorded reason for leaving it out. Appendix A lists every requirement with its section and its strength as written. Requirement IDs carry a prefix: W Wrappers, E errors, T tracking, M macros, X XML and parsing, P ad pods, A verification, I icons, S media selection, R robustness. P04 is unassigned.

\textbf{Duty groups.} Table 2 gives the nine duty groups by which results are reported. They follow the ID prefixes with one exception. The Wrapper requirements are split by what they grade: W04 and W06 to W13 (error codes and Error URIs along the chain) join error reporting, and W14 to W18 (tracking and clicks along the chain) join impressions and tracking.

\begingroup\small\setlength{\tabcolsep}{4pt}\Needspace{10\baselineskip}

\begin{longtable}[]{@{}
  >{\raggedright\arraybackslash}p{(\columnwidth - 10\tabcolsep) * \real{0.4643}}
  >{\raggedright\arraybackslash}p{(\columnwidth - 10\tabcolsep) * \real{0.1786}}
  >{\raggedright\arraybackslash}p{(\columnwidth - 10\tabcolsep) * \real{0.0595}}
  >{\raggedright\arraybackslash}p{(\columnwidth - 10\tabcolsep) * \real{0.0893}}
  >{\raggedright\arraybackslash}p{(\columnwidth - 10\tabcolsep) * \real{0.0446}}
  >{\raggedright\arraybackslash}p{(\columnwidth - 10\tabcolsep) * \real{0.1637}}@{}}
\caption{\textbf{Table 2.} Requirements by duty group and strength. ``observation'' counts robustness requirements that are not VAST clauses (Appendix A).}\tabularnewline
\toprule\noalign{}
\begin{minipage}[b]{\linewidth}\raggedright
Duty group
\end{minipage} & \begin{minipage}[b]{\linewidth}\raggedright
Requirements
\end{minipage} & \begin{minipage}[b]{\linewidth}\raggedright
must
\end{minipage} & \begin{minipage}[b]{\linewidth}\raggedright
should
\end{minipage} & \begin{minipage}[b]{\linewidth}\raggedright
may
\end{minipage} & \begin{minipage}[b]{\linewidth}\raggedright
observation
\end{minipage} \\
\midrule\noalign{}
\endfirsthead
\multicolumn{6}{@{}l}{\textbf{Table 2} (continued)}\tabularnewline
\toprule\noalign{}
\begin{minipage}[b]{\linewidth}\raggedright
Duty group
\end{minipage} & \begin{minipage}[b]{\linewidth}\raggedright
Requirements
\end{minipage} & \begin{minipage}[b]{\linewidth}\raggedright
must
\end{minipage} & \begin{minipage}[b]{\linewidth}\raggedright
should
\end{minipage} & \begin{minipage}[b]{\linewidth}\raggedright
may
\end{minipage} & \begin{minipage}[b]{\linewidth}\raggedright
observation
\end{minipage} \\
\midrule\noalign{}
\endhead
\bottomrule\noalign{}
\endlastfoot
Wrapper resolution (W01 to W03, W05, W19 to W23) & 9 & 3 & 5 & 1 & 0 \\
Error reporting (W04, W06 to W13, E01 to E09) & 18 & 14 & 4 & 0 & 0 \\
Impressions and tracking (W14 to W18, T01 to T08, T10 to T23) & 27 & 18 & 9 & 0 & 0 \\
Macros (M01 to M12) & 12 & 7 & 5 & 0 & 0 \\
Ad pods (P01 to P03, P05 to P07) & 6 & 3 & 3 & 0 & 0 \\
XML and parsing (X01 to X03, X05, X06, X08 to X19) & 17 & 7 & 10 & 0 & 0 \\
Media selection (S01, S02) & 2 & 0 & 1 & 1 & 0 \\
Verification and icons (A01 to A05, I01 to I03) & 8 & 6 & 2 & 0 & 0 \\
Robustness observations (R01 to R06, T09, X04, X07) & 9 & 0 & 0 & 0 & 9 \\*
\textbf{Total} & \textbf{108} & \textbf{58} & \textbf{39} & \textbf{2} & \textbf{9} \\
\end{longtable}

\endgroup

\textbf{Strength.} One rule assigns strength. A requirement is must where the text puts a must or ``required'' on the player for that duty, in the body or in an implementation note for media players. It is also must where the macro list marks a macro Required in the context tested. It is should where the text says should, or where the duty is read from a definition, an attribute or an optional macro with no modal verb. Where the body and a note conflict, the body governs, and the other reading is tested in a sensitivity analysis (Section 5.1.6). Appendix A marks the 23 requirements whose graded strength is not the literal word of the text.

\textbf{Macro list version.} The IAB macro list is maintained outside the specification and grows. The study uses the list at commit 99d5d51 of IAB Tech Lab's VAST repository (\protect\hyperlink{ref-iab2026vastmacros}{IAB Technology Laboratory 2026}), which has 59 macros. The 50 whose introduction version is 4.2 or earlier are the graded set (the ``4.3 set''). When VAST 4.3 was published, the list held 54 macros: these 50 and four marked ``Next VAST release'' (STOREID, STOREURL, PLAYBACKMETHODS and CONTENTCAT), which the list now marks ``4.x''; the 4.3 text itself adds a value to {[}PLAYBACKMETHODS{]}. Five macros were added in 2023 and 2024: GPPSTRING and GPPSECTIONID, also marked ``4.x'', and three macros for the EU Digital Services Act (DSA), marked ``Next VAST release''. The catalog's macro requirements use only the 50; the macro grid reports them apart from the other nine.

\textbf{Vectors.} A generator writes 69 vectors with 109 documents (a vector with Wrappers has several). Of the 69, 52 serve one requirement, 9 serve two and 8 serve three or more; the largest, a fully played InLine ad, serves nine requirements (T01 to T08 and E08). Every URL carries a label naming the document and element it came from (for example \texttt{w1/\cbr{}clicktracking}), so an oracle can tell which tracker fired. Requirements whose duty concerns the InLine ad, or no document level in particular, use vectors that start with an InLine; the Wrapper requirements use Wrapper chains. Documents follow the element order of the IAB VAST 4.2 schema, and 102 of the 109 validate against it. The other seven are invalid on purpose (a malformed document, a document without the VAST namespace, three VAST 2.0 and 3.0 documents, a UniversalAdId with the 4.0 idValue attribute, and an unknown tracking event name). IAB's 4.3 schema file is empty, so validation uses 4.2.

\textbf{Verdicts.} A cell is one requirement on one client configuration in one runtime. Each cell receives one of four verdicts:

\begin{itemize}
\tightlist
\item
  \textbf{conforms} or \textbf{deviates}: the client takes on the duty and does or does not meet the requirement;
\item
  \textbf{delegated}: the client's documentation leaves the duty to its integrator (Table 1);
\item
  \textbf{n/a}: one of five reasons applies:

  \begin{itemize}
  \tightlist
  \item
    the requirement is optional and not implemented;
  \item
    the client declined a feature it may decline (pods, skippable ads);
  \item
    the cell only records a measurement (such as T09, the spread of impression requests);
  \item
    the client exposes no control the script needs;
  \item
    the cell depends on a behavior another requirement already graded as missing.
  \end{itemize}
\end{itemize}

\textbf{Grading rules.} Four rules keep one fault from being counted many times or against the wrong code.

\begin{itemize}
\tightlist
\item
  \emph{G1.} G1 applies only to clients that do not follow Wrappers. Such a client deviates on W01 and W02, the two requirements that grade following a Wrapper, and every other Wrapper-dependent cell is n/a for it. A client that follows Wrappers is graded on every Wrapper cell.
\item
  \emph{G2.} A requirement that reads a value another requirement already failed (a macro left unreplaced, progress events unsupported, a pod not continued) is n/a with a pointer to that requirement.
\item
  \emph{G3.} For error codes, the exact code conforms, and so do the group's general code (100, 200, 300, 400) and 900, which the text names as the permitted minimum; a different specific code deviates. Where the error table can be read to give more than one code, every such code is accepted.
\item
  \emph{G4.} A client built on another client is graded on its own behavior, and each deviation is attributed to the code that produces it. When rmp-vast deviates on a path inside its bundled vast-client-js, the finding is reported under vast-client-js, while rmp-vast's cell still records the deviation, because a user of rmp-vast receives it.
\end{itemize}

\textbf{From cells to findings.} A finding is one cause, located in source. For an open-source client, each deviating cell was traced to the source line in the tested release that produces it, or, for a duty the client never implements, to the absence of any code that reads the relevant element (stated with the search that shows it). Cells with one cause form one finding. Fuzzing failures are merged into root causes (Section 4.4). IMA is closed, so it is analyzed as a black box: each of its findings rests on the requests the server logged and on the error events of its public API, and, where a probe vector isolates a behavior, on that probe; no explanation is drawn from its code. IMA was not swept or fuzzed. Appendix B lists the 86 findings; Online Resource 5 gives every verdict by client and by duty group.

\hypertarget{harness-and-test-execution}{%
\subsection{4.3 Harness and test execution}\label{harness-and-test-execution}}

A recording server serves the test documents and rewrites every absolute URL inside them to point back at itself, so Wrapper fetches, media requests and tracking requests all arrive at the server. The server logs each with its arrival time in milliseconds. Query parameters make it return an HTTP status, delay a response, stall a body after its headers, redirect, compress, or change the content type.

\textbf{Node.js layer.} Each test runs one client in its own Node.js process (Node.js 22.22.0), with browser globals supplied by jsdom 26.1.0, the environment in which vast-client-js and video-ad-sdk run their own test suites. Requests sent through \texttt{new Image()} or \texttt{navigator.\cbrsoft{}send\cbrsoft{}Beacon()} become real HTTP requests. Playback uses a scripted video element whose clock the harness advances in steps and which fires the media events a real element fires.

\textbf{Browser layer.} The same catalog, vectors and oracles run in real browsers through Playwright 1.56.1. The libraries are bundled into one page with esbuild 0.25.10, the scripted element is replaced by a real video element playing real media (a synthetic 10 s ad and 40 s content clip, test patterns with a tone, each as VP9 WebM and H.264 MP4), and the browser's own DOMParser, fetch and Image replace jsdom's. The media set also holds an audio-only pair and a corrupt file. Each test gets a fresh browser context. Nine requirements need real decoding or rendering (media failures, audio ads, pod failure handling, icons) and are graded only in the browser layer.

\textbf{Platforms.} On Linux (x86-64), Chromium 141, Firefox 142 and WebKit 26 run through Playwright. The packaged harness was also run on the author's Apple silicon Mac mini (macOS 27.0.1) with Chromium, Google Chrome 156, Firefox, WebKit and Safari 27.0.1; the whole package took 73 minutes. The five browsers use three engines: Blink (Chromium, Chrome), Gecko (Firefox) and WebKit (WebKit, Safari). Chrome and Safari results come from macOS, because a Linux Chrome entry resolves to the Chromium binary; WebKit is reported from both platforms, because the Linux WebKit build differs from macOS WebKit in media support (Section 5.2). Chromium and Firefox results are the Linux references, which the Mac run reproduced in every open-source cell compared (Section 5.2). IMA did not run in Safari, whose driver refused the proxy setting that serves the archived SDK.

\textbf{Scripts and integration.} A test either loads the tag or plays it with one of eight playback scripts (Table 3). For vast-client-js and vastlint-client, which leave part of a player's work to the integrator, the harness acts as an integrator that calls each library exactly as its documentation describes. This integrator code (\texttt{code/\cbr{}harness/\cbr{}adapters/\cbr{}vastclient.\cbrsoft{}mjs} and \texttt{vastlint.\cbrsoft{}mjs}) is part of the released harness, with its hashes. rmp-vast draws its own controls, so the harness builds the page layout its README prescribes and acts as a user would, clicking its video and skip button. IMA is driven through its public API and its own controls (Online Resource 4). It exposes no seek control, so its rewind cells are n/a. Its icon cells I01 and I03 are graded from the requests, and I02, which needs a click on the icon, is not drivable.

\begingroup\small\setlength{\tabcolsep}{4pt}\Needspace{7\baselineskip}

\begin{longtable}[]{@{}ll@{}}
\caption{\textbf{Table 3.} Load and playback scripts.}\tabularnewline
\toprule\noalign{}
Script & What the harness does \\
\midrule\noalign{}
\endfirsthead
\multicolumn{2}{@{}l}{\textbf{Table 3} (continued)}\tabularnewline
\toprule\noalign{}
Script & What the harness does \\
\midrule\noalign{}
\endhead
\bottomrule\noalign{}
\endlastfoot
load & resolves the tag; no playback \\
full & plays the ad to the end \\
partial & plays to 60 percent and stops the ad \\
click & clicks the ad at 30 percent \\
user actions & pauses, resumes, mutes, unmutes and changes the volume while playing \\
skip & probes for the skip control at 1.5 s and skips at 6 s \\
rewind & seeks back 2 s, from 6 s to 4 s \\
icon & clicks a displayed icon \\*
pod & plays every ad of a pod \\
\end{longtable}

\endgroup

\textbf{Oracles.} Each requirement has one oracle, a function that reads the request log of its cell and, for some requirements, what the client's API returned (the resolved ad, its error code). The log is never de-duplicated: a request sent twice is two entries. Impressions, start and quartile events, progress, pause, resume, mute, skip, rewind and loaded events, pod impressions and completions, and the Impression, tracking and click-tracking requests of the InLine and the Wrappers (W14, W15, W18) are graded exactly once, so a second request deviates. Unmute is graded as fired (T15), and an extra unmute after a volume change is graded by T23. Error requests are graded on presence and code. Timing is graded only where the text implies it: quartile events must arrive in order; a pod ad's completion must arrive before the next ad's impression, with 50 ms of tolerance. An icon's view request must not precede the start request by more than 50 ms, and a timed icon's must arrive at least 800 ms after start for a 1 s offset. The simultaneity of impressions is measured, not graded. After a script ends, the harness waits 300 ms (Node.js) or 600 ms (browsers), longer for the full players, then grades both presence and absence on the log. For example, W13 serves a chain of two Wrappers and an InLine whose only media file has an unplayable type, plays it, and grades the fan-out: the InLine's and both Wrappers' Error URIs must be requested with one consistent code. The code itself is graded by E03 (rule G2), so a client that sends one wrong code to all three is graded on that code once. X04 and X07, which are not VAST clauses, are graded as observations: the client conforms when it parses the document or follows the URI.

\textbf{Timing constants.} The harness uses fixed thresholds, chosen as engineering settings rather than derived from the specification: a 50 ms ordering tolerance (a pod ad's completion against the next ad's impression, an icon's view against start; the quartile order check has none), settle windows of 300 ms under Node.js and 600 ms in browsers before an absent request is graded, an 800 ms threshold for an icon with a 1 s offset, and, in fuzzing, a 2 s client timeout, a 30 s wall limit and a slow-load limit of three times the client's timeout plus 5 s (11 s). Measured margins: every conforming progress request in the browser full players arrives within 310 ms of its nominal offset; the slowest fuzz run that finished took 9.6 s; and one cell sits on a threshold, IMA's I01. Its icon view precedes start in every run on every engine and platform, by 36 to 237 ms in 60 repeated Linux runs (20 per engine), and the 50 ms tolerance falls inside the Linux Chromium spread (36 to 68 ms; beyond it in 6 of 20 runs), so I01 is the one cell whose verdict depends on the run. On macOS, in 10 repeated runs per browser, the view preceded start by 11 to 12 ms in Chromium, 10 to 13 ms in Chrome and 13 to 50 ms in Firefox (never beyond the tolerance) and by 51 to 58 ms in WebKit (beyond it in all 10) (Table 6).

\hypertarget{generated-inputs}{%
\subsection{4.4 Generated inputs}\label{generated-inputs}}

\textbf{Error grid.} The catalog tests each error requirement with one vector; the error grid enumerates the space those vectors sample. For every chain depth \(d\) from 2 to 6 and every failing level \(L\) from 2 to \(d\) (15 pairs), and each of 12 failures, the document at level \(L\) fails. Levels 1 to \(L - 1\) are Wrappers the client has received, whose Error URIs the client must request with a code from the error table. The failures fall into three families:

\begin{itemize}
\tightlist
\item
  network failures (HTTP 404, HTTP 500, a timeout, headers followed by a stalled body, a refused connection; code 301);
\item
  content failures (an empty 200 response, a document that is not well-formed XML, an HTML page, XML whose root is not VAST; 100, 101 or 303);
\item
  chain failures (a document with no ad, one with only a document-level Error element, a Wrapper with no VASTAdTagURI; 303, or 101, 100 or 301 for the last).
\end{itemize}

With seven baselines that check that chains of one to six documents resolve and that a chain of six Wrappers is accepted or rejected with 302, that gives \(15 \times 12 + 7 = 187\) cases per client. The grid runs at the Node.js layer for the four Node.js clients (748 cells). iab-vast-loader has no tracker, so the code it determines is graded instead of its requests, and vastlint-client determines no code, so its cells are delegated unless it resolves an ad it should not.

\textbf{Macro grid.} The macro grid places each of the 56 macros (of the list's 59) whose contexts include all tracking pixels, VAST request URIs or both in those contexts. An InLine carries one Impression for each of the 45 pixel macros (40 of them in the 4.3 set). A Wrapper's VASTAdTagURI carries every one of the 50 request macros (41 in the 4.3 set), once inside a CDATA section with literal ampersands and once as escaped text with \texttt{\&\cbr{}amp;}, two forms that are equally valid XML and must produce the same request. Each value a client sends is classified as left literally, empty or absent, \(-1\) or \(-2\), or a value. A literal or empty value deviates, because §6.1 makes the requesting party responsible for replacement and requires \(-1\) or \(-2\) for a listed macro it does not fill. Of the other three macros, {[}ERRORCODE{]} is graded by E04 and the error grid, {[}REASON{]} by A02, and {[}CLICKPOS{]}, which applies only to click-tracking URLs, is not tested. The grid runs at the Node.js layer for the same four clients (580 cells).

\textbf{Mutation fuzzing.} The seeds are 67 of the 69 crafted vectors (the other two, \texttt{a\_\cbrsoft{}verification\_\cbrsoft{}inline} and \texttt{m\_\cbrsoft{}optional\_\cbrsoft{}noflv}, are single-change variants of seeded vectors), each with all its documents, and the 72 IAB sample documents of VAST 1.0 to 4.2 (139 seeds). Each case takes one seed, applies one to three random mutations to one of its documents, serves the result, and runs every Node.js client on it with the suite's scripts. Mutations are of three kinds:

\begin{itemize}
\tightlist
\item
  structural (deleting, duplicating, moving or copying an element; changing an attribute; replacing text; adding a comment; changing the case of an element name);
\item
  textual (truncation, a byte order mark, a changed encoding declaration, entity expansion, a namespace prefix, an inserted character, removed whitespace, a changed version attribute);
\item
  network (an HTTP status, a delay, a stalled body, compression or a changed content type on a Wrapper's target).
\end{itemize}

The oracle records only robustness failures, of seven kinds:

\begin{itemize}
\tightlist
\item
  a process that dies or prints no result;
\item
  a run over the 30 s wall limit;
\item
  an error thrown outside any promise;
\item
  an unhandled rejection;
\item
  an exception from the client's documented API;
\item
  a load that takes more than three times the client's timeout plus 5 s;
\item
  an internal JavaScript error returned as the client's error.
\end{itemize}

A failure class is the client, the failure kind and the normalized message. Every class was reproduced three times from its first example, attributed to a source line by a separate script that calls the library directly, and reduced to a minimal document. It was then merged with the other classes of its cause into a root cause, labeled C1 to C9 (Appendix D). A differential pass groups cases by which clients resolved an ad, separates disagreements that a mutation induced from those its seed already showed, and probes the induced groups with single-change documents; its findings are labeled D1 to D4. The campaign drew from a single seeded pseudorandom stream in rounds of 300 draws, until a round added no new class, with at least 4 rounds. A draw whose mutations cannot apply (a network mutation on a document with no Wrapper target) is skipped. The campaign's seed (20261017) regenerates all 1,442 cases in order, and a replay mode reruns any stored case.

\hypertarget{version-sweep}{%
\subsection{4.5 Version sweep}\label{version-sweep}}

The sweep grades 93 requirements: the Node.js catalog except A04, A05, P07, X17, X18 and X19 (the nine browser-only requirements run as n/a), with A01, A02, R04, R05 and M06 to M09 on their Wrapper-first or FLV-first vector variants, on the last patch release of each minor release. The swept releases are vast-client-js 3.3.0 to 6.5.0 (9 releases), video-ad-sdk 2.1.0 to 4.0.1 (9), iab-vast-loader 2.0.0 to 2.5.1 (6, in both configurations) and vastlint-client 0.4.24 to 0.15.0 (12), 36 releases in all. Every verdict change between two releases was checked against a source diff of the two installed releases and, where available, the release notes. The sweep bounds each change between two swept releases; it does not cover the browser-only requirements, rmp-vast, Shaka or IMA. Old releases ran with dependencies resolved in October 2026.

\hypertarget{validity-checks}{%
\subsection{4.6 Validity checks}\label{validity-checks}}

Online Resource 2 documents each check in full.

Five checks rest on deterministic evidence that a reader can verify (Table 4). Three further checks were automated consistency audits: separate automated reviewers, each given only the inputs listed in Table 4b and none of the author's reasoning, reviewed the catalog, re-derived the open-source deviations from client code, and re-derived IMA's from logged requests, public API events and the archived SDK files. These reviewers were separate sessions of the generative AI tool named in the Use of generative AI statement (Statements and Declarations), each started with only those inputs; the author checked and resolved their outputs. Table 4b reports what they found, all of which the reported results address. These audits catch errors; they are not a measure of inter-rater reliability, and no human second rater has reviewed the catalog or the verdicts (Section 7). The verdicts rest on the evidence of Table 4: Appendix A gives the specification basis of every requirement and Online Resource 5 quotes the sentence behind it, Appendix B traces every open-source finding to a source location in the tested release, and every robustness defect was reproduced by a separate script outside the harness.

\begingroup\small\setlength{\tabcolsep}{4pt}\Needspace{11\baselineskip}

\begin{longtable}[]{@{}
  >{\raggedright\arraybackslash}p{(\columnwidth - 6\tabcolsep) * \real{0.2500}}
  >{\raggedright\arraybackslash}p{(\columnwidth - 6\tabcolsep) * \real{0.2500}}
  >{\raggedright\arraybackslash}p{(\columnwidth - 6\tabcolsep) * \real{0.2500}}
  >{\raggedright\arraybackslash}p{(\columnwidth - 6\tabcolsep) * \real{0.2500}}@{}}
\caption{\textbf{Table 4.} Validity checks.}\tabularnewline
\toprule\noalign{}
\begin{minipage}[b]{\linewidth}\raggedright
Check
\end{minipage} & \begin{minipage}[b]{\linewidth}\raggedright
Inputs
\end{minipage} & \begin{minipage}[b]{\linewidth}\raggedright
Scope
\end{minipage} & \begin{minipage}[b]{\linewidth}\raggedright
Outcome
\end{minipage} \\
\midrule\noalign{}
\endfirsthead
\multicolumn{4}{@{}l}{\textbf{Table 4} (continued)}\tabularnewline
\toprule\noalign{}
\begin{minipage}[b]{\linewidth}\raggedright
Check
\end{minipage} & \begin{minipage}[b]{\linewidth}\raggedright
Inputs
\end{minipage} & \begin{minipage}[b]{\linewidth}\raggedright
Scope
\end{minipage} & \begin{minipage}[b]{\linewidth}\raggedright
Outcome
\end{minipage} \\
\midrule\noalign{}
\endhead
\bottomrule\noalign{}
\endlastfoot
Source tracing & client source of each tested release & every deviation of an open-source client & every finding located (Appendix B) \\
Standalone reproduction & a script per defect, outside the harness & every robustness root cause & each reproduced 3 of 3 times (Appendix D) \\
Driver audits & each script's request log against each library & every script and library & 8 harness issues resolved \\
Determinism & repeated runs on two platforms & Node.js and browser matrices, grids & identical verdicts except IMA's I01 (Section 5.2) \\*
Access check & the 49 IAB VAST 4.x sample tags & the five open-source clients that load tags & each resolves every sample except those its findings predict (iab-vast-loader: 12) \\
\end{longtable}

\endgroup

\begingroup\small\setlength{\tabcolsep}{4pt}\Needspace{12\baselineskip}

\begin{longtable}[]{@{}
  >{\raggedright\arraybackslash}p{(\columnwidth - 6\tabcolsep) * \real{0.2500}}
  >{\raggedright\arraybackslash}p{(\columnwidth - 6\tabcolsep) * \real{0.2500}}
  >{\raggedright\arraybackslash}p{(\columnwidth - 6\tabcolsep) * \real{0.2500}}
  >{\raggedright\arraybackslash}p{(\columnwidth - 6\tabcolsep) * \real{0.2500}}@{}}
\caption{\textbf{Table 4b.} Automated consistency audits (separate automated reviewers, isolated inputs).}\tabularnewline
\toprule\noalign{}
\begin{minipage}[b]{\linewidth}\raggedright
Audit
\end{minipage} & \begin{minipage}[b]{\linewidth}\raggedright
Inputs
\end{minipage} & \begin{minipage}[b]{\linewidth}\raggedright
Scope
\end{minipage} & \begin{minipage}[b]{\linewidth}\raggedright
Outcome
\end{minipage} \\
\midrule\noalign{}
\endfirsthead
\multicolumn{4}{@{}l}{\textbf{Table 4b} (continued)}\tabularnewline
\toprule\noalign{}
\begin{minipage}[b]{\linewidth}\raggedright
Audit
\end{minipage} & \begin{minipage}[b]{\linewidth}\raggedright
Inputs
\end{minipage} & \begin{minipage}[b]{\linewidth}\raggedright
Scope
\end{minipage} & \begin{minipage}[b]{\linewidth}\raggedright
Outcome
\end{minipage} \\
\midrule\noalign{}
\endhead
\bottomrule\noalign{}
\endlastfoot
Catalog review & specification, vectors, oracles, drivers; no results & every requirement in the catalog at the time of the review & 8 verdict-affecting and 16 wording issues, all resolved \\
Re-derivation, two clients & installed source and neutral questions; no findings & 3 deviations of video-ad-sdk and vast-client-js & 3 of 3 confirmed with its own scripts \\*
Re-derivation, three clients & result matrices of the three clients; open-source client code and standalone scripts; IMA request logs, public API events and archived SDK files; none of the author's reasoning & 46 deviating cells of vastlint-client, rmp-vast and IMA, and 2 claims & 40 cells confirmed, 4 harness issues resolved, 2 borderline (reported as such); both claims confirmed \\
\end{longtable}

\endgroup

The three-client re-derivation did not cover vastlint-client W23, X17, X18 and X19, rmp-vast P07, IMA W17, A05 and I01, or rmp-vast's \texttt{omid\cbrsoft{}Support} configuration.

In the access check, vast-client-js, video-ad-sdk and vastlint-client resolve an InLine ad from all 49 IAB samples, and rmp-vast starts an ad for all 49 (46 play to completion; the other three are overlays, which have no completion event). iab-vast-loader resolves 37 with defaults and 43 with \texttt{no\cbrsoft{}Single\cbrsoft{}Ad\cbrsoft{}Pods}: the six it rejects in both configurations repeat UniversalAdId or ClosedCaptionFile, and the six it rejects only with defaults are the Wrapper samples (Section 5.1.3).

\hypertarget{results}{%
\section{5. Results}\label{results}}

\hypertarget{rq1-conformance}{%
\subsection{5.1 RQ1: Conformance}\label{rq1-conformance}}

\hypertarget{overview}{%
\subsubsection{5.1.1 Overview}\label{overview}}

Table 5 gives the verdicts and the number of findings per configuration, and Figure 3 gives the deviating cells over the cells each client takes on, by duty group. Counts of conforming cells are not comparable between clients that take on different shares of the duties. Within a duty group, each rate in Figure 3 is a conditional failure fraction: the share of the cells a client takes on that deviate. Clients take on different cells, so the rates are not a ranking of client quality. Which cells a client takes on follows each library's documentation (Table 1): a cell is delegated where that documentation leaves the duty to the integrator. Cells are not independent: one finding can produce several deviating cells (video-ad-sdk's 16 must-level cells in Chromium come from 12 findings), so Table 5 reports both. Appendix C lists every deviating cell of the open-source configurations of Table 5 except Shaka 5.3.0, whose cells are in Online Resource 3.

\begingroup\footnotesize\setlength{\tabcolsep}{4pt}\Needspace{15\baselineskip}

\begin{longtable}[]{@{}
  >{\raggedright\arraybackslash}p{(\columnwidth - 16\tabcolsep) * \real{0.2898}}
  >{\raggedright\arraybackslash}p{(\columnwidth - 16\tabcolsep) * \real{0.1136}}
  >{\raggedright\arraybackslash}p{(\columnwidth - 16\tabcolsep) * \real{0.1136}}
  >{\raggedright\arraybackslash}p{(\columnwidth - 16\tabcolsep) * \real{0.0568}}
  >{\raggedright\arraybackslash}p{(\columnwidth - 16\tabcolsep) * \real{0.0852}}
  >{\raggedright\arraybackslash}p{(\columnwidth - 16\tabcolsep) * \real{0.0568}}
  >{\raggedright\arraybackslash}p{(\columnwidth - 16\tabcolsep) * \real{0.1278}}
  >{\raggedright\arraybackslash}p{(\columnwidth - 16\tabcolsep) * \real{0.0426}}
  >{\raggedright\arraybackslash}p{(\columnwidth - 16\tabcolsep) * \real{0.1136}}@{}}
\caption{\textbf{Table 5.} Verdicts per configuration (108 requirements), Node.js layer and Chromium 141, and findings per client (Appendix B). Each row sums to 108. The must, should and obs. columns count deviating cells at must level, at should level and on robustness observations (R01 for vast-client-js and rmp-vast). Google IMA is the closed comparison. Shaka Player, rmp-vast and IMA run in the browser layer only, vastlint-client in the Node.js layer only. Shaka Player 5.3.0, released on the day of the reference runs, was run in Chromium with the same adapter, and an independent repeat agrees on every cell graded from the same vectors (Online Resource 3); it is not in Figure 3 or Appendix B. The nine browser-only requirements are n/a in the Node.js layer, which is why delegated and n/a counts differ between layers.}\tabularnewline
\toprule\noalign{}
\begin{minipage}[b]{\linewidth}\raggedright
Configuration
\end{minipage} & \begin{minipage}[b]{\linewidth}\raggedright
Layer
\end{minipage} & \begin{minipage}[b]{\linewidth}\raggedright
conforms
\end{minipage} & \begin{minipage}[b]{\linewidth}\raggedright
must
\end{minipage} & \begin{minipage}[b]{\linewidth}\raggedright
should
\end{minipage} & \begin{minipage}[b]{\linewidth}\raggedright
obs.
\end{minipage} & \begin{minipage}[b]{\linewidth}\raggedright
delegated
\end{minipage} & \begin{minipage}[b]{\linewidth}\raggedright
n/a
\end{minipage} & \begin{minipage}[b]{\linewidth}\raggedright
findings
\end{minipage} \\
\midrule\noalign{}
\endfirsthead
\multicolumn{9}{@{}l}{\textbf{Table 5} (continued)}\tabularnewline
\toprule\noalign{}
\begin{minipage}[b]{\linewidth}\raggedright
Configuration
\end{minipage} & \begin{minipage}[b]{\linewidth}\raggedright
Layer
\end{minipage} & \begin{minipage}[b]{\linewidth}\raggedright
conforms
\end{minipage} & \begin{minipage}[b]{\linewidth}\raggedright
must
\end{minipage} & \begin{minipage}[b]{\linewidth}\raggedright
should
\end{minipage} & \begin{minipage}[b]{\linewidth}\raggedright
obs.
\end{minipage} & \begin{minipage}[b]{\linewidth}\raggedright
delegated
\end{minipage} & \begin{minipage}[b]{\linewidth}\raggedright
n/a
\end{minipage} & \begin{minipage}[b]{\linewidth}\raggedright
findings
\end{minipage} \\
\midrule\noalign{}
\endhead
\bottomrule\noalign{}
\endlastfoot
vast-client-js 6.5.0 & Node.js & 79 & 3 & 2 & 1 & 12 & 11 & 11 \\
& Chromium & 79 & 3 & 2 & 1 & 21 & 2 & \\
video-ad-sdk 4.0.1 & Node.js & 68 & 13 & 8 & 0 & 0 & 19 & 19 \\
& Chromium & 71 & 16 & 9 & 0 & 0 & 12 & \\
iab-vast-loader 2.5.1, defaults & Node.js & 23 & 6 & 1 & 0 & 68 & 10 & 9 \\
& Chromium & 24 & 5 & 1 & 0 & 77 & 1 & \\
iab-vast-loader 2.5.1, \texttt{no\cbrsoft{}Single\cbrsoft{}Ad\cbrsoft{}Pods} & Node.js & 24 & 5 & 1 & 0 & 68 & 10 & \\
& Chromium & 25 & 4 & 1 & 0 & 77 & 1 & \\
vastlint-client 0.15.0 & Node.js & 50 & 11 & 9 & 0 & 22 & 16 & 9 \\
Shaka Player 5.2.12 & Chromium & 35 & 21 & 8 & 0 & 0 & 44 & 15 \\
Shaka Player 5.3.0 & Chromium & 44 & 15 & 7 & 0 & 0 & 42 & - \\
rmp-vast 17.2.0, defaults & Chromium & 80 & 12 & 9 & 1 & 0 & 6 & 15 \\
rmp-vast 17.2.0, \texttt{omid\cbrsoft{}Support} & Chromium & 84 & 11 & 8 & 1 & 0 & 4 & \\*
Google IMA HTML5 SDK 3.794.0 & Chromium & 93 & 4 & 4 & 0 & 0 & 7 & 7 \\
\end{longtable}

\endgroup

A client's findings include robustness and macro-grid findings outside the cells of Table 5. IMA's count is its Chromium findings; Firefox adds one (8 across browsers). rmp-vast's count is its own findings; its cells that come from the bundled vast-client-js are counted under vast-client-js (rule G4).

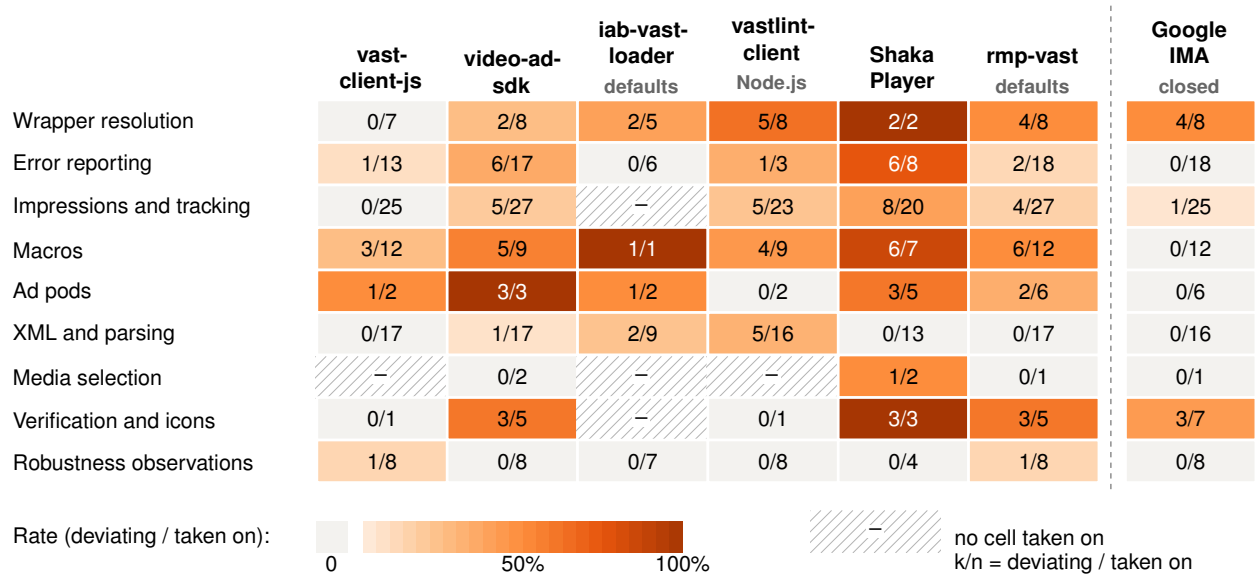
\begin{figure}[t]
\centering
\begin{tikzpicture}[
  font=\sffamily\fontsize{8.5}{10}\selectfont,
  x=1pt, y=1pt,
  cell/.style={draw=white, line width=1pt, inner sep=0pt, minimum width=49pt, minimum height=16pt, anchor=south west},
  colhead/.style={anchor=south, align=center, inner sep=0pt, text width=47pt}
]
\def\C#1#2#3#4#5{\node[cell, fill=#3] at (#1,#2) {\color{#5}#4};}
\def\E#1#2{\node[cell, fill=white] at (#1,#2) {};
  \begin{scope}\clip (#1,#2) rectangle +(49,16);
  \foreach \h in {-16,-12,...,49} \draw[black!30, line width=0.3pt] ([xshift=\h pt]#1,#2) -- +(16,16);\end{scope}
  \node[fill=white, inner sep=1pt] at (#1+24.5,#2+8.0) {--};}
\node[colhead] at (138.5,147) {\bfseries vast-\\ \bfseries client-js};
\node[colhead] at (187.5,147) {\bfseries video-ad-\\ \bfseries sdk};
\node[colhead] at (236.5,147) {\bfseries iab-vast-\\ \bfseries loader\\[1pt] {\fontsize{7.5}{9}\selectfont\color{black!60}defaults}};
\node[colhead] at (285.5,147) {\bfseries vastlint-\\ \bfseries client\\[1pt] {\fontsize{7.5}{9}\selectfont\color{black!60}Node.js}};
\node[colhead] at (334.5,147) {\bfseries Shaka\\ \bfseries Player};
\node[colhead] at (383.5,147) {\bfseries rmp-vast\\[1pt] {\fontsize{7.5}{9}\selectfont\color{black!60}defaults}};
\node[colhead] at (442.5,147) {\bfseries Google\\ \bfseries IMA\\[1pt] {\fontsize{7.5}{9}\selectfont\color{black!60}closed}};
\node[anchor=west, inner sep=0pt] at (0,136.0) {Wrapper resolution};
\C{114}{128}{{rgb,255:red,243;green,242;blue,239}}{0/7}{black}
\C{163}{128}{{rgb,255:red,253;green,190;blue,133}}{2/8}{black}
\C{212}{128}{{rgb,255:red,253;green,161;blue,89}}{2/5}{black}
\C{261}{128}{{rgb,255:red,242;green,113;blue,36}}{5/8}{black}
\C{310}{128}{{rgb,255:red,166;green,54;blue,3}}{2/2}{white}
\C{359}{128}{{rgb,255:red,253;green,141;blue,60}}{4/8}{black}
\C{418}{128}{{rgb,255:red,253;green,141;blue,60}}{4/8}{black}
\node[anchor=west, inner sep=0pt] at (0,120.0) {Error reporting};
\C{114}{112}{{rgb,255:red,254;green,223;blue,195}}{1/13}{black}
\C{163}{112}{{rgb,255:red,253;green,170;blue,103}}{6/17}{black}
\C{212}{112}{{rgb,255:red,243;green,242;blue,239}}{0/6}{black}
\C{261}{112}{{rgb,255:red,253;green,174;blue,109}}{1/3}{black}
\C{310}{112}{{rgb,255:red,230;green,85;blue,13}}{6/8}{white}
\C{359}{112}{{rgb,255:red,254;green,216;blue,182}}{2/18}{black}
\C{418}{112}{{rgb,255:red,243;green,242;blue,239}}{0/18}{black}
\node[anchor=west, inner sep=0pt] at (0,104.0) {Impressions and tracking};
\C{114}{96}{{rgb,255:red,243;green,242;blue,239}}{0/25}{black}
\C{163}{96}{{rgb,255:red,253;green,202;blue,156}}{5/27}{black}
\E{212}{96}
\C{261}{96}{{rgb,255:red,253;green,196;blue,145}}{5/23}{black}
\C{310}{96}{{rgb,255:red,253;green,161;blue,89}}{8/20}{black}
\C{359}{96}{{rgb,255:red,253;green,209;blue,169}}{4/27}{black}
\C{418}{96}{{rgb,255:red,254;green,229;blue,208}}{1/25}{black}
\node[anchor=west, inner sep=0pt] at (0,88.0) {Macros};
\C{114}{80}{{rgb,255:red,253;green,190;blue,133}}{3/12}{black}
\C{163}{80}{{rgb,255:red,248;green,129;blue,50}}{5/9}{black}
\C{212}{80}{{rgb,255:red,166;green,54;blue,3}}{1/1}{white}
\C{261}{80}{{rgb,255:red,253;green,152;blue,76}}{4/9}{black}
\C{310}{80}{{rgb,255:red,203;green,72;blue,9}}{6/7}{white}
\C{359}{80}{{rgb,255:red,253;green,141;blue,60}}{6/12}{black}
\C{418}{80}{{rgb,255:red,243;green,242;blue,239}}{0/12}{black}
\node[anchor=west, inner sep=0pt] at (0,72.0) {Ad pods};
\C{114}{64}{{rgb,255:red,253;green,141;blue,60}}{1/2}{black}
\C{163}{64}{{rgb,255:red,166;green,54;blue,3}}{3/3}{white}
\C{212}{64}{{rgb,255:red,253;green,141;blue,60}}{1/2}{black}
\C{261}{64}{{rgb,255:red,243;green,242;blue,239}}{0/2}{black}
\C{310}{64}{{rgb,255:red,244;green,119;blue,41}}{3/5}{black}
\C{359}{64}{{rgb,255:red,253;green,174;blue,109}}{2/6}{black}
\C{418}{64}{{rgb,255:red,243;green,242;blue,239}}{0/6}{black}
\node[anchor=west, inner sep=0pt] at (0,56.0) {XML and parsing};
\C{114}{48}{{rgb,255:red,243;green,242;blue,239}}{0/17}{black}
\C{163}{48}{{rgb,255:red,254;green,226;blue,201}}{1/17}{black}
\C{212}{48}{{rgb,255:red,253;green,195;blue,143}}{2/9}{black}
\C{261}{48}{{rgb,255:red,253;green,178;blue,115}}{5/16}{black}
\C{310}{48}{{rgb,255:red,243;green,242;blue,239}}{0/13}{black}
\C{359}{48}{{rgb,255:red,243;green,242;blue,239}}{0/17}{black}
\C{418}{48}{{rgb,255:red,243;green,242;blue,239}}{0/16}{black}
\node[anchor=west, inner sep=0pt] at (0,40.0) {Media selection};
\E{114}{32}
\C{163}{32}{{rgb,255:red,243;green,242;blue,239}}{0/2}{black}
\E{212}{32}
\E{261}{32}
\C{310}{32}{{rgb,255:red,253;green,141;blue,60}}{1/2}{black}
\C{359}{32}{{rgb,255:red,243;green,242;blue,239}}{0/1}{black}
\C{418}{32}{{rgb,255:red,243;green,242;blue,239}}{0/1}{black}
\node[anchor=west, inner sep=0pt] at (0,24.0) {Verification and icons};
\C{114}{16}{{rgb,255:red,243;green,242;blue,239}}{0/1}{black}
\C{163}{16}{{rgb,255:red,244;green,119;blue,41}}{3/5}{black}
\E{212}{16}
\C{261}{16}{{rgb,255:red,243;green,242;blue,239}}{0/1}{black}
\C{310}{16}{{rgb,255:red,166;green,54;blue,3}}{3/3}{white}
\C{359}{16}{{rgb,255:red,244;green,119;blue,41}}{3/5}{black}
\C{418}{16}{{rgb,255:red,253;green,155;blue,81}}{3/7}{black}
\node[anchor=west, inner sep=0pt] at (0,8.0) {Robustness observations};
\C{114}{0}{{rgb,255:red,254;green,214;blue,178}}{1/8}{black}
\C{163}{0}{{rgb,255:red,243;green,242;blue,239}}{0/8}{black}
\C{212}{0}{{rgb,255:red,243;green,242;blue,239}}{0/7}{black}
\C{261}{0}{{rgb,255:red,243;green,242;blue,239}}{0/8}{black}
\C{310}{0}{{rgb,255:red,243;green,242;blue,239}}{0/4}{black}
\C{359}{0}{{rgb,255:red,254;green,214;blue,178}}{1/8}{black}
\C{418}{0}{{rgb,255:red,243;green,242;blue,239}}{0/8}{black}
\draw[black!50, line width=0.5pt, dash pattern=on 2pt off 2pt] (413.0,-2) -- (413.0,180);
\node[anchor=west, inner sep=0pt] at (0,-20) {Rate (deviating / taken on):};
\fill[fill={rgb,255:red,254;green,233;blue,215}] (132,-26) rectangle +(5,12);
\fill[fill={rgb,255:red,254;green,225;blue,200}] (137,-26) rectangle +(5,12);
\fill[fill={rgb,255:red,254;green,217;blue,185}] (142,-26) rectangle +(5,12);
\fill[fill={rgb,255:red,253;green,210;blue,170}] (147,-26) rectangle +(5,12);
\fill[fill={rgb,255:red,253;green,202;blue,155}] (152,-26) rectangle +(5,12);
\fill[fill={rgb,255:red,253;green,194;blue,140}] (157,-26) rectangle +(5,12);
\fill[fill={rgb,255:red,253;green,186;blue,127}] (162,-26) rectangle +(5,12);
\fill[fill={rgb,255:red,253;green,178;blue,115}] (167,-26) rectangle +(5,12);
\fill[fill={rgb,255:red,253;green,170;blue,103}] (172,-26) rectangle +(5,12);
\fill[fill={rgb,255:red,253;green,161;blue,90}] (177,-26) rectangle +(5,12);
\fill[fill={rgb,255:red,253;green,153;blue,78}] (182,-26) rectangle +(5,12);
\fill[fill={rgb,255:red,253;green,145;blue,66}] (187,-26) rectangle +(5,12);
\fill[fill={rgb,255:red,251;green,136;blue,56}] (192,-26) rectangle +(5,12);
\fill[fill={rgb,255:red,247;green,127;blue,48}] (197,-26) rectangle +(5,12);
\fill[fill={rgb,255:red,243;green,118;blue,40}] (202,-26) rectangle +(5,12);
\fill[fill={rgb,255:red,240;green,108;blue,33}] (207,-26) rectangle +(5,12);
\fill[fill={rgb,255:red,236;green,99;blue,25}] (212,-26) rectangle +(5,12);
\fill[fill={rgb,255:red,232;green,90;blue,17}] (217,-26) rectangle +(5,12);
\fill[fill={rgb,255:red,225;green,82;blue,12}] (222,-26) rectangle +(5,12);
\fill[fill={rgb,255:red,214;green,77;blue,10}] (227,-26) rectangle +(5,12);
\fill[fill={rgb,255:red,203;green,72;blue,9}] (232,-26) rectangle +(5,12);
\fill[fill={rgb,255:red,193;green,67;blue,7}] (237,-26) rectangle +(5,12);
\fill[fill={rgb,255:red,182;green,62;blue,6}] (242,-26) rectangle +(5,12);
\fill[fill={rgb,255:red,171;green,57;blue,4}] (247,-26) rectangle +(5,12);
\fill[fill={rgb,255:red,243;green,242;blue,239}] (114,-26) rectangle +(12,12);
\node[anchor=north, inner sep=0pt] at (120,-27) {0};
\node[anchor=north, inner sep=0pt] at (192,-27) {50\%};
\node[anchor=north, inner sep=0pt] at (252,-27) {100\%};
\E{300}{-26}
\node[anchor=west, inner sep=0pt] at (354,-20) {no cell taken on};
\node[anchor=west, inner sep=0pt] at (354,-30) {k/n = deviating / taken on};
\end{tikzpicture}
\vspace{6pt}
\caption{\textbf{Figure 3.} Deviating cells over cells taken on, by duty group and client (Chromium; vastlint-client under Node.js): a conditional failure fraction on the cells each client takes on, not a ranking of client quality. Color depth is the rate, and a gray cell means none of the cells taken on deviates; each cell shows k/n. A hatched cell with a dash means the client takes on no cell of that group. IMA is a closed comparison.}
\end{figure}

\hypertarget{what-every-client-gets-right}{%
\subsubsection{5.1.2 What every client gets right}\label{what-every-client-gets-right}}

Every client that follows Wrappers terminates a Wrapper loop and follows HTTP redirects, compressed responses and text/plain responses. Every client that determines error codes gives 303 for a no-ad response behind a Wrapper. Every client accepts a byte order mark and VAST 2.0 and 3.0 documents, and all but iab-vast-loader accept repeated UniversalAdId elements. Every client with a tracker requests all Impression URIs, fires start, the three quartile events and complete once each and in order, fires the pause, mute, unmute and skip trackers, and fires the impression once across user actions. Each also stops at 60 percent without thirdQuartile or complete (Shaka has no API to stop an ad and is not graded), ignores an unknown event name, and keeps tracking after a tracking URL answers HTTP 500. Each except vastlint-client, which leaves skip availability to the integrator, offers the skip control only after a time skipoffset. Those with a tracker that follow Wrappers also fire every Wrapper's Impression, tracking and click-tracking URLs. In Figure 3, XML and parsing duties are wrong mainly in the two clients that parse XML their own way (iab-vast-loader, through iab-vast-parser, and vastlint-client).

\hypertarget{where-the-clients-deviate}{%
\subsubsection{5.1.3 Where the clients deviate}\label{where-the-clients-deviate}}

Five findings matter most. Every tested Wrapper-following open-source client leaves the required macros in a Wrapper's request unreplaced (M12, Macros). vast-client-js's request timeout can end a Node.js process (R01, Robustness). Shaka credits a finishing pod ad's completion to the next ad (P03, Ad pods), and video-ad-sdk sends wrong error codes along Wrapper chains (W07, E03, Error reporting). Whether some clients report a malformed Wrapper target correctly depends on the runtime's XML parser (W11, Section 5.2).

The paragraphs below take the duty groups in the order of the harm they do to the companies that count on the requests, naming the clients that deviate and the mechanism behind the main findings; Appendix B lists every finding with its source location and release range.

\textbf{Macros.} \textbf{None of the five Wrapper-following open-source clients replaces {[}CACHEBUSTING{]} or {[}TIMESTAMP{]} in a Wrapper's VASTAdTagURI (M12, must).} The macro list gives ``VAST request URIs'' as a context for both required macros, and §6.1 places the replacement on the party that makes the request, which is the player. vast-client-js only resolves relative URLs before the request; video-ad-sdk, iab-vast-loader and vastlint-client request the URI as written; rmp-vast expands macros only in the root tag it passes to vast-client-js. A downstream ad server sees literal brackets where it expected a cache buster and a time, so it cannot use the macros to tell a fresh request from a cached or repeated one. The macro grid extends the finding to every request macro in the list (Section 5.3), and the version sweep finds it in every swept release of the four swept clients (Section 5.4). The archived IMA 3.794.0 configuration tested fills both macros (Section 5.1.5), so the duty is implementable. vast-client-js fills or marks as \(-1\) every macro of the 4.3 set, but leaves {[}PLAYBACKMETHODS{]}, which VAST 4.3 mentions and the list marks 4.x, literal (Section 5.3). video-ad-sdk and vastlint-client fill only CACHEBUSTING and TIMESTAMP and send {[}LIMITADTRACKING{]} and the other listed macros literally (M03, M06); Shaka 5.2.12 replaces no macro (M01 to M03, M06, M11); and rmp-vast skips replacement on Error URLs (M11). Two clients get the required format wrong: video-ad-sdk's cache buster has 1 to 10 digits and rmp-vast's is eight alphanumeric characters, where the list specifies an 8-digit integer (M01).

\textbf{Error reporting.} Error requests are how an ad server learns why its ad did not play, and the clients that determine codes often choose the wrong one. video-ad-sdk reports a failed Wrapper fetch as 502, a code for NonLinear (overlay) ads, instead of 301 (W07), at every depth and level of the error grid. It reports a missing playable media file as 401 instead of 403, to the InLine and to every Wrapper (E03, E09). It also reports a media file that answers 404 or cannot be decoded with the code \texttt{undefined}, followed 3 s later by a timeout code (E05, E06), because its ad unit sets a code only for JavaScript Error objects. Shaka never requests a no-ad response's document-level Error URIs, sends a literal {[}ERRORCODE{]} for every media failure, after the impression has already fired, and never reports a media file that does not answer (E01 to E03, E05 to E07). rmp-vast maps an undecodable file to 401 instead of 405 (E06). A Wrapper target that is not well-formed XML must be reported with 100 (W11). vast-client-js reports a malformed Wrapper target as 301 or 303, never 100, and whether video-ad-sdk and iab-vast-loader conform depends on the runtime (Section 5.2). vastlint-client determines no code; its error cells are delegated.

\textbf{Ad pods.} Shaka credits each pod ad's completion request to the next ad: when an ad ends, its \texttt{complete()} first makes the next ad current and only then fires the complete event, whose URLs come from the current ad (P03). This is the case of Section 1. video-ad-sdk sorts pod ads with a comparator that returns 0 for every pair, so a pod in document order 3, 1, 2 plays the ad with sequence 3 first, in the root document and behind a Wrapper alike (P01, P07). vast-client-js sorts a pod in the root document but not one behind a Wrapper, because its nested parse skips the sort (P07), and rmp-vast inherits that. iab-vast-loader checks only the ads outside a pod after each Wrapper, so it reports a pod behind a Wrapper as 303, a document with no ad (P07, the cause of its W02 deviation); Shaka plays pods in document order (P01). video-ad-sdk plays only the first ad of a pod (P06), and rmp-vast and Shaka do not substitute the stand-alone ad when a pod ad fails, or do so only after the pod ends (P05). Shaka 5.3.0, released on the day of the reference runs, fixes the completion crediting (P03) and seven macro and tracking deviations (M01 to M03, M06, M11, T10, T23), percent-encodes macro values and fills {[}ADPLAYHEAD{]} (M04, M08, n/a in 5.2.12 because no macro was replaced), and adds one regression: unknown macros are replaced with \(-1\) (M05). It still sends {[}ADSERVINGID{]} as \(-1\) (M07). Its row in Table 5 shows the net change; the rest of the tables report 5.2.12, the release graded in every browser and on both platforms, and Online Resource 3 gives every 5.3.0 cell.

\textbf{Impressions and tracking.} The deviations are in progress events, user actions and click handling. In the tested setup, progress events at a percentage offset never fire in video-ad-sdk (T07). The vector gives the ad's Duration as 00:00:10, so a percentage offset, which §3.14.3 defines against the ad's progress, has a defined time, and the media element reports the same duration once the media is loaded. The library converts each percentage offset to milliseconds once, from the media element's duration before that duration is set, so every offset becomes NaN. vastlint-client's documented controller fires quartiles but no progress, loaded or rewind events (T05 to T07, T20, T21). Shaka fires no progress, rewind or loaded event, and rmp-vast loses every progress event in macOS WebKit (Section 5.2). rmp-vast fires the resume tracker at ad start, because it fires on every play event including the first (T13), and both rmp-vast and Shaka fire unmute on any volume change (T23). video-ad-sdk and IMA ignore a ClickThrough that only the Wrapper carries (W17, a VAST 4.2 duty), so a click on such an ad opens no landing page; IMA also sends no click-tracking request for it.

\textbf{Wrapper resolution.} Three clients reject a chain of five Wrappers, which VAST 4 players are required to accept (W03): rmp-vast, whose default of four redirects meets an off-by-one in vast-client-js's limit check; vastlint-client, whose depth limit counts documents instead of Wrappers; and IMA, whose own error message gives its limit as four Wrappers. With its defaults, iab-vast-loader never follows a lone Wrapper with \texttt{sequence=\cbr{}\textquotedbl{}1\textquotedbl{}}, which it files as a one-ad pod, and ends the chain with no ad and no error (W02); every IAB VAST 4.x Wrapper sample has that shape. Shaka ignores Wrappers altogether (W01, W02). The Wrapper attributes are honored unevenly: rmp-vast overrides allowMultipleAds with \texttt{true}, vastlint-client and iab-vast-loader never read followAdditionalWrappers, and video-ad-sdk defaults allowMultipleAds to true (W19 to W21). vastlint-client classifies a response that holds a Wrapper ad and a stand-alone InLine ad as an InLine document, so it never follows that Wrapper (W23).

\textbf{XML and parsing.} iab-vast-parser rejects valid VAST 4.2 documents as a whole, with error 101, when a creative has two UniversalAdId elements or a MediaFiles element has two ClosedCaptionFile elements (X05, X06), and fuzzing found four more repeatable elements that cause the same rejection (differential finding D3). Its table of collections lists them as single. vastlint-client reads XML with regular expressions, so it requests an Impression that sits entirely inside an XML comment (X17), does not request an Impression whose element also holds an XML comment (X02), and requests tracking URLs inside Extensions (X09). It also truncates an entity-escaped VASTAdTagURI at the first \texttt{\&\cbr{}amp;} (X18, X19). video-ad-sdk keeps only the first text node of an element, so a URI split across CDATA sections is truncated (X11).

\textbf{Verification and icons.} video-ad-sdk never reads AdVerifications, so a verification vendor whose code it does not run is never told (A01, A03). Shaka never parses AdVerifications (A01), and rmp-vast handles verification only with its \texttt{omid\cbrsoft{}Support} setting. When a Wrapper and its InLine each carry a vendor, IMA reports only the InLine's, and it does not report the tested verification resource that answers 404 (A03, A05; Section 5.1.5). Icons with a display offset never appear in video-ad-sdk, which draws icons once before the duration is known, and appear at once and never leave in rmp-vast, which ignores offset and duration (I03); Shaka draws no icons (I01). IMA requests an icon's view URL before the ad's start tracking request (I01; Section 5.1.5).

\textbf{Media selection.} Shaka sorts media files by height and plays the first without checking its type, so an unsupported file listed first fails the ad (S01).

\textbf{Robustness.} Three defects threaten the process that runs the library instead of the counts. \textbf{vast-client-js's request timeout throws an exception that no caller can catch (R01).} On timeout, its URL handler aborts the request and throws inside the timer callback, outside the request's promise. Under Node.js, which the library supports, the exception ends the process with exit code 1. A reproduction outside the harness under Node.js 20, 22 and 24 crashed in 49 of 49 fault runs and none of 21 controls. With 50 concurrent requests and one hanging Wrapper, none of the 49 healthy requests completed, and an HTTP service that resolves one tag per request stopped accepting connections. A server that resolves VAST chains for server-side ad insertion can therefore be stopped by one slow Wrapper; in a browser the same error is an uncaught page error, and rmp-vast inherits it. The same library clears its timer when the response headers arrive and then reads the body with no timeout, so a Wrapper target that stalls its body makes the client wait at every hop (fuzzing root cause C4). With a 2 s timeout and a 6 s stall, the ad arrived after the stall ended, and a self-referencing Wrapper ran nine stalled hops in 54 s. video-ad-sdk discards the promise of an ad whose MP4 file is marked as VPAID, producing an unhandled rejection that ends a Node.js process without a handler, and then reports an error and an impression for the same play (C3). The other robustness findings return internal TypeErrors where a VAST error code belongs. vast-client-js does so on an ad element with neither InLine nor Wrapper, which also loses a valid sibling ad (C2), and iab-vast-loader on empty, non-XML and non-VAST responses (C5 to C7) and on a lower-case element name (C8).

\hypertarget{integration-and-delegation}{%
\subsubsection{5.1.4 Integration and delegation}\label{integration-and-delegation}}

Three libraries take on only part of a player's work, and the split decides what their verdicts mean. Where vast-client-js and vastlint-client leave a duty to the integrator, the harness performed it as their documentation describes, so their conforming cells show that correct behavior is possible through the documented API. A real integration can do worse: rmp-vast, a player built on vast-client-js, gives up two behaviors. It overrides allowMultipleAds (W20, W21), which vast-client-js honors through its documented API, and it never falls back to the stand-alone ads that vast-client-js keeps in reserve for its integrator (W23). In vastlint-client, a duty can be delegated and still be impossible to carry out. After a fetch failure in the middle of a chain, its session snapshot keeps only the root, so an integrator cannot reach the Error URIs of the intermediate Wrappers that the integrator is supposed to request. Its cells stay delegated, because the harness should not invent the code the library leaves open, but an integrator limited to the documented post-failure snapshot cannot conform on them.

\hypertarget{a-closed-comparison-google-ima-html5-sdk-3.794.0}{%
\subsubsection{5.1.5 A closed comparison: Google IMA HTML5 SDK 3.794.0}\label{a-closed-comparison-google-ima-html5-sdk-3.794.0}}

IMA is Google's client-side ad SDK for web publishers, so it serves as a commercial reference point. It is proprietary, runs in browsers only and loads server-assigned configuration, so two live runs can load different code. Every file the SDK loaded while playing the vectors was recorded once from Google's servers (Chromium, Linux, October 16, 2026), and every later run was answered from that archive, with Google's telemetry answered locally. IMA runs against an HTTPS copy of the recording server, because on a plain-HTTP page it raises AdError 1005 without requesting the tag (Online Resource 4). Its results describe this one archived configuration. Unlike the open-source clients, IMA is analyzed as a black box: its verdicts and findings rest only on the requests the server logged, the events and error codes of its public API, and three probe vectors that isolate a behavior; no explanation in this paper is drawn from its code. Table 5 gives its Chromium verdicts; Linux WebKit gives the same deviations, Firefox adds one (W11, below), and in the macOS run the icon cell I01 conforms in all four browsers, although repeated macOS WebKit runs deviate (Table 6).

IMA fills {[}CACHEBUSTING{]} and {[}TIMESTAMP{]} in a Wrapper's VASTAdTagURI (M12) and sorts a pod behind a Wrapper (P07), which vast-client-js, rmp-vast, video-ad-sdk and iab-vast-loader do not. Its 8 deviating cells in Chromium come from 7 findings, each visible on the wire; Online Resource 4 gives the requests, API events and probes behind each. It rejects a chain of five Wrappers after requesting all five Error URIs with code 302, and its error message gives four Wrappers as its limit (W03). When only the Wrapper carries a ClickThrough, a click on the ad sends no click-tracking request and opens no landing page (W17), a VAST 4.2 duty outside the 4.2 features Google lists as supported. It rejects a Wrapper target that holds a pod and a stand-alone ad with its own error code 2, which is not a VAST code, and after a Wrapper that leads to no ad it reports 303 but does not fall back to another ad (W20, W21, W23). It reports the InLine's verification vendor but not the Wrapper's, sending at most one verificationNotExecuted request per ad in these probes, and does not report the tested verification resource that answers 404 (A03, A05). In every run on every engine and platform it requests an icon's view URL before the ad's start request (I01). On Linux the gap exceeds the 50 ms tolerance in all 20 repeated runs in Firefox (54 to 111 ms) and in WebKit (165 to 237 ms); in Chromium it straddles the tolerance (36 to 68 ms, beyond it in 6 of 20 repeated runs and in both reference runs, 90 and 122 ms), so the Chromium verdict in Table 5 is the verdict of the reference runs and depends on the run. On macOS, in 10 repeated runs per browser, the view preceded start by 11 to 12 ms in Chromium, 10 to 13 ms in Chrome and 13 to 50 ms in Firefox (never beyond the tolerance) and by 51 to 58 ms in WebKit (beyond it in all 10) (Table 6). In Firefox it reports 303 for a malformed Wrapper target, where the other browsers give 100 (W11).

\hypertarget{sensitivity-to-strength-labels}{%
\subsubsection{5.1.6 Sensitivity to strength labels}\label{sensitivity-to-strength-labels}}

M12, like M01 to M03 and M11, rests on the macro list's Required label and on §6.1's responsibility sentence rather than on a modal verb in the 4.3 text. P06's should sits beside a conflicting must in the Ad Pods implementation note (Appendix A). P01 and P07 rest on a literal must. Each configuration is counted once here: the seven open-source configurations in Chromium and vastlint-client under Node.js. Their must-level deviations total 83. If M12 were graded should, the total would fall to 76: vast-client-js would keep 2 must-level deviations (P07, W11), iab-vast-loader 3 or 4, and every other configuration 10 or more. If M12, P01 and P07 were all graded should, the total would fall to 68, and vast-client-js would keep 1 (W11). If every macro requirement that rests on the list's Required label were graded should (M01 to M03, M11, M12), the total would fall to 63. If P06 were graded must, following the implementation note, only video-ad-sdk would change (17 must in Chromium). M03 accepts -1 or -2 for {[}LIMITADTRACKING{]} because a browser has no device advertising ID, although §6.1 reserves those values for optional macros; graded strictly to 0 or 1, M03 would add one must-level deviation and one finding each to vast-client-js (total 84) and IMA (5), and one to Shaka 5.3.0. IMA otherwise keeps 4 in every scenario. No other reading adds or removes a finding. The order of the configurations by must-level deviations does not change when only M12 is relabeled; under the other two readings it changes only between rmp-vast and vastlint-client, which stay within one and two deviations of each other. Online Resource 5 gives the counts per configuration.

\begin{quote}
\textbf{Answer to RQ1.} The open-source clients handle the common path well and diverge where other companies count and diagnose ads: macros in redirect requests, error codes, pod order and pod tracking, progress events and verification. Every tested Wrapper-following open-source client fails to replace the required macros in a Wrapper's request; the archived IMA 3.794.0 configuration tested replaces them and deviates on 8 of the 101 cells it takes on in Chromium (8 percent), from 7 findings, and on 9 of 102 in Firefox, from 8 findings across browsers. The full players carry the most findings (video-ad-sdk 19, rmp-vast 15, Shaka 15) and deviate on 19 to 45 percent of the cells they take on in Chromium (from 20 of 104 for rmp-vast with \texttt{omid\cbrsoft{}Support} to 29 of 64 for Shaka; video-ad-sdk 25 of 96). The libraries that take on only part of a player's work deviate on 7 to 29 percent of the cells they take on (vast-client-js 6 of 85, iab-vast-loader 6 of 30, vastlint-client 20 of 70) and leave duties that integrators can get wrong, as rmp-vast shows. The most serious defects are outside the specification: an uncaught timeout exception that ends a Node.js process, and a timeout that a stalled response body defeats.
\end{quote}

\hypertarget{rq2-runtime-and-platform}{%
\subsection{5.2 RQ2: Runtime and platform}\label{rq2-runtime-and-platform}}

Most verdicts do not depend on the runtime. One of the two Linux WebKit runs graded nine requirements (A01, A02, A04, R04, R05, M06 to M09) on different vectors, so its comparisons leave those nine out; the macOS run used the released package and is compared on every cell. Repeated runs were identical except for IMA's I01 (Section 4.3). They covered the Node.js matrix (540 cells) in four runs on two platforms, the full Chromium matrix (864 cells) in two runs, WebKit on two Linux machines (792 cells), and the error grid (748 cells) and macro grid (580 cells) on both platforms. Between Node.js and the browsers, the four configurations that run in both layers agree in 393 of the 396 cells graded in both in Chromium, Chrome, macOS WebKit and Safari. They agree in all 396 in Firefox and in 392 of 396 in Linux WebKit. Between Linux and macOS, the open-source configurations agree in every cell in Chromium and Firefox and differ in 9 cells in WebKit; IMA's icon cell I01 differs in all three engines (Table 6). Of their 756 cells, the seven open-source browser configurations differ from Chromium in 5 in Firefox and 3 in Linux WebKit; against macOS Chromium in the same Mac run, they differ in 2 of 756 in Chrome and in Safari and in 10 in macOS WebKit. Every difference is in Table 6 and has one of four causes: the runtime's XML parser (W11), the media support of the browser build (S01 in Linux WebKit; S02, through the Playwright Chromium build's missing AAC decoder), engine timing (T05 to T07, T19) and run-to-run timing (IMA's I01). The XML parser matters most.

\begingroup\small\setlength{\tabcolsep}{4pt}\Needspace{15\baselineskip}

\begin{longtable}[]{@{}
  >{\raggedright\arraybackslash}p{(\columnwidth - 12\tabcolsep) * \real{0.1393}}
  >{\raggedright\arraybackslash}p{(\columnwidth - 12\tabcolsep) * \real{0.1820}}
  >{\raggedright\arraybackslash}p{(\columnwidth - 12\tabcolsep) * \real{0.1393}}
  >{\raggedright\arraybackslash}p{(\columnwidth - 12\tabcolsep) * \real{0.1393}}
  >{\raggedright\arraybackslash}p{(\columnwidth - 12\tabcolsep) * \real{0.1393}}
  >{\raggedright\arraybackslash}p{(\columnwidth - 12\tabcolsep) * \real{0.1393}}
  >{\raggedright\arraybackslash}p{(\columnwidth - 12\tabcolsep) * \real{0.1214}}@{}}
\caption{\textbf{Table 6.} Verdicts that differ between layers or browsers (open-source clients, and IMA's one timing-dependent cell; codes in parentheses; ms: icon view before start; repeats: runs beyond the 50 ms tolerance among repeated runs of the cell alone).}\tabularnewline
\toprule\noalign{}
\begin{minipage}[b]{\linewidth}\raggedright
Req.
\end{minipage} & \begin{minipage}[b]{\linewidth}\raggedright
Client
\end{minipage} & \begin{minipage}[b]{\linewidth}\raggedright
Node.js (jsdom)
\end{minipage} & \begin{minipage}[b]{\linewidth}\raggedright
Blink: Chromium, Chrome
\end{minipage} & \begin{minipage}[b]{\linewidth}\raggedright
Gecko: Firefox
\end{minipage} & \begin{minipage}[b]{\linewidth}\raggedright
WebKit (Linux / macOS)
\end{minipage} & \begin{minipage}[b]{\linewidth}\raggedright
Safari
\end{minipage} \\
\midrule\noalign{}
\endfirsthead
\multicolumn{7}{@{}l}{\textbf{Table 6} (continued)}\tabularnewline
\toprule\noalign{}
\begin{minipage}[b]{\linewidth}\raggedright
Req.
\end{minipage} & \begin{minipage}[b]{\linewidth}\raggedright
Client
\end{minipage} & \begin{minipage}[b]{\linewidth}\raggedright
Node.js (jsdom)
\end{minipage} & \begin{minipage}[b]{\linewidth}\raggedright
Blink: Chromium, Chrome
\end{minipage} & \begin{minipage}[b]{\linewidth}\raggedright
Gecko: Firefox
\end{minipage} & \begin{minipage}[b]{\linewidth}\raggedright
WebKit (Linux / macOS)
\end{minipage} & \begin{minipage}[b]{\linewidth}\raggedright
Safari
\end{minipage} \\
\midrule\noalign{}
\endhead
\bottomrule\noalign{}
\endlastfoot
W11 (must) & video-ad-sdk & conforms (100) & deviates (ad accepted) & conforms (100) & deviates / deviates & deviates \\
W11 (must) & iab-vast-loader (both) & deviates (ad accepted) & conforms (100) & deviates (101) & conforms / conforms & conforms \\
T05 to T07 (must), T19 (should) & rmp-vast (both) & not run & conforms & conforms & conforms / deviates (T19 n/a) & conforms \\
S01 (should) & video-ad-sdk & conforms & conforms & conforms & deviates (FLV chosen) / conforms & conforms \\
S02 (may) & rmp-vast (both) & not run & Chromium: n/a (401); Chrome: conforms & conforms & conforms / conforms & conforms \\*
I01 (must) & Google IMA & not run & Chromium, Linux: deviates (122, 90 ms; repeats 6/20, 36 to 68); Chromium, macOS: conforms (9 ms; repeats 0/10, 11 to 12); Chrome: conforms (10 ms; repeats 0/10, 10 to 13) & Linux: deviates (88 ms; repeats 20/20, 54 to 111); macOS: conforms (13 ms; repeats 0/10, 13 to 50) & deviates (367 ms; repeats 20/20, 165 to 237) / conforms (27 ms; repeats 10/10, 51 to 58) & not run \\
\end{longtable}

\endgroup

\textbf{The XML parser (W11).} A Wrapper whose target is not well-formed XML must be rejected with error 100. For a given library version, the verdict follows the XML parser the library uses in that runtime, and there are three parser behaviors.

\begin{itemize}
\tightlist
\item
  \emph{Node.js.} jsdom's DOMParser, like Gecko's, returns a document whose root element is \texttt{parsererror}; video-ad-sdk detects it and reports 100. iab-vast-parser does not use it under Node.js: it parses with xmldom, which recovers a partial tree, and the loader returns an ad.
\item
  \emph{Blink and WebKit.} The browser's DOMParser returns a document that contains a \texttt{parsererror} element alongside the parts it could read. iab-vast-parser checks for that element and reports 100; video-ad-sdk does not check, takes the partial tree as a VAST response, accepts an ad from it, and reports 401 when it finds no media.
\item
  \emph{Gecko.} The DOMParser returns a document whose root element is \texttt{parsererror}. video-ad-sdk then finds no VAST content and reports 100; iab-vast-parser sees a document that is not VAST and reports 101.
\end{itemize}

vast-client-js and rmp-vast deviate in every runtime, with a code that changes (301 under Node.js and in Firefox, 303 elsewhere), and vastlint-client, which uses no XML parser, accepts the ad.
\textbf{Media support (S01, S02).} The S01 vector lists an FLV media file before WebM and MP4 files. The Linux WebKit build reports FLV as playable, so video-ad-sdk chooses it, fails to play it and reports \texttt{undefined} and then 402; S01 deviates. Every other browser, including WebKit and Safari on macOS, lets video-ad-sdk skip the FLV file. The Playwright Chromium build has no AAC decoder, so rmp-vast reports 401 for the optional audio ad without starting it; Chrome, Firefox, WebKit and Safari play it.

\textbf{Engine timing (T05 to T07, T19).} rmp-vast fires no progress tracker in WebKit on macOS and fires them in every other browser, Safari included. It builds its list of progress offsets once, on the first \texttt{durationchange} event, and only if the duration is finite. HTML allows a first \texttt{durationchange} with a duration of NaN or Infinity, and the list then stays empty for the whole ad. Dispatching one early \texttt{durationchange} in Chromium reproduces the pattern in 2 of 2 runs, with 2 of 2 controls unaffected. The engine behavior that triggers it is inferred, because the harness does not log the ad element's duration events, and it is reported as an rmp-vast defect that one engine exposes.

\begin{quote}
\textbf{Answer to RQ2.} The XML parser of the runtime decides one requirement. On it, video-ad-sdk conforms under jsdom and in Firefox and deviates in every tested Blink and WebKit browser, and iab-vast-loader does the reverse, so a library's own jsdom-based test suite cannot see the defect its users meet. Media codecs and engine timing decide a handful of other cells. Everything else is stable: verdicts are identical across repeated runs (except IMA's I01), between Linux and macOS for the open-source configurations in Chromium and in Firefox (WebKit differs in 9 cells, and IMA's icon timing in all three engines, Table 6), and between Node.js and the browsers in 392 to 396 of 396 cells.
\end{quote}

\hypertarget{rq3-beyond-the-catalog}{%
\subsection{5.3 RQ3: Beyond the catalog}\label{rq3-beyond-the-catalog}}

\textbf{Macro grid.} Table 7 gives the results.

\begingroup\small\setlength{\tabcolsep}{4pt}\Needspace{12\baselineskip}

\begin{longtable}[]{@{}llllll@{}}
\caption{\textbf{Table 7.} Macro grid (number of macros). ``value'', ``-1 or -2'' and ``literal'' are what a client sends for the 40 tracking-pixel macros of the 4.3 set. ``4.x'' counts the 5 pixel macros marked 4.x (STOREID, STOREURL, PLAYBACKMETHODS, GPPSTRING, GPPSECTIONID) left literal; ``requests'' counts the request-URI macros (50, 41 in the 4.3 set) the client replaces. Online Resource 5 has the full grid.}\tabularnewline
\toprule\noalign{}
Client & value & -1 or -2 & literal & 4.x & requests \\
\midrule\noalign{}
\endfirsthead
\multicolumn{6}{@{}l}{\textbf{Table 7} (continued)}\tabularnewline
\toprule\noalign{}
Client & value & -1 or -2 & literal & 4.x & requests \\
\midrule\noalign{}
\endhead
\bottomrule\noalign{}
\endlastfoot
vast-client-js 6.5.0 & 6 & 34 & 0 & 5 & 0 / 50 \\
video-ad-sdk 4.0.1 & 2 & 0 & 38 & 5 & 0 / 50 \\
iab-vast-loader 2.5.1 & delegated & delegated & delegated & delegated & 0 / 50 \\*
vastlint-client 0.15.0 & 2 & 0 & 38 & 5 & 0 / 50 \\
\end{longtable}

\endgroup

No Node.js client replaces any request macro, of either set, in either XML form; vastlint-client's escaped form drops 49 of the 50 macros, because it truncates the URI at the first \texttt{\&\cbr{}amp;}, and sends {[}TIMESTAMP{]} literally. In tracking pixels, vast-client-js fills six macros with values (TIMESTAMP, CACHEBUSTING, ADTYPE, UNIVERSALADID, ASSETURI, ADSERVINGID), marks the other 34 of the 4.3 set as \(-1\), and leaves only the five pixel macros marked 4.x literal, among them {[}PLAYBACKMETHODS{]}, which the VAST 4.3 text mentions. video-ad-sdk and vastlint-client fill CACHEBUSTING and TIMESTAMP and leave the other 38 of the 4.3 set literal.

\textbf{Error grid.} Table 8 gives the results.

For the four Node.js clients, each client gives the same code at every depth and level for a given failure, with two exceptions, vast-client-js on a stalled document and video-ad-sdk on a missing VASTAdTagURI, described below. Separately, the grid's malformed document (a truncated file) differs from the catalog's (a mismatched end tag), and on it iab-vast-loader reports 303 instead of accepting an ad and vastlint-client resolves no ad. vast-client-js accepts the ad when the stalled document is the InLine (the stalled-body defect, C4) and reports 301 when it is a Wrapper. video-ad-sdk reports 502 for a missing VASTAdTagURI except at the sixth level, where it reports 302. For a Wrapper without a VASTAdTagURI, vast-client-js reports 303 to the Wrappers above it and nothing to that Wrapper itself, at every level.

\begingroup\small\setlength{\tabcolsep}{4pt}\Needspace{15\baselineskip}

\begin{longtable}[]{@{}
  >{\raggedright\arraybackslash}p{(\columnwidth - 10\tabcolsep) * \real{0.1339}}
  >{\raggedright\arraybackslash}p{(\columnwidth - 10\tabcolsep) * \real{0.1719}}
  >{\raggedright\arraybackslash}p{(\columnwidth - 10\tabcolsep) * \real{0.1719}}
  >{\raggedright\arraybackslash}p{(\columnwidth - 10\tabcolsep) * \real{0.1786}}
  >{\raggedright\arraybackslash}p{(\columnwidth - 10\tabcolsep) * \real{0.1719}}
  >{\raggedright\arraybackslash}p{(\columnwidth - 10\tabcolsep) * \real{0.1719}}@{}}
\caption{\textbf{Table 8.} Error grid: code and verdict per client, by failure (15 cells per failure per client, 7 baselines; 187 cases and 748 cells in all). \(d\) is the number of documents in the chain: \(d - 1\) Wrappers and the InLine. A cell with two verdicts differs by depth or level.}\tabularnewline
\toprule\noalign{}
\begin{minipage}[b]{\linewidth}\raggedright
Failure
\end{minipage} & \begin{minipage}[b]{\linewidth}\raggedright
Accepted codes
\end{minipage} & \begin{minipage}[b]{\linewidth}\raggedright
vast-client-js
\end{minipage} & \begin{minipage}[b]{\linewidth}\raggedright
video-ad-sdk
\end{minipage} & \begin{minipage}[b]{\linewidth}\raggedright
iab-vast-loader (code determined)
\end{minipage} & \begin{minipage}[b]{\linewidth}\raggedright
vastlint-client
\end{minipage} \\
\midrule\noalign{}
\endfirsthead
\multicolumn{6}{@{}l}{\textbf{Table 8} (continued)}\tabularnewline
\toprule\noalign{}
\begin{minipage}[b]{\linewidth}\raggedright
Failure
\end{minipage} & \begin{minipage}[b]{\linewidth}\raggedright
Accepted codes
\end{minipage} & \begin{minipage}[b]{\linewidth}\raggedright
vast-client-js
\end{minipage} & \begin{minipage}[b]{\linewidth}\raggedright
video-ad-sdk
\end{minipage} & \begin{minipage}[b]{\linewidth}\raggedright
iab-vast-loader (code determined)
\end{minipage} & \begin{minipage}[b]{\linewidth}\raggedright
vastlint-client
\end{minipage} \\
\midrule\noalign{}
\endhead
\bottomrule\noalign{}
\endlastfoot
(none) & resolve \(d \le 6\); \(d = 7\) may 302 & conforms & conforms & conforms & deviates at \(d = 6\); rejects \(d = 7\) with no code \\
http404 & 301 & 301; conforms & 502; wrong code & 301; conforms & delegated (no code) \\
http500 & 301 & 301; conforms & 502; wrong code & 301; conforms & delegated (no code) \\
timeout & 301 & 301; conforms; uncaught exception & 301; conforms & 301; conforms & delegated (no code) \\
stall & 301 & ad accepted or 301; deviates or conforms & 301; conforms & 301; conforms & delegated (no code) \\
refused & 301 & 301; conforms & 502; wrong code & 301; conforms & delegated (no code) \\
empty & 100/303 & 301; wrong code & 100; conforms & none; no code; internal error & delegated (no code) \\
html & 100/303 & 301; wrong code & 100; conforms & 101; wrong code; internal error & delegated (no code) \\
malformed & 100 & 301; wrong code & 100; conforms & 303; wrong code & delegated (no code) \\
novast & 101/100/303 & 301; wrong code & 303; conforms & 101; conforms; internal error & delegated (no code) \\
noad & 303 & 303; conforms & 303; conforms & 303; conforms & delegated (no code) \\
noad\_err & 303 & 303; conforms & 303; conforms & 303; conforms & delegated (no code) \\*
nouri & 101/100/301/303 & 303; partial & 502 or 302; wrong code & 301; conforms & delegated (no code) \\
\end{longtable}

\endgroup

\textbf{Mutation fuzzing.} The campaign ran 5 rounds (289, 291, 290, 283 and 289 cases; 1,442 in all, of which 1,437 were unique) and found 12 failure classes from 9 root causes: 8 library defects and 1 environment artifact (Appendix D). By client: vast-client-js 3 (C1, C2, C4), video-ad-sdk 1 (C3), iab-vast-loader 4 (C5 to C8), vastlint-client 0. The artifact, C9, is an unhandled rejection in video-ad-sdk's icon code that occurs only because Node.js 22's own CustomEvent reaches jsdom's event target; the affected icon cells are browser-only, so no verdict changes. Five of the 12 classes are raised by an unmutated seed as well; the mutations alone contributed C2, C4, C5 and C8.

New classes per round were 8, 2, 1, 1 and 0, and new root causes 8, 1, 0, 0 and 0. The empty final round is weak evidence that classes are exhausted: 3 classes occur in 1 case each, none in exactly 2, and the bias-corrected Chao1 estimate (\protect\hyperlink{ref-chao1984nonparametric}{Chao 1984}, \protect\hyperlink{ref-chao1987capture}{1987}; \protect\hyperlink{ref-bohme2018stads}{Böhme 2018}) is 15.0 classes (95 percent interval 12.4 to 37.3) against 12 observed. Classes are message strings, and one root cause (C8) accounts for four of them, so root causes are the unit reported. The last new root cause appeared at unique case 360, and rounds 3 to 5 (858 cases not seen before) added none. Under this generator, a root cause with a per-case rate above 0.35 percent would have shown with 95 percent probability. The bound assumes independent cases and holds only for this generator's seeds and operators (Section 7).

\textbf{Differential pass.} The differential pass found four conformance differences the robustness oracle cannot see. They are malformed XML accepted by iab-vast-loader under Node.js and by vastlint-client (D1), 303 from vast-client-js for an InLine with no usable media file (D2), four more repeatable elements rejected by iab-vast-parser (D3), and the stalled body (D4, the same defect as C4).

\begin{quote}
\textbf{Answer to RQ3.} For the error requirements, mostly. Across 187 generated chains per client, each client gives the same code at every depth and level for a given failure, with two level-dependent exceptions. For a malformed Wrapper target, iab-vast-loader and vastlint-client respond differently to the grid's truncated file than to the catalog's file with a mismatched end tag. The macro grid extends the Wrapper-macro finding from the two macros the catalog grades to all 41 request macros of the 4.3 set. Fuzzing found 8 library defects, 7 of them outside the catalog (C1 is R01, rediscovered): two end a Node.js process (C1, C3) and one defeats the integrator's timeout (C4); the differential pass added four conformance differences. It gives no evidence that the space of such defects is exhausted.
\end{quote}

\Needspace{14\baselineskip}

\hypertarget{rq4-history}{%
\subsection{5.4 RQ4: History}\label{rq4-history}}

Table 9 lists every verdict change between consecutive swept releases; the bullets explain the main changes and the deviations that never changed.

\begingroup\small\setlength{\tabcolsep}{4pt}\Needspace{11\baselineskip}

\begin{longtable}[]{@{}llll@{}}
\caption{\textbf{Table 9.} Version sweep: verdict changes between swept releases (Node.js layer, 93 requirements, last patch of each minor release).}\tabularnewline
\toprule\noalign{}
Client & Req. & Between releases & Verdict \\
\midrule\noalign{}
\endfirsthead
\multicolumn{4}{@{}l}{\textbf{Table 9} (continued)}\tabularnewline
\toprule\noalign{}
Client & Req. & Between releases & Verdict \\
\midrule\noalign{}
\endhead
\bottomrule\noalign{}
\endlastfoot
vast-client-js & M09 & 3.3.0 and 4.0.1 & conforms to deviates \\
vast-client-js & P01 & 6.1.0 and 6.2.0 & deviates to conforms \\
vast-client-js & T02 & 4.0.1 and 5.0.0 & deviates to conforms \\
vast-client-js & R01 & 5.0.0 and 6.0.2 & conforms to deviates \\
video-ad-sdk & W04 & 2.3.0 and 2.4.3 & deviates to conforms \\
video-ad-sdk & W06 & 2.3.0 and 2.4.3 & deviates to conforms \\
video-ad-sdk & E02 & 2.3.0 and 2.4.3 & deviates to conforms \\
video-ad-sdk & T05 & 2.3.0 and 2.4.3 & deviates to conforms \\
video-ad-sdk & T06 & 2.3.0 and 2.4.3 & deviates to conforms \\
video-ad-sdk & M11 & 2.5.1 and 2.6.3 & n/a to conforms \\*
video-ad-sdk & P01 & 2.7.2 and 3.0.0 & conforms to deviates \\
\end{longtable}

\endgroup

\begin{itemize}
\tightlist
\item
  \textbf{The Node.js crash dates from vast-client-js's move to fetch.} Release 6.0.0 replaced the Node.js URL handler, which aborted the request and passed a 408 error to its callback, with a fetch-based handler that throws inside the timer callback. Under plain Node.js, 6.0.0 to 6.0.2 cannot make any request (each fails on an undefined \texttt{window}) and 6.1.0 to 6.5.0 crash; with the harness's browser globals, 6.0.2 throws as well. The release notes place the change at 6.0.0.
\item
  \textbf{vast-client-js 4.0.1 began sending every UniversalAdId.} Release 3.3.0 fills {[}UNIVERSALADID{]} with the first ID only, which conforms because several IDs are optional; from 4.0.1 the library joins both IDs and percent-encodes the separating comma (M09).
\item
  \textbf{Pod order was fixed in one library and broken in the other.} vast-client-js sorts root pods from 6.2.0. video-ad-sdk sorted them through 2.7.2; its 3.0.0 rewrite introduced the comparator that always returns 0.
\item
  \textbf{video-ad-sdk 2.4.x fixed 5 cells} (the 302 Wrapper-limit code, every document-level Error URI and time-offset progress events). Releases up to 2.5.1 report nothing for unplayable media; the release notes place the change at 2.6.1, and the sweep bounds it between 2.5.1 and 2.6.3.
\item
  \textbf{No change in iab-vast-loader or vastlint-client.} iab-vast-loader's verdicts are the same from 2.0.0 (2018) to 2.5.1 (2021), and vastlint-client's JavaScript did not change across its 12 swept releases; only its validation core did.
\item
  \textbf{Long-standing behavior.} M12 deviates in every swept release of the four swept clients; vast-client-js reports 301 for a malformed Wrapper target in all 9 releases; video-ad-sdk's 502 code, Wrapper ClickThrough, cache-buster length, LIMITADTRACKING and percentage-progress deviations are present in all 9.
\end{itemize}

Four verdict changes in the sweep's results are not in Table 9, because none is a change in library behavior. video-ad-sdk's empty-Impression cell (X13) is a harness artifact: rerun on its own, it conforms before 3.0.0 as after. video-ad-sdk 2.1.0's T18 could not be driven, because that release has no skip method, and video-ad-sdk's T19 follows its T05 fix as a dependent cell. video-ad-sdk's W13 becomes n/a at 2.6.3 because, from then on, it sends one consistent wrong code that E03 grades (rule G2); before, it requested no Error URI.

\begin{quote}
\textbf{Answer to RQ4.} Most deviations are as old as the swept history: M12 is in every swept release of the four swept clients, and several of video-ad-sdk's are in all 9 of its releases. Three changes made a client deviate: vast-client-js's UniversalAdId encoding at 4.0.1, its Node.js crash at its move to fetch, and video-ad-sdk's pod order at its 3.0.0 rewrite. Maintainers have fixed others (vast-client-js's root pod order at 6.2.0, 5 video-ad-sdk cells at 2.4.x).
\end{quote}

\hypertarget{discussion}{%
\section{6. Discussion}\label{discussion}}

\hypertarget{what-the-deviations-cost}{%
\subsection{6.1 What the deviations cost}\label{what-the-deviations-cost}}

A completion request credited to the next ad in a pod leaves the pod's total right and the per-ad counts wrong. An error coded \texttt{undefined}, or 502 for a failed fetch, sends an ad server's diagnosis to the wrong cause. A Wrapper request with literal macros defeats the cache-busting that ad servers rely on to tell a fresh request from a cached one. A verification vendor whose code was never run receives no verificationNotExecuted request, so its data shows an ad that was not measured. The robustness defects threaten availability rather than counts: a server that resolves VAST can be stopped by one slow Wrapper or delayed past the ad break by a stalled one (Section 5.1.3).

A plausible reason these deviations persist across releases (Section 5.4) is that most of them do not break playback. The explanation applies to the tracking defects: each produces a request that looks normal, or a missing request that nobody notices, and the party that would notice runs neither the library nor its tests. It does not apply to the deviations a viewer can see, such as a rejected chain of five Wrappers (W03) or a valid document rejected as a whole (X05, X06). The study does not test this explanation, but it is consistent with the sweep and with the maintainers' issue trackers, where several of these defects were reported years ago (Section 3).

A conformance claim about a player built on one of these libraries depends on which duties its integrator wrote; the delegated verdict makes that split visible (Section 5.1.4).

\hypertarget{implications}{%
\subsection{6.2 Implications}\label{implications}}

\textbf{For library maintainers.} The checks with the highest yield are cheap:

\begin{itemize}
\tightlist
\item
  replace {[}CACHEBUSTING{]} and {[}TIMESTAMP{]} in every Wrapper request;
\item
  bound both the headers and the body of a request by the same timeout, and reject rather than throw when it expires;
\item
  sort pods after every Wrapper hop, not only at the root;
\item
  map each failure to its own code from the error table;
\item
  run the test suite in at least one Blink or WebKit browser as well as under jsdom, because the XML parser decides some verdicts.
\end{itemize}

\textbf{For integrators and player builders.} A library that delegates a duty is conformant only together with the code that performs it. Before relying on one of these libraries, an integrator should know:

\begin{itemize}
\tightlist
\item
  which duties it leaves open (Table 1);
\item
  whether those duties can be carried out through its API (vastlint-client's failed chains show that they sometimes cannot);
\item
  which of its defaults change the outcome (rmp-vast's redirect limit and allowMultipleAds override).
\end{itemize}

\textbf{For IAB Tech Lab.} A published set of player vectors with expected behavior (which requests, with which codes, in which order) would let library maintainers test themselves; the 69 vectors and the two grids here are a starting point. A machine-checkable 4.3 schema would also help. The file \texttt{schemas/\cbr{}vast\_\cbrsoft{}4.\cbrsoft{}3.\cbrsoft{}xsd} in IAB Tech Lab's VAST repository (commit 99d5d51) is empty, and the published schemas declare each content model as an ordered sequence, so a document whose InLine children follow the order in which §3.4 lists them fails validation. Two strength statements are hard to grade. The pod body text says a player should play as many ads as possible while its implementation note says all ads must be played, and the macros' request-URI obligation is stated as a responsibility rather than a must.

\textbf{For researchers.} Ad-tech clients, small independent implementations of one specification whose failures land on third parties, are a natural fit for specification-derived testing. Clients take on different duties and share code, so agreement is a poor oracle. The catalog, harness and grading rules can be reused for other ad formats and for HLS and DASH interstitial players.

\hypertarget{reporting-to-maintainers}{%
\subsection{6.3 Reporting to maintainers}\label{reporting-to-maintainers}}

The vast-client-js timeout exception was public before this study: it had been reported in the project's issue tracker as a browser error (\protect\hyperlink{ref-8bitsten2026issue510}{8bitsten 2026}), and while the study was under way another contributor opened a public pull request that removes the throw (\protect\hyperlink{ref-justinjeong52026pr514}{justinjeong5 2026}). Neither addresses the stalled body, which was not public. On October 16, 2026, the author reported both defects and the Node.js consequence privately to the project's security contact, offering a fix, and had received no reply by October 28. On October 28, 2026, the author opened a public pull request that complements the existing one by bounding the response body read with the same timeout, with tests (\protect\hyperlink{ref-chaudhary2026pr515}{Chaudhary 2026}), so the stalled body was disclosed together with a proposed fix; both pull requests were open on October 28. Under the patched build the crash and the stalled-body wait disappear in every reproduction, and the project's test suite passes. The remaining findings concern conformance; the author will report them as public issues to each active project (vast-client-js, rmp-vast, vastlint-client, Shaka Player for the findings that still hold in its 5.3 line) after the preprint is posted, each first checked against the project's latest release, and the replication repository's README will list each issue.

\hypertarget{threats-to-validity}{%
\section{7. Threats to validity}\label{threats-to-validity}}

\textbf{Construct validity.} The catalog operationalizes VAST 4.3 as 108 requirements extracted from its obligation lines and the macro list; another extraction could cut the duties differently or assign different strengths. The extraction inventory and coverage map are released, Appendix A gives the basis and literal modal of every requirement, one rule assigns strength, and the sensitivity analysis (Section 5.1.6) shows that no finding depends on the labels in dispute and that the order of clients by must-level deviations changes only between rmp-vast and vastlint-client. Deviations are not weighted by severity, so a must-level icon deviation counts like a must-level error code. Clients that claim older versions are still graded against 4.3 (Section 4.1). For libraries that delegate duties, a conforming cell measures whether correct behavior is possible through the documented API with the harness as a careful integrator. It is an upper bound on what a real integration achieves, and rmp-vast measures one real integration. Cells are not independent, so findings are reported alongside them (Section 5.1.1). Several couplings remain. One vector can serve several requirements, so a defect in that vector or in the client's handling of it moves all of them (for example, the vector of M10 lists an FLV file first, which ties M10 to media selection). Some n/a verdicts depend on another outcome (G1, G2), so a client that fails early has fewer cells graded. The macro requirements depend on which macros the list marks Required and in which contexts, and on the list commit used (Section 4.2). How cells are grouped into findings is a judgment; a finding is one cause, but another analyst could split or merge some. Download counts measure use, not deployment in players, so they do not weigh the findings by the viewers they reach.

\textbf{Internal validity.} The checks of Table 4 guard against reporting a harness artifact as library behavior, and every robustness defect was reproduced by a separate script outside the harness. The Node.js layer supplies browser globals through jsdom, which changes the code path of some releases (vast-client-js 6.x takes its browser branch) and produced one false failure, traced and excluded (C9). IMA runs over HTTPS with a browser security check disabled, while the other clients run over HTTP, and its telemetry is answered locally. If any IMA behavior depends on a telemetry response or server-assigned state, the archived run cannot show it.

\textbf{External validity.} The tested clients are a purposive sample. They cover the three integration styles the survey found and include the library behind IAB Tech Lab's tester and a player built on another client. However, four eligible small players and the derivatives of vast-client-js were not tested, and native mobile and TV SDKs are outside a JavaScript harness. Findings that trace to vast-client-js apply to players that bundle an affected release. The browser layer covers desktop browsers on Linux and macOS with short synthetic clips from a local server; mobile WebViews, smart-TV browsers, adaptive streams and production network conditions are not covered. The clients keep moving: Shaka Player 5.3.0 changes several verdicts (Section 5.1.3). The releases current on 27 October 2026 give the tested verdicts: vastlint-client 0.15.3 and 0.16.2 match 0.15.0 in all 108 Node.js cells; Shaka Player 5.2.15 matches 5.2.12 in all 108 cells in Chromium, Firefox and WebKit; and Shaka Player 5.3.2 matches 5.3.0 in Chromium and shows the same 11 changes from 5.2.12 in Firefox and WebKit. The other clients had no later release. The fuzzer ran at the Node.js layer only, so Shaka, rmp-vast and IMA were not fuzzed; network mutations were 5 percent of the mutations applied; and the fuzzing bound holds only for this generator. The version sweep ran old releases with October 2026 dependencies, so their XML-handling verdicts reflect current parser versions.

\textbf{Reliability.} Verdicts are deterministic: repeated runs and both platforms gave identical verdicts except for the reported engine differences (Section 5.2), and on a second machine and platform the packaged harness reproduced the Node.js and Chromium rows of Table 5 for every pinned open-source configuration (Shaka 5.3.0 was run separately), Tables 7 and 8 and the access check. The sweep and the fuzzing campaign were run once. The requirements, vectors and oracles were written by one author, who works in the industry these clients serve, and the catalog was built while outcomes were already visible, so the cut of duties and the strengths could have been influenced by them; the extraction inventory and the strength rule are released so that this can be checked. The timing thresholds (Section 4.3) are fixed choices, and a different tolerance could move a cell near its boundary. The consistency audits (Table 4b) were automated; they catch errors but are not a measure of inter-rater reliability. No human second rater has reviewed the catalog or the verdicts, and that is the main remaining gap.

\hypertarget{conclusion}{%
\section{8. Conclusion}\label{conclusion}}

Open-source VAST clients get the common path right and diverge on the requests that ad servers and verification vendors rely on (RQ1). None of the tested open-source clients that follow Wrappers fills the required macros in a Wrapper's request, while the archived configuration of Google's closed IMA SDK 3.794.0 that was tested does. One library ends a Node.js process on a slow Wrapper and lets a stalled body outlast its timeout, and a player built on it inherits both; pods play out of order in several clients, and Shaka Player 5.2.12 credits a finishing pod ad's completion to the next ad. A library can conform under jsdom and deviate in every tested Blink and WebKit browser because of the runtime's XML parser, while almost every other verdict is stable across runs and platforms (RQ2). Generated grids reproduce the catalog's error-code verdicts across depths and levels, with the exceptions of Section 5.3, and extend the macro finding to every request macro, and fuzzing finds robustness defects the specification does not name (RQ3). Most deviations are as old as the swept history (RQ4). Each open-source finding traces to a source location or to a missing feature, and the released harness can check a fix. Native SDKs, server-side insertion and the remaining small players are open.

\hypertarget{statements-and-declarations}{%
\section{Statements and Declarations}\label{statements-and-declarations}}

\textbf{Funding.} No funding was received for this work.

\textbf{Competing interests.} The author has no competing interests to declare. The author works in advertising technology and has no affiliation with the maintainers of the tested clients, IAB Tech Lab or Google.

\textbf{Ethics.} The study uses no human participants and no production data. All test traffic in the reported runs went to local servers. For IMA, the SDK files were recorded once from Google's servers on October 16, 2026, and served from that local archive afterward, with Google's telemetry hosts answered locally; Online Resource 4 documents the setup. IMA is graded only from its network requests and public API. The process-ending defect was disclosed privately to the maintainers before publication (Section 6.3).

\textbf{Data availability.} The harness, oracles, fuzzer, packaged runner and other code (Apache-2.0) and the test vectors, synthetic test media, requirement catalog with its extraction inventory and coverage map, grids and all results (CC BY 4.0) are at \url{https://github.com/methodtrace/2026-vast-client-conformance} (tag v1.0.0), archived at Zenodo (\url{https://doi.org/10.5281/zenodo.23581408}), following open-science practice for empirical software engineering (\protect\hyperlink{ref-mendezfernandez2019openscience}{Méndez Fernández et al. 2019}). Online Resources 1 to 5 are in that archive and can be browsed at \url{https://methodtrace.github.io/2026-vast-client-conformance}. One command installs a pinned toolchain, runs every layer and seals the results with hash manifests; on the author's Mac it took 73 minutes. The IAB sample tags and specification text are third-party material and are fetched by script. The archived IMA SDK files are proprietary and are kept privately; the archive's hashes and the recording script are released. The package runs every open-source client without them, and runs IMA only when the archive is supplied.

\textbf{Use of generative AI.} During the preparation of this work the author used Claude Opus 5.5 (Anthropic) to create the test harness, trace deviations, run the automated consistency audits of Section 4.6, and help with drafting the text. After using this tool, the author reviewed and edited the content as needed and takes full responsibility for the content of this manuscript.

\appendix

\hypertarget{appendix-a.-requirement-catalog}{%
\section{Appendix A. Requirement catalog}\label{appendix-a.-requirement-catalog}}

Each row gives the requirement as graded, the VAST 4.3 section or macro list entry it rests on, and the obligation as the text writes it next to the strength graded. In the last column, ``required'' is the text's own word; ``list'' means the IAB macro list marks the macro Required; ``none'' means a definition or attribute with no modal verb; ``imper.'' an instruction with no modal verb; ``n/a'' a robustness observation, which is not a VAST clause; ``obs.'' a requirement graded as an observation. (a) marks a graded strength that differs from the literal word; the rule of Section 4.2 decides it, and Online Resource 5 quotes the sentence behind every row.

\begingroup\small\setlength{\tabcolsep}{4pt}\Needspace{13\baselineskip}

\begin{longtable}[]{@{}
  >{\raggedright\arraybackslash}p{(\columnwidth - 6\tabcolsep) * \real{0.0430}}
  >{\raggedright\arraybackslash}p{(\columnwidth - 6\tabcolsep) * \real{0.4731}}
  >{\raggedright\arraybackslash}p{(\columnwidth - 6\tabcolsep) * \real{0.3440}}
  >{\raggedright\arraybackslash}p{(\columnwidth - 6\tabcolsep) * \real{0.1399}}@{}}
\caption{\textbf{Table A1.} Requirement catalog: requirement as graded, VAST 4.3 basis, and strength as written and as graded.}\tabularnewline
\toprule\noalign{}
\begin{minipage}[b]{\linewidth}\raggedright
ID
\end{minipage} & \begin{minipage}[b]{\linewidth}\raggedright
Requirement
\end{minipage} & \begin{minipage}[b]{\linewidth}\raggedright
VAST 4.3 basis
\end{minipage} & \begin{minipage}[b]{\linewidth}\raggedright
Written / graded
\end{minipage} \\
\midrule\noalign{}
\endfirsthead
\multicolumn{4}{@{}l}{\textbf{Table A1} (continued)}\tabularnewline
\toprule\noalign{}
\begin{minipage}[b]{\linewidth}\raggedright
ID
\end{minipage} & \begin{minipage}[b]{\linewidth}\raggedright
Requirement
\end{minipage} & \begin{minipage}[b]{\linewidth}\raggedright
VAST 4.3 basis
\end{minipage} & \begin{minipage}[b]{\linewidth}\raggedright
Written / graded
\end{minipage} \\
\midrule\noalign{}
\endhead
\bottomrule\noalign{}
\endlastfoot
W01 & Follows a Wrapper to its InLine ad. & §2.2; §2.3.5 & must / must \\
W02 & Follows a lone Wrapper that carries \texttt{sequence=\cbr{}\textquotedbl{}1\textquotedbl{}} (a one-ad pod) to its InLine ad, or declines it with an error (pod support is optional). & §2.2; The Ad Element (sequence); Ad Pods & must / must \\
W03 & Accepts a chain of five Wrappers before the InLine ad (players are required to accept five). & §3.19.1 & required / must \\
W04 & With six Wrappers the player may reject; if it does, it determines 302 (and, with a tracker, reports it to the Error URI of every Wrapper it received). & §2.3.5.1; §3.19.1 & should / should \\
W05 & Terminates a Wrapper that points to itself (stops requesting it). & §2.3.5.1 & none / should (a) \\
W06 & Reports 302 (Wrapper limit) for a self-referencing Wrapper loop. & §2.3.5.1; Error Reporting table (302) & should / should \\
W07 & When the VASTAdTagURI is unavailable (HTTP 404), determines error 301. & Error Reporting (Media Player Details; table 301) & must / must \\
W08 & When the VASTAdTagURI times out, determines error 301. & Error Reporting (table 301) & must / must \\
W09 & When a Wrapper leads to a no-ad response and the player determines a code, the code is 303. & Error Reporting (No Ad Response); Error (VAST) & should / should \\
W10 & Requests the VAST-level Error URI of a no-ad response reached through a Wrapper (with 303 for {[}ERRORCODE{]}). & §3.2.1 Error (VAST); Error Reporting (No Ad Response) & must / must \\
W11 & When the Wrapper target is not well-formed XML, determines error 100 (and does not use an ad from it). & Error Reporting (table 100); §2.3.2 (well-formed XML) & must / must \\
W12 & Requests the Wrapper Error URI when its VASTAdTagURI fails. & Error Reporting (Media Player Details) & must / must \\
W13 & An InLine error after a chain of Wrappers is reported to the InLine and to every Wrapper Error URI, with the same code (403 here). & Error Reporting (Media Player Details); §3.4.11 & must / must \\
W14 & At impression time, requests the Impression URIs of the InLine ad and of every Wrapper before it. & §3.4.4; §2.3.4 & required / must \\
W15 & Fires the Wrapper's Linear tracking events (start, quartiles, complete) together with the InLine ones. & §3.19 (tracking elements in Wrappers) & required / must \\
W16 & When both the InLine and a Wrapper carry a ClickThrough, the InLine one is used. & §2.3.5.2 & must / must \\
W17 & When the InLine has no ClickThrough (no VideoClicks), the ClickThrough of the Wrapper closest to the InLine is used. & §2.3.5.2 & must / must \\
W18 & On a click, requests the ClickTracking URIs of the InLine and of every Wrapper. & §3.10.2 ClickTracking; §3.19 & required / must \\
W19 & With \texttt{follow\cbrsoft{}Additional\cbrsoft{}Wrappers=\cbr{}\textquotedbl{}false\textquotedbl{}}, a Wrapper received next is ignored rather than followed. & §3.19 followAdditionalWrappers & should / should \\
W20 & With \texttt{allow\cbrsoft{}Multiple\cbrsoft{}Ads=\cbr{}\textquotedbl{}false\textquotedbl{}}, only the first stand-alone Ad of the response is used (pod ads are dropped). & §3.19 (allowMultipleAds) & none / should (a) \\
W21 & Without the attribute, allowMultipleAds defaults to false, so only the first stand-alone Ad is used. & §3.19 (allowMultipleAds default false) & none / should (a) \\
W22 & Optional: rejects an InLine whose Category is blocked upstream; a player that does reject it must report 205. & §3.19.2 BlockedAdCategories & may / may \\
W23 & With \texttt{fallback\cbrsoft{}On\cbrsoft{}No\cbrsoft{}Ad=\cbr{}\textquotedbl{}true\textquotedbl{}}, when the Wrapper's target returns no ad, a stand-alone ad from the response is used. & §3.19 (fallbackOnNoAd) & should / should \\
E01 & For a no-ad response, requests the VAST-level Error URI. & §3.2.1 Error (VAST) & must / must \\
E02 & Requests every VAST-level Error URI when several are given. & §3.2.1 Error (VAST), bounded 0+ & must / must \\
E03 & With no supported MediaFile, requests every InLine Error URI with 403. & §3.4.11; Error Reporting (table 403) & must / must \\
E04 & Replaces {[}ERRORCODE{]} in Error URIs with a numeric code. & Error Reporting (Media Player Details); macro list ERRORCODE & must / must \\
E05 & When the MediaFile URI returns HTTP 404, requests the InLine Error URI with 401 (file not found). & Error Reporting (table 401) & must / must \\
E06 & When a MediaFile of a supported type cannot be decoded, requests the InLine Error URI with 405. & Error Reporting (table 405) & must / must \\
E07 & When the MediaFile URI answers only after 30 s, requests the InLine Error URI with 402 (media timeout). & Error Reporting (table 402) & must / must \\
E08 & Does not request the Error URI when the ad plays without a problem. & §3.4.11 (Error is for ads that cannot be served) & none / should (a) \\
E09 & Given a NonLinear-only ad, either displays it (impression or creativeView) or reports 200/201 (unexpected ad type or linearity). & §2.3.1; Error Reporting (table 200, 201) & must / must \\
T01 & Requests every Impression URI of the ad. & §2.3.4; §3.4.4 & required / must \\
T02 & Does not dispatch an Impression whose URI is \texttt{about:\cbrsoft{}blank}. & §3.4.4 (\texttt{about:\cbrsoft{}blank}) & should / should \\
T03 & Fires start, firstQuartile, midpoint, thirdQuartile and complete exactly once each, in that order. & §3.14.1 (Linear metrics) & required / must \\
T04 & Fires every Tracking element for the same event (two start trackers for two parties). & §3.14.3 (multiple Tracking elements per event) & required / must \\
T05 & Fires a progress tracker with an HH:MM:SS offset. & §3.14.1 (progress, HH:MM:SS offset) & required / must \\
T06 & Fires a progress tracker with an HH:MM:SS.mmm offset. & §3.14.1 (progress, HH:MM:SS.mmm offset) & required / must \\
T07 & Fires progress trackers with percentage offsets (10\% and 80\%). & §3.14.1 (progress, n\% offset) & required / must \\
T08 & Stopping at 60\% fires midpoint but neither thirdQuartile nor complete. & §3.14.1 (thirdQuartile, complete definitions) & required / must \\
T09 & Spread between the first and last Impression request at impression time (ms, measured at the server). & §2.3.4; §3.4.4 (same time) & required / obs. (a) \\
T10 & Does not request the Impression of an ad whose media never plays (no supported MediaFile). & §2.3.4 (IAB Digital Video Measurement Guidelines); §3.4.4 & none / should (a) \\
T11 & Requests the Impression exactly once across pause, resume, mute and unmute. & §3.4.4 (notify when the impression occurs); §2.3.4 & none / should (a) \\
T12 & Fires the pause tracker once when the user pauses. & §3.14.1 (pause) & required / must \\
T13 & Fires the resume tracker once when the user resumes. & §3.14.1 (resume) & required / must \\
T14 & Fires the mute tracker once when the user mutes. & §3.14.1 (mute) & required / must \\
T15 & Fires the unmute tracker when the user unmutes. & §3.14.1 (unmute) & required / must \\
T16 & Skipping at 6 s (skipoffset 00:00:03) fires skip and neither thirdQuartile nor complete (skippable support is an optional ad type). & §2.3.1 (Skippable Linear); §3.8 skipoffset; §3.14.1 (skip) & required / must \\
T17 & The skip control is not available at 1.5 s and is available at 6 s (skipoffset 00:00:03). & §3.8 (skipoffset: when skip controls are made available) & none / should (a) \\
T18 & A percentage skipoffset (30\% of 10 s) makes skip available at 3 s, and skipping fires skip. & §3.8 skipoffset; VAST 4.2 XSD (skipoffset pattern allows n\%) & none / should (a) \\
T19 & Progress offsets passed before the skip (2 s and 5 s) are fired. & §3.14.3 (progress counts a view even if the ad is skipped) & none / should (a) \\
T20 & Fires the loaded tracker once when the media is ready to play. & §3.14.1 (loaded) & should / should \\
T21 & Fires the rewind tracker when the user seeks back 2 s (from 60\% to 40\%). & §3.14.1 (rewind) & required / must \\
T22 & After rewinding from 6 s to 4 s, midpoint and the 5 s progress tracker are not fired a second time when playback passes them again. & §3.14.1 (quartiles: played continuously) & none / should (a) \\
T23 & A volume change that is not an unmute (0.5 while unmuted) does not fire unmute. & §3.14.1 (unmute: the user unmuted) & none / should (a) \\
M01 & Replaces {[}CACHEBUSTING{]} with a random 8-digit integer. & IAB macro list: CACHEBUSTING (required) & list / must (a) \\
M02 & Replaces {[}TIMESTAMP{]} with an ISO 8601 date-time (milliseconds and zone as in the macro list example 2016-01-17T8:15:07.127-05). & IAB macro list: TIMESTAMP (required) & list / must (a) \\
M03 & Replaces {[}LIMITADTRACKING{]} with 0 or 1 (the harness also accepts -1/-2, since a browser has no device advertising ID; see detail). & IAB macro list: LIMITADTRACKING (required) & list / must (a) \\
M04 & Percent-encodes macro values (the colons in TIMESTAMP). & §6.1 (encodeURIComponent on values) & must / must \\
M05 & Does not write -1 into a macro it does not know ({[}NOTAREALMACRO{]}). & §6.1 (do not replace unknown macros with -1) & imper. / should (a) \\
M06 & Replaces each of eight listed optional macros ({[}ADPLAYHEAD{]}, {[}ADSERVINGID{]}, {[}UNIVERSALADID{]}, {[}PODSEQUENCE{]}, {[}ASSETURI{]}, {[}BREAKPOSITION{]}, {[}IFA{]}, {[}CLIENTUA{]}) with a plausible value or -1/-2; none left as macro or empty. & §6.1 (unprovided optional macros must become -1 or -2) & must / must \\
M07 & Fills {[}ADSERVINGID{]} with the ad's AdServingId (in-serving). & §3.4.3 (player fills {[}ADSERVINGID{]} from AdServingId) & none / should (a) \\
M08 & When it fills {[}ADPLAYHEAD{]}, the midpoint value is HH:MM:SS.mmm and about 5 s. & IAB macro list: ADPLAYHEAD (HH:MM:SS.mmm) & none / should (a) \\
M09 & Fills {[}UNIVERSALADID{]} with one or more of the Creative's IDs as ``registry value'' pairs, each percent-encoded; several IDs are joined by a raw comma. & §6.1 (Array values: encode each value, commas unencoded); IAB macro list: UNIVERSALADID & should / should \\
M10 & When it fills {[}ASSETURI{]}, the value is the MediaFile actually chosen (an unsupported FLV is listed first). & IAB macro list: ASSETURI (the media file being played) & none / should (a) \\
M11 & Replaces {[}CACHEBUSTING{]} and {[}TIMESTAMP{]} in an Error URI as in any other tracking URI. & IAB macro list: CACHEBUSTING, TIMESTAMP (contexts: all tracking pixels, including Error) & list / must (a) \\
M12 & Replaces {[}CACHEBUSTING{]} and {[}TIMESTAMP{]} in a Wrapper's VASTAdTagURI before requesting it. & IAB macro list: CACHEBUSTING, TIMESTAMP (contexts: VAST request URIs) & list / must (a) \\
X01 & Trims the whitespace around a CDATA-wrapped Impression URI before handing it to the network (checked on the client side, since URL parsers also strip it). & §2.3.3 (whitespace around a CDATA-wrapped URI) & none / should (a) \\
X02 & Ignores XML comments inside an element when reading its URI. & §2.2 (XML comments) & must / must \\
X03 & Decodes character entities (\texttt{\&\cbr{}amp;}) in a URI given as plain text (a document that breaks §2.3.3's CDATA rule, which real tags still do). & §2.3.2 (character entities should be honored) & should / should \\
X04 & Parses a document that omits \texttt{xmlns=\cbr{}\textquotedbl{}http:\cbrsoft{}/\cbr{}/\cbr{}www.\cbrsoft{}iab.\cbrsoft{}com/\cbr{}VAST\textquotedbl{}}. & (robustness; a 4.x document without the VAST namespace is not schema-valid) & n/a / obs. \\
X05 & Accepts a Creative with two UniversalAdId elements. & §2.2; §3.7.1 (UniversalAdId 1+; 4.2 XSD unbounded) & must / must \\
X06 & Accepts two ClosedCaptionFile elements. & §2.2 (play the ad in a VAST response); §3.9.5 (ClosedCaptionFile 0+) & must / must \\
X07 & Follows a VASTAdTagURI whose CDATA is surrounded by whitespace and line breaks (on the wire; URL parsers also strip the whitespace). & §2.3.3; §3.19.1 VASTAdTagURI & n/a / obs. \\
X08 & Never requests a Tracking element whose event name is not a VAST event (notARealEvent). & §3.14.3 (event values are an enumeration) & none / should (a) \\
X09 & Does not request Impression or Tracking URIs that appear inside an Extension. & §3.18 Extensions (custom XML for the publisher, not VAST elements) & should / should \\
X10 & Parses a document that starts with a UTF-8 byte order mark. & §2.2; §2.3.2 (XML 1.0: a UTF-8 byte order mark is allowed) & must / must \\
X11 & Reassembles an Impression URI split across two CDATA sections (query s=x{]}{]}\textgreater x). & §2.3.3 (a URI containing ``{]}{]}\textgreater{}'' split across CDATA sections) & required / must \\
X12 & Accepts a UniversalAdId that still carries the VAST 4.0 idValue attribute. & §3.7.1 (VAST 4.0 idValue attribute: players should not fail) & should / should \\
X13 & An empty Impression element does not stop the other Impression from being requested. & §3.4.4 (Impression) & required / must \\
X14 & Accepts a VAST 3.0 InLine document. & §4.3 Media Players (accept older VAST versions) & should / should \\
X15 & Accepts a VAST 2.0 InLine document. & §4.3 Media Players (accept older VAST versions) & should / should \\
X16 & Follows a VAST 3.0 Wrapper to a VAST 4.3 InLine. & §4.3 Media Players (accept older VAST versions) & should / should \\
X17 & Does not request an Impression element that sits entirely inside an XML comment (next to a real Impression). & §2.2 (acknowledge XML comments); §2.3.2 & must / must \\
X18 & Requests a VASTAdTagURI written as escaped text, with \texttt{\&\cbr{}amp;} between query parameters, with all parameters intact (a=1, b=2, c=3). & §2.3.2 (character entities should be honored); §3.19.1 VASTAdTagURI & should / should \\
X19 & Requests a VASTAdTagURI written as escaped text, with the character reference \texttt{\&\cbr{}\#\cbr{}38;} between query parameters, with all parameters intact (a=1, b=2, c=3). & §2.3.2 (character entities should be honored); §3.19.1 VASTAdTagURI & should / should \\
P01 & Orders pod ads by their sequence attribute (document order is 3, 1, 2): the ad with sequence 1 comes first (or the pod is declined with an error; pod support is optional). & Ad Pods and Stand-Alone Ads & must / must \\
P02 & When the second pod ad cannot play, playback continues with another ad (the stand-alone ad or the third pod ad). & Ad Pods (an ad fails: move on) & should / should \\
P03 & Playing a three-ad pod, each ad gets one impression and one complete, fired between its own impression and the next ad's. & §2.3.4 (fire each tracker at the appropriate time); Ad Pods & required / must \\
P05 & The failed pod ad is replaced by the stand-alone ad from the response, after the first pod ad and before the third. & Ad Pods (substitute an un-played stand-alone ad) & should / should \\
P06 & Plays all three ads of a pod. & Ad Pods (play the pod in sequence; play as many ads as possible) & should / should \\
P07 & Orders the pod ads of a Wrapper's target by their sequence attribute (document order 3, 1, 2; the Wrapper sets \texttt{allow\cbrsoft{}Multiple\cbrsoft{}Ads=\cbr{}\textquotedbl{}true\textquotedbl{}}): the ad with sequence 1 comes first (or the pod is declined with an error; pod support is optional). & Ad Pods and Stand-Alone Ads (play the pod in the prescribed sequence); §3.19 & must / must \\
A01 & When it does not execute the InLine's verification resource (a JavaScript resource with an apiFramework it does not implement), requests its verificationNotExecuted URI. & §3.17; §3.17.4 verificationNotExecuted & must / must \\
A02 & Fills {[}REASON{]} in verificationNotExecuted with a reason code (1, 2 or 3). & §3.17.4 {[}REASON{]}; IAB macro list: REASON (required) & must / must \\
A03 & Handles the Wrapper's verification too: requests its verificationNotExecuted URI. & Viewability with Ad Verification Services (resources from intermediary Wrappers) & should / should \\
A04 & Reports a verification resource whose apiFramework it does not implement with reason 2 (verification not supported). & §3.17.4 (reason code 2); §3.17.1; Viewability with Ad Verification Services & must / must \\
A05 & When an OMID verification resource fails to load (HTTP 404), requests its verificationNotExecuted URI with reason 3 (error during resource load). & §3.17.4 (reason code 3: 404 or other failed response codes); Viewability with Ad Verification Services & must / must \\
I01 & Displays the Icon after the ad starts and requests its IconViewTracking URI. & Industry Icon Support; §3.11.2 IconViewTracking & must / must \\
I02 & On a click on the Icon, requests its IconClickTracking URI. & §3.11.5 IconClickTracking & required / must \\
I03 & Displays an Icon with offset 00:00:01 and duration 00:00:08 once playback passes 1 s, requesting its IconViewTracking URI. & §3.11.1 Icon (offset, duration) & should / should \\
S01 & Skips a MediaFile of an unsupported type listed first (video/x-flv) and plays a supported one. & MediaFiles (choose by attributes, not document order) & should / should \\
S02 & Optional: plays an audio ad (adType=``audio'', Opus WebM and AAC files). & §1.4; §3.4 (adType=``audio'') & may / may \\
R01 & A timed-out Wrapper request does not escape as an uncaught exception (which terminates a Node.js process). & (engineering; not a VAST clause) & n/a / obs. \\
R02 & A failed Wrapper request does not escape as an uncaught exception. & (engineering; not a VAST clause) & n/a / obs. \\
R03 & Follows an HTTP 302 redirect on a VASTAdTagURI. & (HTTP; not a VAST clause) & n/a / obs. \\
R04 & Parses a VAST response sent with Content-Encoding: gzip. & (HTTP; not a VAST clause) & n/a / obs. \\
R05 & Parses a VAST response served as text/plain rather than an XML media type. & (HTTP; not a VAST clause) & n/a / obs. \\*
R06 & A tracking URI that answers HTTP 500 does not stop later trackers (quartiles, complete). & (HTTP; not a VAST clause) & n/a / obs. \\
\end{longtable}

\endgroup

\hypertarget{appendix-b.-findings}{%
\section{Appendix B. Findings}\label{appendix-b.-findings}}

One row per finding (one cause, located in source for the open-source clients). Requirements are catalog IDs with their strength (must, should or observation) as graded in \texttt{code/\cbr{}harness/\cbr{}suite/\cbr{}requirements.\cbrsoft{}mjs}. Locations are in the tested release, relative to the package root: vast-client-js 6.5.0 \texttt{src/}, video-ad-sdk 4.0.1 \texttt{dist/} (the code that ran), iab-vast-loader 2.5.1 and its parser iab-vast-parser 1.2.1 \texttt{lib/}, vastlint-client 0.15.0 \texttt{dist/}, Shaka Player 5.2.12 \texttt{lib/\cbr{}ads/}, rmp-vast 17.2.0 \texttt{src/\cbr{}js/} at commit 5975ead. ``absent:'' means the feature is not implemented and says what was searched. IMA is closed source and is analyzed as a black box: Table B7 gives, in place of a location, the evidence the server logged and the events of IMA's public API, from the Chromium reference run (Firefox for F86; the three Linux engines and the macOS run for F85); the probes are in the replication package. rmp-vast cells that come from its bundled vast-client are attributed to vast-client-js (F01, F02, F03, F05).

Release range is the span of vast-client-js, video-ad-sdk, iab-vast-loader and vastlint-client releases (Node.js sweep) in which the finding appears; Shaka, rmp-vast and IMA were not swept.

\begingroup\small\setlength{\tabcolsep}{4pt}\Needspace{14\baselineskip}

\begin{longtable}[]{@{}
  >{\raggedright\arraybackslash}p{(\columnwidth - 8\tabcolsep) * \real{0.0438}}
  >{\raggedright\arraybackslash}p{(\columnwidth - 8\tabcolsep) * \real{0.2298}}
  >{\raggedright\arraybackslash}p{(\columnwidth - 8\tabcolsep) * \real{0.2629}}
  >{\raggedright\arraybackslash}p{(\columnwidth - 8\tabcolsep) * \real{0.2336}}
  >{\raggedright\arraybackslash}p{(\columnwidth - 8\tabcolsep) * \real{0.2298}}@{}}
\caption{\textbf{Table B1.} Findings in vast-client-js 6.5.0.}\tabularnewline
\toprule\noalign{}
\begin{minipage}[b]{\linewidth}\raggedright
ID
\end{minipage} & \begin{minipage}[b]{\linewidth}\raggedright
Requirements
\end{minipage} & \begin{minipage}[b]{\linewidth}\raggedright
Symptom
\end{minipage} & \begin{minipage}[b]{\linewidth}\raggedright
Location
\end{minipage} & \begin{minipage}[b]{\linewidth}\raggedright
Releases
\end{minipage} \\
\midrule\noalign{}
\endfirsthead
\multicolumn{5}{@{}l}{\textbf{Table B1} (continued)}\tabularnewline
\toprule\noalign{}
\begin{minipage}[b]{\linewidth}\raggedright
ID
\end{minipage} & \begin{minipage}[b]{\linewidth}\raggedright
Requirements
\end{minipage} & \begin{minipage}[b]{\linewidth}\raggedright
Symptom
\end{minipage} & \begin{minipage}[b]{\linewidth}\raggedright
Location
\end{minipage} & \begin{minipage}[b]{\linewidth}\raggedright
Releases
\end{minipage} \\
\midrule\noalign{}
\endhead
\bottomrule\noalign{}
\endlastfoot
F01 & M12 (must) & Requests a Wrapper's VASTAdTagURI with {[}CACHEBUSTING{]} and {[}TIMESTAMP{]} still in brackets. & \texttt{src/\cbr{}parser/\cbr{}vast\_\cbrsoft{}parser.\cbrsoft{}js:\cbrsoft{}391}; \texttt{src/\cbr{}parser/\cbr{}vast\_\cbrsoft{}parser.\cbrsoft{}js:\cbrsoft{}403-\cbr{}405} & all 9 (3.3.0 to 6.5.0) \\
F02 & P07 (must) & A pod in a Wrapper's target is returned in document order (3, 1, 2). & \texttt{src/\cbr{}parser/\cbr{}vast\_\cbrsoft{}parser.\cbrsoft{}js:\cbrsoft{}258-\cbr{}315} (\texttt{parse()}) & not in the swept catalog \\
F03 & R01 (obs.) & A Wrapper request timeout throws outside any promise: uncaught exception, Node.js process exits with code 1. & \texttt{src/\cbr{}fetcher/\cbr{}url\_\cbrsoft{}handler.\cbrsoft{}js:\cbrsoft{}57-\cbr{}64} & 6.0.2 to 6.5.0 (of 3.3.0 to 6.5.0) \\
F04 & none: robustness (fuzz C4, D4) & A target that sends headers and stalls its body is waited on with no timeout at every hop. & \texttt{src/\cbr{}fetcher/\cbr{}url\_\cbrsoft{}handler.\cbrsoft{}js:\cbrsoft{}66-\cbr{}72} & tested release only \\
F05 & W11 (must) & Non-well-formed Wrapper target reported as 301 (Node.js, Firefox) or 303 (Chromium, Chrome, WebKit, Safari), never 100. & \texttt{src/\cbr{}parser/\cbr{}vast\_\cbrsoft{}parser.\cbrsoft{}js:\cbrsoft{}442,\cbr{} 170-\cbr{}187,\cbr{} 467-\cbr{}477} & all 9 (3.3.0 to 6.5.0) \\
F06 & none: robustness (fuzz C2) & An \texttt{\textless{}Ad\textgreater{}} with neither InLine nor Wrapper makes \texttt{get()} throw a TypeError; no Error URI; sibling ads lost. & \texttt{src/\cbr{}parser/\cbr{}ad\_\cbrsoft{}parser.\cbrsoft{}js:\cbrsoft{}20-\cbr{}59}; \texttt{src/\cbr{}parser/\cbr{}vast\_\cbrsoft{}parser.\cbrsoft{}js:\cbrsoft{}226} & tested release only \\
F07 & none: macro grid (4.x macros) & Five pixel macros marked 4.x ({[}STOREID{]}, {[}STOREURL{]}, {[}PLAYBACKMETHODS{]}, {[}GPPSTRING{]}, {[}GPPSECTIONID{]}) sent literally. & \texttt{src/\cbr{}util/\cbr{}macros.\cbrsoft{}js:\cbrsoft{}1-\cbr{}49}; \texttt{src/\cbr{}util/\cbr{}util.\cbrsoft{}js:\cbrsoft{}72-\cbr{}74}; absent: STOREID, STOREURL, PLAYBACKMETHODS, GPPSTRING, GPPSECTIONID anywhere under src/ & tested release only \\
F08 & M10 (should) & {[}ASSETURI{]} carries the first MediaFile, overwriting the value the integrator passes; not the one played. & \texttt{src/\cbr{}vast\_\cbrsoft{}tracker.\cbrsoft{}js:\cbrsoft{}899-\cbr{}906} & all 9 (3.3.0 to 6.5.0) \\
F09 & M09 (should) & {[}UNIVERSALADID{]} percent-encodes the separating comma (\%2C). & \texttt{src/\cbr{}vast\_\cbrsoft{}tracker.\cbrsoft{}js:\cbrsoft{}913-\cbr{}917}; \texttt{src/\cbr{}util/\cbr{}util.\cbrsoft{}js:\cbrsoft{}42} & 4.0.1 to 6.5.0 (of 3.3.0 to 6.5.0) \\
F10 & none: outside catalog & \texttt{wrapper\cbrsoft{}Limit} accepts one Wrapper fewer than its value (decides rmp-vast's W03). & \texttt{src/\cbr{}parser/\cbr{}vast\_\cbrsoft{}parser.\cbrsoft{}js:\cbrsoft{}369,\cbr{} :\cbrsoft{}382,\cbr{} :\cbrsoft{}74} & tested release only \\*
F11 & none: outside catalog (fuzz D2) & An InLine with no usable MediaFile is reported with 303. & \texttt{src/\cbr{}parser/\cbr{}vast\_\cbrsoft{}parser.\cbrsoft{}js:\cbrsoft{}467-\cbr{}477} & tested release only \\
\end{longtable}

\endgroup

\begingroup\small\setlength{\tabcolsep}{4pt}\Needspace{14\baselineskip}

\begin{longtable}[]{@{}
  >{\raggedright\arraybackslash}p{(\columnwidth - 6\tabcolsep) * \real{0.0430}}
  >{\raggedright\arraybackslash}p{(\columnwidth - 6\tabcolsep) * \real{0.2833}}
  >{\raggedright\arraybackslash}p{(\columnwidth - 6\tabcolsep) * \real{0.3440}}
  >{\raggedright\arraybackslash}p{(\columnwidth - 6\tabcolsep) * \real{0.3297}}@{}}
\caption{\textbf{Table B2.} Findings in rmp-vast 17.2.0 (not swept).}\tabularnewline
\toprule\noalign{}
\begin{minipage}[b]{\linewidth}\raggedright
ID
\end{minipage} & \begin{minipage}[b]{\linewidth}\raggedright
Requirements
\end{minipage} & \begin{minipage}[b]{\linewidth}\raggedright
Symptom
\end{minipage} & \begin{minipage}[b]{\linewidth}\raggedright
Location
\end{minipage} \\
\midrule\noalign{}
\endfirsthead
\multicolumn{4}{@{}l}{\textbf{Table B2} (continued)}\tabularnewline
\toprule\noalign{}
\begin{minipage}[b]{\linewidth}\raggedright
ID
\end{minipage} & \begin{minipage}[b]{\linewidth}\raggedright
Requirements
\end{minipage} & \begin{minipage}[b]{\linewidth}\raggedright
Symptom
\end{minipage} & \begin{minipage}[b]{\linewidth}\raggedright
Location
\end{minipage} \\
\midrule\noalign{}
\endhead
\bottomrule\noalign{}
\endlastfoot
F12 & W03 (must) & Stops a five-Wrapper chain after four documents with 302 (API reports 303). & \texttt{src/\cbr{}js/\cbr{}helpers/\cbr{}utils.\cbrsoft{}js:\cbrsoft{}223}; \texttt{src/\cbr{}js/\cbr{}index.\cbrsoft{}js:\cbrsoft{}591}; bundled vast-client 6.4.2 \texttt{src/\cbr{}parser/\cbr{}vast\_\cbrsoft{}parser.\cbrsoft{}js:\cbrsoft{}362,\cbr{} 375} \\
F13 & W20 (should), W21 (should) & Passes \texttt{allow\cbrsoft{}Multiple\cbrsoft{}Ads:\cbrsoft{} true}, overriding the Wrapper attribute. & \texttt{src/\cbr{}js/\cbr{}index.\cbrsoft{}js:\cbrsoft{}593}; bundled vast-client 6.4.2 \texttt{src/\cbr{}parser/\cbr{}vast\_\cbrsoft{}parser.\cbrsoft{}js:\cbrsoft{}389-\cbr{}392} \\
F14 & W23 (should), P05 (should) & Never falls back to the stand-alone ad (fallbackOnNoAd, failed pod ad). & absent: fallbackOnNoAd, remainingAds, getNextAds not used in rmp-vast's own modules (bundle search); \texttt{src/\cbr{}js/\cbr{}index.\cbrsoft{}js:\cbrsoft{}364-\cbr{}367}; bundled vast-client 6.4.2 \texttt{src/\cbr{}parser/\cbr{}ad\_\cbrsoft{}parser.\cbrsoft{}js:\cbrsoft{}223-\cbr{}234}, \texttt{src/\cbr{}parser/\cbr{}vast\_\cbrsoft{}parser.\cbrsoft{}js:\cbrsoft{}304-\cbr{}315} \\
F15 & M01 (must) & {[}CACHEBUSTING{]} is eight alphanumeric characters. & \texttt{src/\cbr{}js/\cbr{}framework/\cbr{}fw.\cbrsoft{}js:\cbrsoft{}194-\cbr{}201} \\
F16 & M03 (must), M06 (must) & {[}LIMITADTRACKING{]} only if the integrator passes it; {[}PODSEQUENCE{]}, {[}IFA{]}, {[}CLIENTUA{]} get no -1. & \texttt{src/\cbr{}js/\cbr{}index.\cbrsoft{}js:\cbrsoft{}652-\cbr{}655}; \texttt{src/\cbr{}js/\cbr{}helpers/\cbr{}tracking.\cbrsoft{}js:\cbrsoft{}359-\cbr{}362,\cbr{} 438-\cbr{}442}; absent: no rmp-vast code names {[}IFA{]} or {[}CLIENTUA{]} (only params.macros, tracking.js:214-222) \\
F17 & M11 (must) & Error URIs skip macro replacement except {[}ERRORCODE{]}. & \texttt{src/\cbr{}js/\cbr{}helpers/\cbr{}tracking.\cbrsoft{}js:\cbrsoft{}462-\cbr{}482} \\
F18 & M09 (should) & {[}UNIVERSALADID{]} starts with a comma and encodes the separators. & \texttt{src/\cbr{}js/\cbr{}helpers/\cbr{}tracking.\cbrsoft{}js:\cbrsoft{}338-\cbr{}351} \\
F19 & E06 (must) & Undecodable media reported as 401, not 405. & \texttt{src/\cbr{}js/\cbr{}creatives/\cbr{}linear.\cbrsoft{}js:\cbrsoft{}147-\cbr{}176} \\
F20 & T13 (must) & Resume tracker fires on every play event, including ad start. & \texttt{src/\cbr{}js/\cbr{}helpers/\cbr{}tracking.\cbrsoft{}js:\cbrsoft{}176-\cbr{}180} \\
F21 & T23 (should) & Any volume change while unmuted fires unmute. & \texttt{src/\cbr{}js/\cbr{}helpers/\cbr{}tracking.\cbrsoft{}js:\cbrsoft{}112-\cbr{}123} \\
F22 & T02 (should) & An \texttt{about:\cbrsoft{}blank} Impression is requested. & \texttt{src/\cbr{}js/\cbr{}index.\cbrsoft{}js:\cbrsoft{}460-\cbr{}466}; \texttt{src/\cbr{}js/\cbr{}helpers/\cbr{}tracking.\cbrsoft{}js:\cbrsoft{}82-\cbr{}110} \\
F23 & T05 (must), T06 (must), T07 (must) & macOS WebKit only: no progress tracker fires (quartiles do); T19 then n/a. & \texttt{src/\cbr{}js/\cbr{}creatives/\cbr{}linear.\cbrsoft{}js:\cbrsoft{}67-\cbr{}108}, :74-76 and :292; \texttt{src/\cbr{}js/\cbr{}framework/\cbr{}fw.\cbrsoft{}js:\cbrsoft{}84-\cbr{}88}; \texttt{src/\cbr{}js/\cbr{}players/\cbr{}ad-\cbr{}player.\cbrsoft{}js:\cbrsoft{}53-\cbr{}58} \\
F24 & T21 (must) & No rewind tracker when the element is seeked back (borderline: no seek control by default). & absent: no rewind tracker or handler in rmp-vast's own modules (bundle search for rewind, seeked, seeking); \texttt{show\cbrsoft{}Controls\cbrsoft{}For\cbrsoft{}Ad\cbrsoft{}Player} is false by default (\texttt{src/\cbr{}js/\cbr{}helpers/\cbr{}utils.\cbrsoft{}js:\cbrsoft{}230}, used at \texttt{src/\cbr{}js/\cbr{}players/\cbr{}ad-\cbr{}player.\cbrsoft{}js:\cbrsoft{}139}) \\
F25 & I03 (should) & Icon drawn on first playing event; offset and duration never read. & \texttt{src/\cbr{}js/\cbr{}creatives/\cbr{}icons.\cbrsoft{}js:\cbrsoft{}186-\cbr{}190} \\*
F26 & A01 (must), A03 (should) & With defaults, no AdVerifications handling: no verificationNotExecuted. & \texttt{src/\cbr{}js/\cbr{}index.\cbrsoft{}js:\cbrsoft{}527-\cbr{}530} (verification handled only if \texttt{omid\cbrsoft{}Support} is set); \texttt{src/\cbr{}js/\cbr{}verification/\cbr{}omsdk.\cbrsoft{}js:\cbrsoft{}41-\cbr{}54}; \texttt{src/\cbr{}js/\cbr{}helpers/\cbr{}utils.\cbrsoft{}js:\cbrsoft{}245} \\
\end{longtable}

\endgroup

\begingroup\footnotesize\setlength{\tabcolsep}{4pt}\Needspace{12\baselineskip}

\begin{longtable}[]{@{}
  >{\raggedright\arraybackslash}p{(\columnwidth - 8\tabcolsep) * \real{0.0394}}
  >{\raggedright\arraybackslash}p{(\columnwidth - 8\tabcolsep) * \real{0.1714}}
  >{\raggedright\arraybackslash}p{(\columnwidth - 8\tabcolsep) * \real{0.3154}}
  >{\raggedright\arraybackslash}p{(\columnwidth - 8\tabcolsep) * \real{0.3023}}
  >{\raggedright\arraybackslash}p{(\columnwidth - 8\tabcolsep) * \real{0.1714}}@{}}
\caption{\textbf{Table B3.} Findings in video-ad-sdk 4.0.1.}\tabularnewline
\toprule\noalign{}
\begin{minipage}[b]{\linewidth}\raggedright
ID
\end{minipage} & \begin{minipage}[b]{\linewidth}\raggedright
Requirements
\end{minipage} & \begin{minipage}[b]{\linewidth}\raggedright
Symptom
\end{minipage} & \begin{minipage}[b]{\linewidth}\raggedright
Location
\end{minipage} & \begin{minipage}[b]{\linewidth}\raggedright
Releases
\end{minipage} \\
\midrule\noalign{}
\endfirsthead
\multicolumn{5}{@{}l}{\textbf{Table B3} (continued)}\tabularnewline
\toprule\noalign{}
\begin{minipage}[b]{\linewidth}\raggedright
ID
\end{minipage} & \begin{minipage}[b]{\linewidth}\raggedright
Requirements
\end{minipage} & \begin{minipage}[b]{\linewidth}\raggedright
Symptom
\end{minipage} & \begin{minipage}[b]{\linewidth}\raggedright
Location
\end{minipage} & \begin{minipage}[b]{\linewidth}\raggedright
Releases
\end{minipage} \\
\midrule\noalign{}
\endhead
\bottomrule\noalign{}
\endlastfoot
F27 & P01 (must), P07 (must) & Pods play in document order: \texttt{compare\cbrsoft{}By\cbrsoft{}Sequence} returns 0 for every pair. & \texttt{dist/\cbr{}vast\cbrsoft{}Selectors/\cbr{}index.\cbrsoft{}js:\cbrsoft{}13-\cbr{}18}; :103; \texttt{dist/\cbr{}vast\cbrsoft{}Request/\cbr{}helpers/\cbr{}get\cbrsoft{}Next\cbrsoft{}Ad.\cbrsoft{}js:\cbrsoft{}3-\cbr{}5} & P01: 3.0.0 to 4.0.1 (of 2.1.0 to 4.0.1); P07: not in the swept catalog \\
F28 & P06 (should) & Only the first ad of a pod plays; no error. & \texttt{dist/\cbr{}runner/\cbr{}run\cbrsoft{}Waterfall.\cbrsoft{}js:\cbrsoft{}90} & all 9 (2.1.0 to 4.0.1) \\
F29 & E05 (must), E06 (must) & Error URI requested with code=undefined for media 404 or decode failure, then 402. & \texttt{dist/\cbr{}ad\cbrsoft{}Unit/\cbr{}Vast\cbrsoft{}Ad\cbrsoft{}Unit.\cbrsoft{}js:\cbrsoft{}34} & browser-only, not swept \\
F30 & E03 (must), E09 (must) & No playable MediaFile (or NonLinear-only ad) reported as 401 to InLine and every Wrapper. & \texttt{dist/\cbr{}runner/\cbr{}helpers/\cbr{}start\cbrsoft{}Video\cbrsoft{}Ad.\cbrsoft{}js:\cbrsoft{}39} & all 9 (2.1.0 to 4.0.1); 2.1.0 to 2.5.1 request no Error URI, 401 from 2.6.3 \\
F31 & I03 (should) & Icon with an offset is never drawn; no view tracker. & \texttt{dist/\cbr{}ad\cbrsoft{}Unit/\cbr{}helpers/\cbr{}icons/\cbr{}add\cbrsoft{}Icons.\cbrsoft{}js:\cbrsoft{}15} & browser-only, not swept \\
F32 & A01 (must), A03 (should) & AdVerifications never read; no verificationNotExecuted. & absent: no occurrence of ``verification'' (any case) in video-ad-sdk 4.0.1 dist/ or src/ & all 9 (2.1.0 to 4.0.1) \\
F33 & X11 (must) & A URI split across CDATA sections is truncated to the first part. & \texttt{dist/\cbr{}xml/\cbr{}helpers/\cbr{}xml\cbrsoft{}Selectors.\cbrsoft{}js:\cbrsoft{}36-\cbr{}38} & all 9 (2.1.0 to 4.0.1) \\
F34 & T07 (must) & Percentage progress offsets computed from the media element's duration before it is set become NaN; never fire in the tested setup. & \texttt{dist/\cbr{}ad\cbrsoft{}Unit/\cbr{}helpers/\cbr{}metrics/\cbr{}handlers/\cbr{}on\cbrsoft{}Progress.\cbrsoft{}js:\cbrsoft{}22-\cbr{}23,\cbr{} :\cbrsoft{}19} and :29 & all 9 (2.1.0 to 4.0.1) \\
F35 & W07 (must) & Failed Wrapper fetch (404, 500, refused) reported as 502. & \texttt{dist/\cbr{}vast\cbrsoft{}Request/\cbr{}request\cbrsoft{}Ad.\cbrsoft{}js:\cbrsoft{}14-\cbr{}26} & all 9 (2.1.0 to 4.0.1) \\
F36 & W17 (must) & Wrapper ClickThrough not used when the InLine has none. & \texttt{dist/\cbr{}ad\cbrsoft{}Unit/\cbr{}helpers/\cbr{}metrics/\cbr{}setup\cbrsoft{}Metric\cbrsoft{}Handlers.\cbrsoft{}js}:8; \texttt{dist/\cbr{}vast\cbrsoft{}Selectors/\cbr{}index.\cbrsoft{}js:\cbrsoft{}327} & all 9 (2.1.0 to 4.0.1) \\
F37 & M01 (must) & {[}CACHEBUSTING{]} has 1 to 10 digits. & \texttt{dist/\cbr{}tracker/\cbr{}helpers/\cbr{}parse\cbrsoft{}Macro.\cbrsoft{}js}:9 and :22-24 & all 9 (2.1.0 to 4.0.1) \\
F38 & M03 (must), M06 (must), M07 (should) & \texttt{parse\cbrsoft{}Macro} fills only CACHEBUSTING and TIMESTAMP; other listed macros literal. & \texttt{dist/\cbr{}tracker/\cbr{}helpers/\cbr{}parse\cbrsoft{}Macro.\cbrsoft{}js:\cbrsoft{}19-\cbr{}35} & all 9 (2.1.0 to 4.0.1) \\
F39 & M12 (must) & Requests a Wrapper's VASTAdTagURI as written. & \texttt{dist/\cbr{}vast\cbrsoft{}Request/\cbr{}request\cbrsoft{}Ad.\cbrsoft{}js:\cbrsoft{}118} (the Wrapper's \texttt{get\cbrsoft{}VASTAd\cbrsoft{}Tag\cbrsoft{}URI} is requested unchanged) & all 9 (2.1.0 to 4.0.1) \\
F40 & T18 (should) & Percentage skipoffset ignored. & \texttt{dist/\cbr{}ad\cbrsoft{}Unit/\cbr{}helpers/\cbr{}metrics/\cbr{}handlers/\cbr{}on\cbrsoft{}Skip.\cbrsoft{}js:\cbrsoft{}22}; \texttt{dist/\cbr{}vast\cbrsoft{}Selectors/\cbr{}helpers/\cbr{}parse\cbrsoft{}Offset.\cbrsoft{}js:\cbrsoft{}2-\cbr{}9}; \texttt{dist/\cbr{}vast\cbrsoft{}Selectors/\cbr{}index.\cbrsoft{}js:\cbrsoft{}368-\cbr{}372} & 2.2.1 to 4.0.1 (of 2.1.0 to 4.0.1) \\
F41 & T20 (should) & No loaded event. & \texttt{dist/\cbr{}tracker/\cbr{}linear\cbrsoft{}Events.\cbrsoft{}js:\cbrsoft{}12-\cbr{}176}; absent: no loaded event or handler in dist/tracker/ or dist/adUnit/helpers/metrics/ & all 9 (2.1.0 to 4.0.1) \\
F42 & W20 (should), W21 (should) & allowMultipleAds defaults to true; false rejects the response with 203. & \texttt{dist/\cbr{}vast\cbrsoft{}Request/\cbr{}request\cbrsoft{}Ad.\cbrsoft{}js:\cbrsoft{}51} and :57-61 & all 9 (2.1.0 to 4.0.1) \\
F43 & T02 (should) & An \texttt{about:\cbrsoft{}blank} Impression is passed to the pixel tracker. & \texttt{dist/\cbr{}tracker/\cbr{}helpers/\cbr{}create\cbrsoft{}Vast\cbrsoft{}Event\cbrsoft{}Tracker.\cbrsoft{}js:\cbrsoft{}13} & all 9 (2.1.0 to 4.0.1) \\
F44 & W11 (must) & Browser DOMParser partial tree (parsererror) accepted as VAST; ad accepted, then 401. & \texttt{dist/\cbr{}xml/\cbr{}helpers/\cbr{}xml2js.\cbrsoft{}js}:5; sane-domparser-error 0.2.0 index.js:18-23 & none in sweep \\*
F45 & none: robustness (fuzz C3) & VPAID-marked MP4: unhandled rejection; 901 to Error URI, then impression and playback. & \texttt{dist/\cbr{}runner/\cbr{}helpers/\cbr{}start\cbrsoft{}Video\cbrsoft{}Ad.\cbrsoft{}js:\cbrsoft{}33} (adUnit.start promise not handled; src startVideoAd.ts:74); \texttt{dist/\cbr{}ad\cbrsoft{}Unit/\cbr{}helpers/\cbr{}vpaid/\cbr{}load\cbrsoft{}Creative.\cbrsoft{}js}:7 & tested release only \\
\end{longtable}

\endgroup

\begingroup\footnotesize\setlength{\tabcolsep}{4pt}\Needspace{18\baselineskip}

\begin{longtable}[]{@{}
  >{\raggedright\arraybackslash}p{(\columnwidth - 8\tabcolsep) * \real{0.0394}}
  >{\raggedright\arraybackslash}p{(\columnwidth - 8\tabcolsep) * \real{0.1649}}
  >{\raggedright\arraybackslash}p{(\columnwidth - 8\tabcolsep) * \real{0.3154}}
  >{\raggedright\arraybackslash}p{(\columnwidth - 8\tabcolsep) * \real{0.3154}}
  >{\raggedright\arraybackslash}p{(\columnwidth - 8\tabcolsep) * \real{0.1649}}@{}}
\caption{\textbf{Table B4.} Findings in iab-vast-loader 2.5.1.}\tabularnewline
\toprule\noalign{}
\begin{minipage}[b]{\linewidth}\raggedright
ID
\end{minipage} & \begin{minipage}[b]{\linewidth}\raggedright
Requirements
\end{minipage} & \begin{minipage}[b]{\linewidth}\raggedright
Symptom
\end{minipage} & \begin{minipage}[b]{\linewidth}\raggedright
Location
\end{minipage} & \begin{minipage}[b]{\linewidth}\raggedright
Releases
\end{minipage} \\
\midrule\noalign{}
\endfirsthead
\multicolumn{5}{@{}l}{\textbf{Table B4} (continued)}\tabularnewline
\toprule\noalign{}
\begin{minipage}[b]{\linewidth}\raggedright
ID
\end{minipage} & \begin{minipage}[b]{\linewidth}\raggedright
Requirements
\end{minipage} & \begin{minipage}[b]{\linewidth}\raggedright
Symptom
\end{minipage} & \begin{minipage}[b]{\linewidth}\raggedright
Location
\end{minipage} & \begin{minipage}[b]{\linewidth}\raggedright
Releases
\end{minipage} \\
\midrule\noalign{}
\endhead
\bottomrule\noalign{}
\endlastfoot
F46 & W02 (must), P07 (must) & The loader checks only \texttt{vast.\cbrsoft{}ads} after a Wrapper: a lone \texttt{\textless{}Ad sequence=\cbr{}\textquotedbl{}1\textquotedbl{}\textgreater{}} Wrapper, filed as a pod, is never followed (defaults only; no ad, no error), and a pod behind a Wrapper is reported as 303. & \texttt{lib/\cbr{}loader.\cbrsoft{}js:\cbrsoft{}58-\cbr{}65}; iab-vast-parser \texttt{lib/\cbr{}factory/\cbr{}vast.\cbrsoft{}js:\cbrsoft{}53-\cbr{}55} & W02: all 6 (2.0.0 to 2.5.1); P07: not in the swept catalog \\
F47 & X05 (must), X06 (must) & Valid 4.2 documents with repeated UniversalAdId or ClosedCaptionFile rejected with 101 (and four more elements, fuzz D3). & iab-vast-parser \texttt{lib/\cbr{}vast/\cbr{}schema.\cbrsoft{}js:\cbrsoft{}13-\cbr{}33}; \texttt{lib/\cbr{}xml/\cbr{}unmarshaler.\cbrsoft{}js:\cbrsoft{}113-\cbr{}118} & all 6 (2.0.0 to 2.5.1) \\
F48 & W19 (should) & followAdditionalWrappers never consulted. & absent: followAdditionalWrappers is not read in iab-vast-loader lib/ (5 files); \texttt{lib/\cbr{}loader.\cbrsoft{}js:\cbrsoft{}58-\cbr{}62} & all 6 (2.0.0 to 2.5.1) \\
F49 & W11 (must) & Node.js: xmldom recovers an ad from non-well-formed XML; Firefox: 101. & iab-vast-parser \texttt{lib/\cbr{}index.\cbrsoft{}js:\cbrsoft{}7,\cbr{} 18-\cbr{}20}; \texttt{lib/\cbr{}parse.\cbrsoft{}js:\cbrsoft{}34-\cbr{}45}; xmldom 0.1.31 dom-parser.js:30, 175-183 & all 6 (2.0.0 to 2.5.1) \\
F50 & M12 (must) & Requests a Wrapper's VASTAdTagURI as written. & \texttt{lib/\cbr{}loader.\cbrsoft{}js:\cbrsoft{}61,\cbr{} :\cbrsoft{}81,\cbr{} :\cbrsoft{}14} and :88 & all 6 (2.0.0 to 2.5.1) \\
F51 & none: robustness (fuzz C5) & Empty response (HTTP 204 or empty 200): an internal TypeError with no VAST code. & iab-vast-parser \texttt{lib/\cbr{}parse.\cbrsoft{}js:\cbrsoft{}39} & tested release only \\
F52 & none: robustness (fuzz C6) & A response that is not XML: rejected with 101 and an internal TypeError, where 100 applies. & iab-vast-parser \texttt{lib/\cbr{}xml/\cbr{}unmarshaler.\cbrsoft{}js:\cbrsoft{}75} & tested release only \\
F53 & none: robustness (fuzz C7) & A root element other than VAST: rejected with 101 and an internal TypeError. & iab-vast-parser \texttt{lib/\cbr{}factory/\cbr{}vast.\cbrsoft{}js:\cbrsoft{}38} & tested release only \\*
F54 & none: robustness (fuzz C8) & A lower-case element name such as \texttt{\textless{}creatives\textgreater{}} rejects the whole response with an internal TypeError. & iab-vast-parser \texttt{lib/\cbr{}inherit/\cbr{}ad.\cbrsoft{}js:\cbrsoft{}57,\cbr{} 85,\cbr{} 88}; \texttt{lib/\cbr{}xml/\cbr{}unmarshaler.\cbrsoft{}js:\cbrsoft{}19-\cbr{}25,\cbr{} 85} & tested release only \\
\end{longtable}

\endgroup

\begingroup\small\setlength{\tabcolsep}{4pt}\Needspace{15\baselineskip}

\begin{longtable}[]{@{}
  >{\raggedright\arraybackslash}p{(\columnwidth - 8\tabcolsep) * \real{0.0438}}
  >{\raggedright\arraybackslash}p{(\columnwidth - 8\tabcolsep) * \real{0.1860}}
  >{\raggedright\arraybackslash}p{(\columnwidth - 8\tabcolsep) * \real{0.3505}}
  >{\raggedright\arraybackslash}p{(\columnwidth - 8\tabcolsep) * \real{0.2336}}
  >{\raggedright\arraybackslash}p{(\columnwidth - 8\tabcolsep) * \real{0.1860}}@{}}
\caption{\textbf{Table B5.} Findings in vastlint-client 0.15.0.}\tabularnewline
\toprule\noalign{}
\begin{minipage}[b]{\linewidth}\raggedright
ID
\end{minipage} & \begin{minipage}[b]{\linewidth}\raggedright
Requirements
\end{minipage} & \begin{minipage}[b]{\linewidth}\raggedright
Symptom
\end{minipage} & \begin{minipage}[b]{\linewidth}\raggedright
Location
\end{minipage} & \begin{minipage}[b]{\linewidth}\raggedright
Releases
\end{minipage} \\
\midrule\noalign{}
\endfirsthead
\multicolumn{5}{@{}l}{\textbf{Table B5} (continued)}\tabularnewline
\toprule\noalign{}
\begin{minipage}[b]{\linewidth}\raggedright
ID
\end{minipage} & \begin{minipage}[b]{\linewidth}\raggedright
Requirements
\end{minipage} & \begin{minipage}[b]{\linewidth}\raggedright
Symptom
\end{minipage} & \begin{minipage}[b]{\linewidth}\raggedright
Location
\end{minipage} & \begin{minipage}[b]{\linewidth}\raggedright
Releases
\end{minipage} \\
\midrule\noalign{}
\endhead
\bottomrule\noalign{}
\endlastfoot
F55 & W03 (must) & \texttt{max\cbrsoft{}Wrapper\cbrsoft{}Depth} 5 counts documents: stops before the InLine of a five-Wrapper chain, no code. & \texttt{dist/\cbr{}session.\cbrsoft{}js}:8 & all 12 (0.4.24 to 0.15.0) \\
F56 & W11 (must), X09 (should), X02 (must), X17 (must) & XML read with regular expressions: ads from malformed XML; commented-out URIs requested. & \texttt{dist/\cbr{}resolved.\cbrsoft{}js:\cbrsoft{}106-\cbr{}148}; \texttt{dist/\cbr{}tracking.\cbrsoft{}js:\cbrsoft{}72-\cbr{}141} & W11, X09, X02: all 12 (0.4.24 to 0.15.0); X17: not in the swept catalog \\
F57 & X18 (should), X19 (should) & Entity-escaped VASTAdTagURI truncated at the first \texttt{\&\cbr{}amp;} (or \texttt{\&\cbr{}\#\cbr{}38;}). & vastlint 0.15.0 upstream \texttt{crates/\cbr{}vastlint-\cbr{}core/\cbr{}src/\cbr{}inspect.\cbrsoft{}rs:\cbrsoft{}110-\cbr{}117,\cbr{} 126,\cbr{} 156-\cbr{}158} (compiled into \texttt{vastlint\_\cbrsoft{}wasm\_\cbrsoft{}bg.\cbrsoft{}wasm}); caller \texttt{dist/\cbr{}session.\cbrsoft{}js:\cbrsoft{}461,\cbr{} 477} & not in the swept catalog \\
F58 & W19 (should), W20 (should), W21 (should) & followAdditionalWrappers and allowMultipleAds never read. & \texttt{dist/\cbr{}resolved.\cbrsoft{}js:\cbrsoft{}457-\cbr{}459} & all 12 (0.4.24 to 0.15.0) \\
F59 & W23 (should) & A response that holds a Wrapper ad and a stand-alone InLine ad is classified as an InLine document; the Wrapper is never followed, and its target (noad.xml) is never requested. & \texttt{dist/\cbr{}session.\cbrsoft{}js:\cbrsoft{}461} (\texttt{build\cbrsoft{}Hop} takes the document type from \texttt{inspect\cbrsoft{}Document}) and :649-651 (a hop that is not a Wrapper ends resolution as resolved) & all 12 (0.4.24 to 0.15.0) \\
F60 & T05 (must), T06 (must), T07 (must), T21 (must), T20 (should) & Documented controller fires quartiles only: no progress, loaded or rewind. & \texttt{dist/\cbr{}playback-\cbr{}queue.\cbrsoft{}js:\cbrsoft{}381-\cbr{}389} & all 12 (0.4.24 to 0.15.0) \\
F61 & M03 (must), M06 (must), M07 (should) & Only CACHEBUSTING and TIMESTAMP expanded; controller takes no macros. & \texttt{dist/\cbr{}tracking.\cbrsoft{}js:\cbrsoft{}169-\cbr{}181} & all 12 (0.4.24 to 0.15.0) \\
F62 & M12 (must) & Requests a Wrapper's VASTAdTagURI as written. & \texttt{dist/\cbr{}session.\cbrsoft{}js:\cbrsoft{}659} and :674 (\texttt{fetch\cbrsoft{}Xml\cbrsoft{}From\cbrsoft{}Url} requests the URL as found) & all 12 (0.4.24 to 0.15.0) \\*
F63 & none: delegated limit & After a mid-chain fetch failure the snapshot keeps only the root hop; intermediate Wrapper Error URIs unreachable. & \texttt{dist/\cbr{}session.\cbrsoft{}js:\cbrsoft{}512-\cbr{}514,\cbr{} 629,\cbr{} 674,\cbr{} 682-\cbr{}689,\cbr{} 700-\cbr{}702} & tested release only \\
\end{longtable}

\endgroup

\begingroup\small\setlength{\tabcolsep}{4pt}\Needspace{15\baselineskip}

\begin{longtable}[]{@{}
  >{\raggedright\arraybackslash}p{(\columnwidth - 6\tabcolsep) * \real{0.0430}}
  >{\raggedright\arraybackslash}p{(\columnwidth - 6\tabcolsep) * \real{0.2689}}
  >{\raggedright\arraybackslash}p{(\columnwidth - 6\tabcolsep) * \real{0.3297}}
  >{\raggedright\arraybackslash}p{(\columnwidth - 6\tabcolsep) * \real{0.3584}}@{}}
\caption{\textbf{Table B6.} Findings in Shaka Player 5.2.12 (not swept).}\tabularnewline
\toprule\noalign{}
\begin{minipage}[b]{\linewidth}\raggedright
ID
\end{minipage} & \begin{minipage}[b]{\linewidth}\raggedright
Requirements
\end{minipage} & \begin{minipage}[b]{\linewidth}\raggedright
Symptom
\end{minipage} & \begin{minipage}[b]{\linewidth}\raggedright
Location
\end{minipage} \\
\midrule\noalign{}
\endfirsthead
\multicolumn{4}{@{}l}{\textbf{Table B6} (continued)}\tabularnewline
\toprule\noalign{}
\begin{minipage}[b]{\linewidth}\raggedright
ID
\end{minipage} & \begin{minipage}[b]{\linewidth}\raggedright
Requirements
\end{minipage} & \begin{minipage}[b]{\linewidth}\raggedright
Symptom
\end{minipage} & \begin{minipage}[b]{\linewidth}\raggedright
Location
\end{minipage} \\
\midrule\noalign{}
\endhead
\bottomrule\noalign{}
\endlastfoot
F64 & W01 (must), W02 (must) & Every \texttt{\textless{}Ad\textgreater{}} without an InLine is skipped: no Wrapper request, no ad, no error. & \texttt{lib/\cbr{}ads/\cbr{}vast\_\cbrsoft{}interstitial\_\cbrsoft{}parser.\cbrsoft{}js:\cbrsoft{}29-\cbr{}33} \\
F65 & M01 (must), M02 (must), M03 (must), M06 (must), M07 (should), M11 (must) & No macro replaced in tracking or Error URLs. & absent: no CACHEBUSTING, TIMESTAMP, ERRORCODE, ASSETURI, ADSERVINGID or UNIVERSALADID in lib/ads/ (22 files); \texttt{lib/\cbr{}ads/\cbr{}ad\_\cbrsoft{}utils.\cbrsoft{}js:\cbrsoft{}173-\cbr{}176} \\
F66 & E01 (must), E02 (must) & A no-ad response's document-level Error URIs are never requested. & \texttt{lib/\cbr{}ads/\cbr{}ad\_\cbrsoft{}utils.\cbrsoft{}js:\cbrsoft{}49}; \texttt{lib/\cbr{}ads/\cbr{}vast\_\cbrsoft{}interstitial\_\cbrsoft{}parser.\cbrsoft{}js:\cbrsoft{}29-\cbr{}33}; absent: no read of the \texttt{\textless{}Error\textgreater{}} children of \texttt{\textless{}VAST\textgreater{}} in lib/ads/ \\
F67 & E03 (must), E05 (must), E06 (must) & The error code is never determined: Error URIs are requested with a literal {[}ERRORCODE{]}. & \texttt{lib/\cbr{}ads/\cbr{}ad\_\cbrsoft{}utils.\cbrsoft{}js:\cbrsoft{}149-\cbr{}151,\cbr{} 173-\cbr{}176}; \texttt{lib/\cbr{}ads/\cbr{}interstitial\_\cbrsoft{}ad\_\cbrsoft{}manager.\cbrsoft{}js:\cbrsoft{}1226-\cbr{}1236}; absent: no ERRORCODE in lib/ads/ \\
F68 & E07 (must) & A media file that does not answer is never reported (no media timeout). & absent: no media-load timeout in \texttt{lib/\cbr{}ads/\cbr{}interstitial\_\cbrsoft{}ad\_\cbrsoft{}manager.\cbrsoft{}js} \\
F69 & T10 (should) & Impression and start fire when the ad is set up, before its media loads. & \texttt{lib/\cbr{}ads/\cbr{}interstitial\_\cbrsoft{}ad\_\cbrsoft{}manager.\cbrsoft{}js:\cbrsoft{}1262-\cbr{}1263} \\
F70 & T05 (must), T06 (must), T07 (must), T20 (should), T21 (must) & No progress, loaded or rewind events: the tracking table has no such keys. & \texttt{lib/\cbr{}ads/\cbr{}ad\_\cbrsoft{}utils.\cbrsoft{}js:\cbrsoft{}22-\cbr{}38,\cbr{} 95-\cbr{}109} \\
F71 & T18 (should) & A percentage skipoffset is not parsed; the ad is not skippable. & \texttt{lib/\cbr{}ads/\cbr{}vast\_\cbrsoft{}interstitial\_\cbrsoft{}parser.\cbrsoft{}js:\cbrsoft{}81-\cbr{}87} \\
F72 & T23 (should) & A volume change fires unmute. & \texttt{lib/\cbr{}ads/\cbr{}ad\_\cbrsoft{}utils.\cbrsoft{}js:\cbrsoft{}161} \\
F73 & S01 (should) & MediaFiles sorted by height; first used without checking type (FLV fails). & \texttt{lib/\cbr{}ads/\cbr{}vast\_\cbrsoft{}interstitial\_\cbrsoft{}parser.\cbrsoft{}js:\cbrsoft{}117-\cbr{}138} \\
F74 & P03 (must) & Each pod ad's complete pixel is sent to the next ad. & \texttt{lib/\cbr{}ads/\cbr{}interstitial\_\cbrsoft{}ad\_\cbrsoft{}manager.\cbrsoft{}js:\cbrsoft{}1237-\cbr{}1244,\cbr{} :\cbrsoft{}892} and :2561-2567 \\
F75 & P01 (must) & Pods play in document order: the sequence attribute is never read. & \texttt{lib/\cbr{}ads/\cbr{}vast\_\cbrsoft{}interstitial\_\cbrsoft{}parser.\cbrsoft{}js:\cbrsoft{}29-\cbr{}62} \\
F76 & P05 (should) & A failed pod ad is not replaced by the stand-alone ad; that ad plays after the pod. & \texttt{lib/\cbr{}ads/\cbr{}interstitial\_\cbrsoft{}ad\_\cbrsoft{}manager.\cbrsoft{}js:\cbrsoft{}1133-\cbr{}1140,\cbr{} 1207-\cbr{}1211}; absent: no code in lib/ads/ handles stand-alone ads \\
F77 & I01 (must), I03 (should) & Icons not drawn; trackers never fire. & absent: no ``icon'' (any case) in shaka-player 5.2.12 lib/ads/; \texttt{lib/\cbr{}ads/\cbr{}vast\_\cbrsoft{}interstitial\_\cbrsoft{}parser.\cbrsoft{}js:\cbrsoft{}39-\cbr{}61} reads Linear and NonLinearAds only \\*
F78 & A01 (must) & AdVerifications never parsed: no verificationNotExecuted request. & \texttt{lib/\cbr{}ads/\cbr{}vast\_\cbrsoft{}interstitial\_\cbrsoft{}parser.\cbrsoft{}js:\cbrsoft{}29-\cbr{}45} (an InLine yields Creatives only); \texttt{lib/\cbr{}ads/\cbr{}ad\_\cbrsoft{}utils.\cbrsoft{}js:\cbrsoft{}44-\cbr{}69} (\texttt{create\cbrsoft{}Tracking\cbrsoft{}From\cbrsoft{}Inline} reads Error and Impression only); absent: no ``verification'' in lib/ads/ \\
\end{longtable}

\endgroup

\begingroup\small\setlength{\tabcolsep}{4pt}\Needspace{20\baselineskip}

\begin{longtable}[]{@{}
  >{\raggedright\arraybackslash}p{(\columnwidth - 6\tabcolsep) * \real{0.0430}}
  >{\raggedright\arraybackslash}p{(\columnwidth - 6\tabcolsep) * \real{0.2976}}
  >{\raggedright\arraybackslash}p{(\columnwidth - 6\tabcolsep) * \real{0.3297}}
  >{\raggedright\arraybackslash}p{(\columnwidth - 6\tabcolsep) * \real{0.3297}}@{}}
\caption{\textbf{Table B7.} Findings in Google IMA HTML5 SDK 3.794.0 (closed comparison, black-box; not swept).}\tabularnewline
\toprule\noalign{}
\begin{minipage}[b]{\linewidth}\raggedright
ID
\end{minipage} & \begin{minipage}[b]{\linewidth}\raggedright
Requirements
\end{minipage} & \begin{minipage}[b]{\linewidth}\raggedright
Symptom
\end{minipage} & \begin{minipage}[b]{\linewidth}\raggedright
Observable evidence (requests logged; public API)
\end{minipage} \\
\midrule\noalign{}
\endfirsthead
\multicolumn{4}{@{}l}{\textbf{Table B7} (continued)}\tabularnewline
\toprule\noalign{}
\begin{minipage}[b]{\linewidth}\raggedright
ID
\end{minipage} & \begin{minipage}[b]{\linewidth}\raggedright
Requirements
\end{minipage} & \begin{minipage}[b]{\linewidth}\raggedright
Symptom
\end{minipage} & \begin{minipage}[b]{\linewidth}\raggedright
Observable evidence (requests logged; public API)
\end{minipage} \\
\midrule\noalign{}
\endhead
\bottomrule\noalign{}
\endlastfoot
F79 & W03 (must) & Rejects a chain of five Wrappers. & five Wrapper documents fetched, no InLine; Error URIs of all five requested with code 302; AdError 302 ``The maximum number of VAST wrappers (4) has been reached.'' \\
F80 & W17 (must) & Wrapper ClickThrough not used. & after a click on the ad, no click-tracking request and no landing page; probe PRB3 (InLine without ClickThrough): the click pauses the ad and sends no click-tracking request \\
F81 & W20 (should), W21 (should) & A Wrapper target with a pod and a stand-alone ad is rejected with IMA code 2. & start.xml and pod.xml fetched, no Error URI requested, AdError 2 ``Ad source cannot contain playlist response.'' \\
F82 & W23 (should) & After a Wrapper that leads to no ad (303), no other ad is tried. & Error URIs requested with 303, AdError 900, the stand-alone ad of the same response never requested \\
F83 & A03 (should) & The Wrapper's verification vendor is not reported; at most one verificationNotExecuted per ad in these probes. & InLine verificationNotExecuted requested with reason 2, Wrapper's never; probe PRB1: a Wrapper-only vendor is reported; probe PRB2: of two InLine vendors only the first is reported \\
F84 & A05 (must) & A verification resource that answers 404 is not reported. & resource fetched with status 404, no verificationNotExecuted request \\
F85 & I01 (must) & Icon view requested over 50 ms before the ad's start request: Linux Firefox and WebKit, macOS WebKit, some Linux Chromium runs. & icon view requested before the start request in every run; 20 repeats per Linux engine: 36 to 68 ms in Chromium (6 of 20 beyond the 50 ms tolerance; reference runs 122 and 90 ms), 54 to 111 ms in Firefox and 165 to 237 ms in WebKit (all beyond); macOS, 10 repeats per browser: 11 to 12 ms in Chromium, 10 to 13 ms in Chrome and 13 to 50 ms in Firefox (none beyond), 51 to 58 ms in WebKit (all beyond) \\*
F86 & W11 (must) & Firefox: 303 for a malformed Wrapper target. & Firefox: Error URI requested with code 303; Chromium, Chrome and WebKit: 100 \\
\end{longtable}

\endgroup

\hypertarget{appendix-c.-deviating-cells-per-configuration}{%
\section{Appendix C. Deviating cells per configuration}\label{appendix-c.-deviating-cells-per-configuration}}

Node.js layer and Chromium 141. Requirement texts are in Appendix A; the findings behind the cells are in Appendix B; per-cell details are in Online Resource 5.

\begingroup\small\setlength{\tabcolsep}{4pt}\Needspace{12\baselineskip}

\begin{longtable}[]{@{}
  >{\raggedright\arraybackslash}p{(\columnwidth - 8\tabcolsep) * \real{0.2408}}
  >{\raggedright\arraybackslash}p{(\columnwidth - 8\tabcolsep) * \real{0.1168}}
  >{\raggedright\arraybackslash}p{(\columnwidth - 8\tabcolsep) * \real{0.2408}}
  >{\raggedright\arraybackslash}p{(\columnwidth - 8\tabcolsep) * \real{0.2408}}
  >{\raggedright\arraybackslash}p{(\columnwidth - 8\tabcolsep) * \real{0.1606}}@{}}
\caption{\textbf{Table C1.} Deviating cells per configuration, by graded strength (Node.js layer and Chromium 141).}\tabularnewline
\toprule\noalign{}
\begin{minipage}[b]{\linewidth}\raggedright
Configuration
\end{minipage} & \begin{minipage}[b]{\linewidth}\raggedright
Layer
\end{minipage} & \begin{minipage}[b]{\linewidth}\raggedright
must
\end{minipage} & \begin{minipage}[b]{\linewidth}\raggedright
should
\end{minipage} & \begin{minipage}[b]{\linewidth}\raggedright
observation
\end{minipage} \\
\midrule\noalign{}
\endfirsthead
\multicolumn{5}{@{}l}{\textbf{Table C1} (continued)}\tabularnewline
\toprule\noalign{}
\begin{minipage}[b]{\linewidth}\raggedright
Configuration
\end{minipage} & \begin{minipage}[b]{\linewidth}\raggedright
Layer
\end{minipage} & \begin{minipage}[b]{\linewidth}\raggedright
must
\end{minipage} & \begin{minipage}[b]{\linewidth}\raggedright
should
\end{minipage} & \begin{minipage}[b]{\linewidth}\raggedright
observation
\end{minipage} \\
\midrule\noalign{}
\endhead
\bottomrule\noalign{}
\endlastfoot
vast-client-js 6.5.0 & Node.js & W11, M12, P07 & M09, M10 & R01 \\
video-ad-sdk 4.0.1 & Node.js & W07, W17, E03, T07, M01, M03, P01, E09, A01, M06, M12, X11, P07 & W20, W21, T02, T20, T18, A03, M07, P06 & none \\
iab-vast-loader 2.5.1, defaults & Node.js & W02, W11, X05, X06, M12, P07 & W19 & none \\
iab-vast-loader 2.5.1, \texttt{no\cbrsoft{}Single\cbrsoft{}Ad\cbrsoft{}Pods} & Node.js & W11, X05, X06, M12, P07 & W19 & none \\
vastlint-client 0.15.0 & Node.js & W03, W11, T05, T06, T07, M03, X02, T21, M06, M12, X17 & W19, W20, W21, W23, T20, M07, X09, X18, X19 & none \\
vast-client-js 6.5.0 & Chromium & W11, M12, P07 & M09, M10 & R01 \\
video-ad-sdk 4.0.1 & Chromium & W07, W11, W17, E03, T07, M01, M03, P01, E05, E06, E09, A01, M06, M12, X11, P07 & W20, W21, T02, T20, T18, A03, M07, P06, I03 & none \\
iab-vast-loader 2.5.1, defaults & Chromium & W02, X05, X06, M12, P07 & W19 & none \\
iab-vast-loader 2.5.1, \texttt{no\cbrsoft{}Single\cbrsoft{}Ad\cbrsoft{}Pods} & Chromium & X05, X06, M12, P07 & W19 & none \\
Shaka Player 5.2.12 & Chromium & W01, W02, E01, E02, E03, T05, T06, T07, M01, M02, M03, P01, E05, E06, E07, T21, A01, M06, M11, P03, I01 & T10, T23, T20, T18, S01, M07, P05, I03 & none \\
rmp-vast 17.2.0, defaults & Chromium & W03, W11, M01, M03, E06, T13, T21, A01, M06, M11, M12, P07 & W20, W21, T02, W23, T23, A03, M09, P05, I03 & R01 \\*
rmp-vast 17.2.0, \texttt{omid\cbrsoft{}Support} & Chromium & W03, W11, M01, M03, E06, T13, T21, M06, M11, M12, P07 & W20, W21, T02, W23, T23, M09, P05, I03 & R01 \\
\end{longtable}

\endgroup

In rmp-vast's rows, M12, P07, W11 and R01 are attributed to its bundled vast-client-js (rule G4).

\hypertarget{appendix-d.-mutation-fuzzing-root-causes}{%
\section{Appendix D. Mutation fuzzing: root causes}\label{appendix-d.-mutation-fuzzing-root-causes}}

The campaign used 139 seeds in 5 rounds and produced 1,442 cases (1,437 unique), 12 classes and 9 root causes. video-ad-sdk locations here are in the TypeScript sources shipped in its package (\texttt{src/}), where the fuzzing scripts attributed them; Appendix B gives the matching \texttt{dist/} code that ran. Minimal documents and reproduction scripts are in the replication package.

\begingroup\small\setlength{\tabcolsep}{4pt}\Needspace{13\baselineskip}

\begin{longtable}[]{@{}
  >{\raggedright\arraybackslash}p{(\columnwidth - 12\tabcolsep) * \real{0.1056}}
  >{\raggedright\arraybackslash}p{(\columnwidth - 12\tabcolsep) * \real{0.1820}}
  >{\raggedright\arraybackslash}p{(\columnwidth - 12\tabcolsep) * \real{0.1062}}
  >{\raggedright\arraybackslash}p{(\columnwidth - 12\tabcolsep) * \real{0.1056}}
  >{\raggedright\arraybackslash}p{(\columnwidth - 12\tabcolsep) * \real{0.1365}}
  >{\raggedright\arraybackslash}p{(\columnwidth - 12\tabcolsep) * \real{0.1669}}
  >{\raggedright\arraybackslash}p{(\columnwidth - 12\tabcolsep) * \real{0.1972}}@{}}
\caption{\textbf{Table D1.} Mutation fuzzing root causes, with classes, unique cases and source locations.}\tabularnewline
\toprule\noalign{}
\begin{minipage}[b]{\linewidth}\raggedright
Root cause
\end{minipage} & \begin{minipage}[b]{\linewidth}\raggedright
Client
\end{minipage} & \begin{minipage}[b]{\linewidth}\raggedright
Classes
\end{minipage} & \begin{minipage}[b]{\linewidth}\raggedright
Unique cases
\end{minipage} & \begin{minipage}[b]{\linewidth}\raggedright
Raised by an unmutated seed
\end{minipage} & \begin{minipage}[b]{\linewidth}\raggedright
Attribution
\end{minipage} & \begin{minipage}[b]{\linewidth}\raggedright
Location
\end{minipage} \\
\midrule\noalign{}
\endfirsthead
\multicolumn{7}{@{}l}{\textbf{Table D1} (continued)}\tabularnewline
\toprule\noalign{}
\begin{minipage}[b]{\linewidth}\raggedright
Root cause
\end{minipage} & \begin{minipage}[b]{\linewidth}\raggedright
Client
\end{minipage} & \begin{minipage}[b]{\linewidth}\raggedright
Classes
\end{minipage} & \begin{minipage}[b]{\linewidth}\raggedright
Unique cases
\end{minipage} & \begin{minipage}[b]{\linewidth}\raggedright
Raised by an unmutated seed
\end{minipage} & \begin{minipage}[b]{\linewidth}\raggedright
Attribution
\end{minipage} & \begin{minipage}[b]{\linewidth}\raggedright
Location
\end{minipage} \\
\midrule\noalign{}
\endhead
\bottomrule\noalign{}
\endlastfoot
C1 & vast-client-js 6.5.0 & 1 & 24 & yes & library & \texttt{src/\cbr{}fetcher/\cbr{}url\_\cbrsoft{}handler.\cbrsoft{}js:\cbrsoft{}57-\cbr{}64} \\
C2 & vast-client-js 6.5.0 & 1 & 21 & no & library & \texttt{src/\cbr{}parser/\cbr{}ad\_\cbrsoft{}parser.\cbrsoft{}js:\cbrsoft{}20-\cbr{}59} \\
C3 & video-ad-sdk 4.0.1 & 1 & 18 & yes & library & \texttt{src/\cbr{}ad\cbrsoft{}Unit/\cbr{}helpers/\cbr{}vpaid/\cbr{}load\cbrsoft{}Creative.\cbrsoft{}ts:\cbrsoft{}19} \\
C4 & vast-client-js 6.5.0 & 1 & 1 & no & library & \texttt{src/\cbr{}fetcher/\cbr{}url\_\cbrsoft{}handler.\cbrsoft{}js:\cbrsoft{}66-\cbr{}72} \\
C5 & iab-vast-loader 2.5.1 & 1 & 7 & no & library & \texttt{iab-\cbr{}vast-\cbr{}parser/\cbr{}lib/\cbr{}parse.\cbrsoft{}js:\cbrsoft{}39} \\
C6 & iab-vast-loader 2.5.1 & 1 & 36 & yes & library & \texttt{iab-\cbr{}vast-\cbr{}parser/\cbr{}lib/\cbr{}xml/\cbr{}unmarshaler.\cbrsoft{}js:\cbrsoft{}75} \\
C7 & iab-vast-loader 2.5.1 & 1 & 32 & yes & library & \texttt{iab-\cbr{}vast-\cbr{}parser/\cbr{}lib/\cbr{}factory/\cbr{}vast.\cbrsoft{}js:\cbrsoft{}38} \\
C8 & iab-vast-loader 2.5.1 & 4 & 13 & no & library & \texttt{iab-\cbr{}vast-\cbr{}parser/\cbr{}lib/\cbr{}inherit/\cbr{}ad.\cbrsoft{}js:\cbrsoft{}88} \\*
C9 & video-ad-sdk 4.0.1 & 1 & 30 & yes & environment artifact & \texttt{src/\cbr{}ad\cbrsoft{}Unit/\cbr{}helpers/\cbr{}icons/\cbr{}add\cbrsoft{}Icons.\cbrsoft{}ts:\cbrsoft{}65} \\
\end{longtable}

\endgroup

\hypertarget{references}{%
\section{References}\label{references}}

\hypertarget{refs}{}
\begin{CSLReferences}{1}{0}
\leavevmode\vadjust pre{\hypertarget{ref-8bitsten2026issue510}{}}%
8bitsten (2026) {URLHandler} timeout throws an uncaught exception. GitHub issue \#510, dailymotion/vast-client-js. \url{https://github.com/dailymotion/vast-client-js/issues/510}. Accessed 20 Oct 2026

\leavevmode\vadjust pre{\hypertarget{ref-ahasanuzzaman2020longitudinal}{}}%
Ahasanuzzaman M, Hassan S, Bezemer C-P, Hassan AE (2020) A longitudinal study of popular ad libraries in the {Google} {Play} {Store}. Empirical Software Engineering 25:824--858. \url{https://doi.org/10.1007/s10664-019-09766-x}

\leavevmode\vadjust pre{\hypertarget{ref-ahasanuzzaman2022integration}{}}%
Ahasanuzzaman M, Hassan S, Hassan AE (2022) Studying ad library integration strategies of top free-to-download apps. IEEE Transactions on Software Engineering 48:209--224. \url{https://doi.org/10.1109/TSE.2020.2983399}

\leavevmode\vadjust pre{\hypertarget{ref-andrews2003logfile}{}}%
Andrews JH, Zhang Y (2003) General test result checking with log file analysis. IEEE Transactions on Software Engineering 29:634--648. \url{https://doi.org/10.1109/TSE.2003.1214327}

\leavevmode\vadjust pre{\hypertarget{ref-barr2015oracle}{}}%
Barr ET, Harman M, McMinn P, et al (2015) The oracle problem in software testing: A survey. IEEE Transactions on Software Engineering 41:507--525. \url{https://doi.org/10.1109/TSE.2014.2372785}

\leavevmode\vadjust pre{\hypertarget{ref-bohme2018stads}{}}%
Böhme M (2018) {STADS}: Software testing as species discovery. ACM Transactions on Software Engineering and Methodology 27:1--52. \url{https://doi.org/10.1145/3210309}

\leavevmode\vadjust pre{\hypertarget{ref-bradner1997rfc2119}{}}%
Bradner S (1997) Key words for use in {RFCs} to indicate requirement levels. RFC 2119, BCP 14, Internet Engineering Task Force. \url{https://doi.org/10.17487/RFC2119}

\leavevmode\vadjust pre{\hypertarget{ref-brubaker2014frankencerts}{}}%
Brubaker C, Jana S, Ray B, et al (2014) Using {Frankencerts} for automated adversarial testing of certificate validation in {SSL/TLS} implementations. In: 2014 {IEEE} Symposium on Security and Privacy. IEEE, pp 114--129. \url{https://doi.org/10.1109/SP.2014.15}

\leavevmode\vadjust pre{\hypertarget{ref-chao1984nonparametric}{}}%
Chao A (1984) Nonparametric estimation of the number of classes in a population. Scandinavian Journal of Statistics 11:265--270. \url{https://www.jstor.org/stable/4615964}

\leavevmode\vadjust pre{\hypertarget{ref-chao1987capture}{}}%
Chao A (1987) Estimating the population size for capture-recapture data with unequal catchability. Biometrics 43:783--791. \url{https://doi.org/10.2307/2531532}

\leavevmode\vadjust pre{\hypertarget{ref-chaudhary2026pr515}{}}%
Chaudhary P (2026) Bound the response body read by the request timeout. GitHub pull request \#515, dailymotion/vast-client-js. \url{https://github.com/dailymotion/vast-client-js/pull/515}. Accessed 28 Oct 2026

\leavevmode\vadjust pre{\hypertarget{ref-chaudhary2026trigger}{}}%
Chaudhary P, Khandelwal K (2026) Trigger timing, deadline readiness, and event-aligned accounting for dynamic ad insertion. Preprint arXiv:2609.19899. \url{https://doi.org/10.48550/arXiv.2609.19899}

\leavevmode\vadjust pre{\hypertarget{ref-chen2018rfc}{}}%
Chen C, Tian C, Duan Z, Zhao L (2018) {RFC}-directed differential testing of certificate validation in {SSL/TLS} implementations. In: Proceedings of the 40th International Conference on Software Engineering ({ICSE}). ACM, pp 859--870. \url{https://doi.org/10.1145/3180155.3180226}

\leavevmode\vadjust pre{\hypertarget{ref-chen2016jvm}{}}%
Chen Y, Su T, Sun C, et al (2016) Coverage-directed differential testing of {JVM} implementations. In: Proceedings of the 37th {ACM} {SIGPLAN} Conference on Programming Language Design and Implementation ({PLDI}). ACM, pp 85--99. \url{https://doi.org/10.1145/2908080.2908095}

\leavevmode\vadjust pre{\hypertarget{ref-crussell2014madfraud}{}}%
Crussell J, Stevens R, Chen H (2014) {MAdFraud}: Investigating ad fraud in {Android} applications. In: Proceedings of the 12th Annual International Conference on Mobile Systems, Applications, and Services ({MobiSys}). ACM, pp 123--134. \url{https://doi.org/10.1145/2594368.2594391}

\leavevmode\vadjust pre{\hypertarget{ref-dailymotion2026vastclient}{}}%
Dailymotion (2026) {vast-client-js (@dailymotion/vast-client)}. Version 6.5.0. \url{https://github.com/dailymotion/vast-client-js}. Accessed 20 Oct 2026

\leavevmode\vadjust pre{\hypertarget{ref-davis2019regex}{}}%
Davis JC, Michael IV LG, Coghlan CA, et al (2019) Why aren't regular expressions a lingua franca? {An} empirical study on the re-use and portability of regular expressions. In: Proceedings of the 27th {ACM} Joint Meeting on European Software Engineering Conference and Symposium on the Foundations of Software Engineering ({ESEC/FSE}). ACM, pp 443--454. \url{https://doi.org/10.1145/3338906.3338909}

\leavevmode\vadjust pre{\hypertarget{ref-deruiter2015protocol}{}}%
de Ruiter J, Poll E (2015) Protocol state fuzzing of {TLS} implementations. In: 24th {USENIX} Security Symposium ({USENIX} Security 15). USENIX Association, pp 193--206. \url{https://www.usenix.org/conference/usenixsecurity15/technical-sessions/presentation/de-ruiter}. Accessed 20 Oct 2026

\leavevmode\vadjust pre{\hypertarget{ref-doca2019issue331}{}}%
Doca (2019) {VASTAdTagURI} macros not being replaced. GitHub issue \#331, dailymotion/vast-client-js. \url{https://github.com/dailymotion/vast-client-js/issues/331}. Accessed 20 Oct 2026

\leavevmode\vadjust pre{\hypertarget{ref-dong2018frauddroid}{}}%
Dong F, Wang H, Li L, et al (2018) {FraudDroid}: Automated ad fraud detection for {Android} apps. In: Proceedings of the 26th {ACM} Joint Meeting on European Software Engineering Conference and Symposium on the Foundations of Software Engineering ({ESEC/FSE}). ACM, pp 257--268. \url{https://doi.org/10.1145/3236024.3236045}

\leavevmode\vadjust pre{\hypertarget{ref-google2026ima}{}}%
Google (2026) {Interactive Media Ads (IMA) SDK for HTML5}. Version 3.794.0. Served file archived 16 Oct 2026. \url{https://imasdk.googleapis.com/js/sdkloader/ima3.js}. Accessed 16 Oct 2026

\leavevmode\vadjust pre{\hypertarget{ref-gui2015truth}{}}%
Gui J, McIlroy S, Nagappan M, Halfond WGJ (2015) Truth in advertising: The hidden cost of mobile ads for software developers. In: 2015 {IEEE/ACM} 37th {IEEE} International Conference on Software Engineering ({ICSE}). IEEE, pp 100--110. \url{https://doi.org/10.1109/ICSE.2015.32}

\leavevmode\vadjust pre{\hypertarget{ref-harrand2021json}{}}%
Harrand N, Durieux T, Broman D, Baudry B (2021) The behavioral diversity of {Java} {JSON} libraries. In: 2021 {IEEE} 32nd International Symposium on Software Reliability Engineering ({ISSRE}). IEEE, pp 412--422. \url{https://doi.org/10.1109/ISSRE52982.2021.00050}

\leavevmode\vadjust pre{\hypertarget{ref-herbold2023differential}{}}%
Herbold S, Tunkel S (2023) Differential testing for machine learning: An analysis for classification algorithms beyond deep learning. Empirical Software Engineering 28:34. \url{https://doi.org/10.1007/s10664-022-10273-9}

\leavevmode\vadjust pre{\hypertarget{ref-hwde2019issue70}{}}%
hwde (2019) {VAST} request {URI} macro's defined by {IAB} are not replaced for wrappers. GitHub issue \#70, zentrick/iab-vast-loader. \url{https://github.com/zentrick/iab-vast-loader/issues/70}. Accessed 20 Oct 2026

\leavevmode\vadjust pre{\hypertarget{ref-iab2022vast43}{}}%
IAB Technology Laboratory (2022) {Video Ad Serving Template (VAST) 4.3}. Specification. \url{https://github.com/InteractiveAdvertisingBureau/VAST4.x}. Accessed 20 Oct 2026

\leavevmode\vadjust pre{\hypertarget{ref-iab2026vastmacros}{}}%
IAB Technology Laboratory (2026) {VAST} 4 macros (macros-data.json), in the {VAST} repository, commit 99d5d51. \url{https://github.com/InteractiveAdvertisingBureau/vast}. Accessed 20 Oct 2026

\leavevmode\vadjust pre{\hypertarget{ref-jabiyev2021treqs}{}}%
Jabiyev B, Sprecher S, Onarlioglu K, Kirda E (2021) {T-Reqs}: {HTTP} request smuggling with differential fuzzing. In: Proceedings of the 2021 {ACM} {SIGSAC} Conference on Computer and Communications Security ({CCS}). ACM, pp 1805--1820. \url{https://doi.org/10.1145/3460120.3485384}

\leavevmode\vadjust pre{\hypertarget{ref-justinjeong52026pr514}{}}%
justinjeong5 (2026) {{[}util{]}} {Prevent} uncaught exceptions on request timeout. GitHub pull request \#514, dailymotion/vast-client-js. \url{https://github.com/dailymotion/vast-client-js/pull/514}. Accessed 20 Oct 2026

\leavevmode\vadjust pre{\hypertarget{ref-kallus2024httpgarden}{}}%
Kallus B, Anantharaman P, Locasto M, Smith SW (2024) The {HTTP} {Garden}: Discovering parsing vulnerabilities in {HTTP/1.1} implementations by differential fuzzing of request streams. Preprint arXiv:2405.17737. \url{https://doi.org/10.48550/arXiv.2405.17737}

\leavevmode\vadjust pre{\hypertarget{ref-kropp1998ballista}{}}%
Kropp NP, Koopman PJ, Siewiorek DP (1998) Automated robustness testing of off-the-shelf software components. In: Digest of Papers, Twenty-Eighth Annual International Symposium on Fault-Tolerant Computing ({FTCS}-28). IEEE, pp 230--239. \url{https://doi.org/10.1109/FTCS.1998.689474}

\leavevmode\vadjust pre{\hypertarget{ref-kuchta2018pdf}{}}%
Kuchta T, Lutellier T, Wong E, et al (2018) On the correctness of electronic documents: Studying, finding, and localizing inconsistency bugs in {PDF} readers and files. Empirical Software Engineering 23:3187--3220. \url{https://doi.org/10.1007/s10664-018-9600-2}

\leavevmode\vadjust pre{\hypertarget{ref-lima2021javascript}{}}%
Lima I, Silva J, Miranda B, et al (2021) Exposing bugs in {JavaScript} engines through test transplantation and differential testing. Software Quality Journal 29:129--158. \url{https://doi.org/10.1007/s11219-020-09537-8}

\leavevmode\vadjust pre{\hypertarget{ref-marciel2016viewfraud}{}}%
Marciel M, Cuevas R, Banchs A, et al (2016) Understanding the detection of view fraud in video content portals. In: Proceedings of the 25th International Conference on World Wide Web ({WWW}). International World Wide Web Conferences Steering Committee, pp 357--368. \url{https://doi.org/10.1145/2872427.2882980}

\leavevmode\vadjust pre{\hypertarget{ref-mckeeman1998differential}{}}%
McKeeman WM (1998) Differential testing for software. Digital Technical Journal 10:100--107

\leavevmode\vadjust pre{\hypertarget{ref-mendezfernandez2019openscience}{}}%
Méndez Fernández D, Monperrus M, Feldt R, Zimmermann T (2019) The open science initiative of the {Empirical Software Engineering} journal. Empirical Software Engineering 24:1057--1060. \url{https://doi.org/10.1007/s10664-019-09712-x}

\leavevmode\vadjust pre{\hypertarget{ref-mohajerimoghaddam2019watching}{}}%
Mohajeri Moghaddam H, Acar G, Burgess B, et al (2019) Watching you watch: The tracking ecosystem of over-the-top {TV} streaming devices. In: Proceedings of the 2019 {ACM} {SIGSAC} Conference on Computer and Communications Security ({CCS}). ACM, pp 131--147. \url{https://doi.org/10.1145/3319535.3354198}

\leavevmode\vadjust pre{\hypertarget{ref-petsios2017nezha}{}}%
Petsios T, Tang A, Stolfo S, et al (2017) {NEZHA}: Efficient domain-independent differential testing. In: 2017 {IEEE} Symposium on Security and Privacy. IEEE, pp 615--632. \url{https://doi.org/10.1109/SP.2017.27}

\leavevmode\vadjust pre{\hypertarget{ref-prado2026sbom}{}}%
Prado A, Zendra O, Boinot P, Barais O (2026) Mind the gap: How {SBOM} specification ambiguities lead to divergent software bills of materials. {An} empirical tool study. In: Proceedings of {SCORED} 2026 (to appear). Preprint arXiv:2609.19920. \url{https://doi.org/10.48550/arXiv.2609.19920}

\leavevmode\vadjust pre{\hypertarget{ref-radiant2026rmpvast}{}}%
Radiant Media Player (2026) {rmp-vast}. Version 17.2.0. \url{https://github.com/radiantmediaplayer/rmp-vast}. Accessed 20 Oct 2026

\leavevmode\vadjust pre{\hypertarget{ref-ralph2020standards}{}}%
Ralph P, bin Ali N, Baltes S, et al (2020) Empirical standards for software engineering research. Preprint arXiv:2010.03525. \url{https://doi.org/10.48550/arXiv.2010.03525}

\leavevmode\vadjust pre{\hypertarget{ref-rambler2024videoadsdk}{}}%
Rambler Digital Solutions (2024) {video-ad-sdk}. Version 4.0.1. \url{https://www.npmjs.com/package/video-ad-sdk/v/4.0.1}. Accessed 20 Oct 2026

\leavevmode\vadjust pre{\hypertarget{ref-seeliger2016hbbtv}{}}%
Seeliger R, Silhavy D, Pham S, Arbanowski S (2016) Cross-platform ad-insertion using {HbbTV} and {MPEG-DASH}. In: 2016 Asia Pacific Conference on Multimedia and Broadcasting ({APMediaCast}). IEEE, pp 7--13. \url{https://doi.org/10.1109/APMediaCast.2016.7878174}

\leavevmode\vadjust pre{\hypertarget{ref-segura2016metamorphic}{}}%
Segura S, Fraser G, Sanchez AB, Ruiz-Cortés A (2016) A survey on metamorphic testing. IEEE Transactions on Software Engineering 42:805--824. \url{https://doi.org/10.1109/TSE.2016.2532875}

\leavevmode\vadjust pre{\hypertarget{ref-sekowski2026vastxml}{}}%
Sekowski A (2026a) {VAST} {XML} validation at bid-time scale: Latency analysis and integration patterns for programmatic video pipelines. Preprint, ResearchGate. \url{https://doi.org/10.13140/RG.2.2.11404.27520}

\leavevmode\vadjust pre{\hypertarget{ref-sekowski2026vastlintclient}{}}%
Sekowski A (2026b) {vastlint-client}. Version 0.15.0. \url{https://github.com/aleksUIX/vastlint/tree/v0.15.0}. Accessed 27 Oct 2026

\leavevmode\vadjust pre{\hypertarget{ref-shaka2026player}{}}%
Shaka Project (2026) {Shaka Player}. Version 5.2.12. \url{https://github.com/shaka-project/shaka-player}. Accessed 20 Oct 2026

\leavevmode\vadjust pre{\hypertarget{ref-sivakorn2017hvlearn}{}}%
Sivakorn S, Argyros G, Pei K, et al (2017) {HVLearn}: Automated black-box analysis of hostname verification in {SSL/TLS} implementations. In: 2017 {IEEE} Symposium on Security and Privacy. IEEE, pp 521--538. \url{https://doi.org/10.1109/SP.2017.46}

\leavevmode\vadjust pre{\hypertarget{ref-smith2016ott}{}}%
Smith M (2016) Best practices for {OTT} dynamic ad insertion. In: IBC 2016 Conference. {Institution of Engineering and Technology}, Amsterdam, p 45. \url{https://doi.org/10.1049/ibc.2016.0045}

\leavevmode\vadjust pre{\hypertarget{ref-somorovsky2016tls}{}}%
Somorovsky J (2016) Systematic fuzzing and testing of {TLS} libraries. In: Proceedings of the 2016 {ACM} {SIGSAC} Conference on Computer and Communications Security ({CCS}). ACM, pp 1492--1504. \url{https://doi.org/10.1145/2976749.2978411}

\leavevmode\vadjust pre{\hypertarget{ref-tian2019rfc}{}}%
Tian C, Chen C, Duan Z, Zhao L (2019) Differential testing of certificate validation in {SSL/TLS} implementations: An {RFC}-guided approach. ACM Transactions on Software Engineering and Methodology 28:1--37. \url{https://doi.org/10.1145/3355048}

\leavevmode\vadjust pre{\hypertarget{ref-varmarken2020tv}{}}%
Varmarken J, Le H, Shuba A, et al (2020) The {TV} is smart and full of trackers: Measuring smart {TV} advertising and tracking. Proceedings on Privacy Enhancing Technologies 2020(2):129--154. \url{https://doi.org/10.2478/popets-2020-0021}

\leavevmode\vadjust pre{\hypertarget{ref-wang2026sbom}{}}%
Wang C, Wu J, Lyu H, et al (2026) A large scale empirical analysis on the adherence gap between standards and tools in {SBOM}. ACM Transactions on Software Engineering and Methodology 35:1--37. \url{https://doi.org/10.1145/3788692}

\leavevmode\vadjust pre{\hypertarget{ref-wpt2026}{}}%
web-platform-tests contributors (2026) {web-platform-tests}: A cross-browser test suite for the {Web} platform. \url{https://web-platform-tests.org/}. Accessed 21 Oct 2026

\leavevmode\vadjust pre{\hypertarget{ref-wu2026splash}{}}%
Wu S, Wang B, Zhang Y, et al (2026) When ad networks misbehave: Understanding risks of semi-drive-by splash ads. Preprint arXiv:2609.09574; to appear at {ACM} {CCS} 2026. \url{https://arxiv.org/abs/2609.09574}

\leavevmode\vadjust pre{\hypertarget{ref-yang2011csmith}{}}%
Yang X, Chen Y, Eide E, Regehr J (2011) Finding and understanding bugs in {C} compilers. In: Proceedings of the 32nd {ACM} {SIGPLAN} Conference on Programming Language Design and Implementation ({PLDI}). ACM, pp 283--294. \url{https://doi.org/10.1145/1993498.1993532}

\leavevmode\vadjust pre{\hypertarget{ref-zentrick2021iabvast}{}}%
Zentrick (2021) {iab-vast-loader} 2.5.1, {iab-vast-parser} 1.2.1 and {iab-vast-model} 1.0.2. \url{https://github.com/zentrick/iab-vast-loader}. Accessed 20 Oct 2026

\end{CSLReferences}

\end{document}